\documentclass[aps,prb,reprint,superscriptaddress,showpacs,floatfix,noeprint]{revtex4-2}
\usepackage[english]{babel}
\usepackage{amsmath,amssymb}
\usepackage{graphicx}
\usepackage{color}

\let\vec=\mathbf

\newcommand{\eps}{\epsilon^+}
\newcommand{\abs}[1]{\left| #1 \right|}
\newcommand{\mean}[1]{\left\langle #1 \right\rangle}
\newcommand{\sgn}{\mathrm{sgn}}

\begin{document}

%\title{Subgap spectrum of a 2D Abrikosov vortex in the presence of a point impurity}
\title{Large spectral gap and impurity-induced states in a two-dimensional Abrikosov vortex}

\author{A. A. Bespalov} 
\affiliation{Institute for Physics of Microstructures, Russian Academy of Sciences, 603950 Nizhny Novgorod, GSP-105, Russia}
\affiliation{Sirius University of Science and Technology, 1 Olympic Ave, 354340, Sochi, Russia}
\affiliation{National Research Lobachevsky State University of Nizhny Novgorod, 603950 Nizhny Novgorod, Russia}

\author{V. D. Plastovets}
\affiliation{Institute for Physics of Microstructures, Russian Academy of Sciences, 603950 Nizhny Novgorod, GSP-105, Russia}
%\affiliation{Sirius University of Science and Technology, 1 Olympic Ave, 354340, Sochi, Russia}

\begin{abstract}
We study the subgap spectrum of a 2D Abrikosov vortex in an $s$-wave superconductor in the absence and presence of a point impurity. By solving the Eilenberger equations without impurity for two models of the vortex (including a self-consistent one), we find multiple subgap spectral branches. The number of these branches may be arbitrary large provided that the magnetic field screening length is large enough. The quasiclassical spectrum of the vortex has a local gap with a width of the order of the bulk gap and a spatial extent of several coherence lengths. The existence of such gap is the prerequisite for the appearance of discrete impurity-induced states. Within the Gor'kov equations formalism, we find that a single impurity induces up to four discrete quasiparticle states in the vortex. The energies and wavefunctions of the impurity states are calculated for different parameters. We claim that most of the predicted spectral features can be observed in scanning tunnel spectroscopy experiments.
\end{abstract}

\maketitle

\section{Introduction}
\label{sec:intro}

The existence of stable Abrikosov vortices is a hallmark of type-II superconductivity. Vortices define the thermodynamic and transport properties of superconductors in the mixed state \cite{Kopnin-book,Larkin+Ovchinnikov98PRB,YANG+2001SpecificHeat}. To understand, e.g., the dissipation and behavior of the heat capacity in the mixed state, it is essential to know the quasiparticle spectrum in the vicinity of a vortex. 

Theoretical studies of the spectrum of a vortex started with a pioneering work by Caroli, de Gennes and Matricon (CdGM) \cite{CdGM64}. By solving the Bogoliubov-de Gennes (BdG) equations for a three-dimensional (3D) $s$-wave superconductor, they calculated the spectrum of a vortex at low energies $E$: $\abs{E} \ll \Delta_{\infty}$, where $\Delta_{\infty}$ is the bulk value of the superconducting order parameter. When adapting their result to two-dimensional (2D) systems, e.g., layered or thin-film superconductors, one finds that the vortex spectrum is discrete, and the the low-energy levels are given by $\epsilon = l_z \epsilon_0$, where $l_z$ is a modified angular momentum projection that takes half integer values. The interlevel spacing $\epsilon_0$ can be estimated as $\epsilon_0 \sim \Delta_{\infty}^2/\mu$ at not very low temperatures, where $\mu$ is the chemical potential. This result is valid in the limit $\Delta_{\infty} \ll \mu$, and hence $\epsilon_0 \ll \Delta_{\infty}$. 
Kramer and Pesch \cite{KramerPesch} found that at very low temperatures the core shrinks to a size that is much smaller than the coherence length $\xi$, which results in a significant increase of the interlevel spacing $\epsilon_0$. Still, $\epsilon_0$ remains much smaller than the bulk gap. 

Modifications of the CdGM spectrum in two dimensions by point impurities have been studied by Larkin, Ovchinnikov and Koulakov \cite{Larkin+Ovchinnikov98PRB,Koulakov+Larkin99PRB,Koulakov+Larkin99PRB(2)}. They found that the spectrum at low energies comprises two series of equidistant levels, with level spacing $2\epsilon_0$ within each series. This picture holds for impurity concentrations $c_{\mathrm{imp}}$ up to $\xi^{-2}$, and even somewhat larger. Level statistics in the limit $c_{\mathrm{imp}} \gg \xi^{-2}$ at $\abs{E} \ll \Delta_{\infty}$ have been studied in Ref. \cite{Skvortsov+98JETPL}.

Analytical solutions of the BdG equations in the presence of a vortex at energies of the order of $\Delta_{\infty}$ seem beyond reach (unless some serious simplifying assumptions are made \cite{Bardeen+69PR}), however, for such energies the spectrum has been calculated numerically in a number of papers \cite{Berthod2005PRB,Shore_ZeroBias89,Gygi_ZeroBias90,Gygi+91PRB,Hayashi+98PRL,Kato+2000PTP}. A self-consistent numerical study of the vortex spectrum in the presence of one impurity within a discrete tight-binding model for parameters $\Delta_{\infty} \sim \mu$ can be found in Ref. \cite{Han+2000PRB}. A comprehensive study of the effects of a single impurity on the spectrum of a vortex %in an $s$-wave superconductor 
at energies $E \sim \Delta_{\infty}$ has been missing to date.

A powerful method to study spatially inhomogeneous superconducting systems is provided by the quasiclassical approximation, which is represented mainly by the Eilenberger equations \cite{EilenbergerEquation,Kopnin-book}. This approach allows to reduce a 2D or 3D problem to a set of linear ordinary differential equations on classical straight trajectories, for the solution of which an efficient numerical algorithm exists \cite{Schopohl98}. The applicability condition for the Eilenberger equations is that all spatial scales of the system should be much larger than the Fermi wavelength. This includes the coherence length, hence the condition $\Delta_{\infty} \ll \mu$ arises. The Eilenberger equations do not handle properly individual impurities, however, they allow to calculate measurable quantities (current, density of states) averaged over impurity positions. An important drawback of the quasiclassical approximation is that it does not resolve energy scales of the order of $\Delta_{\infty}^2/\mu$, which may result in a continuous quasiparticle spectrum when it should be discrete, like in the case of a 2D vortex. This happens because the discrete orbital momentum $l_z$ in the quasiclassical approximation becomes a continuous parameter. Then, the spectrum of a vortex is represented by the so-called spectral branches: continuous dependencies of energy vs. $l_z$. These dependencies are easier to perceive in the form of energy vs. impact parameter $d = k_F^{-1} l_z$, where $k_F$ is the Fermi wavenumber. The impact parameter is simply the distance from the vortex center to the classical trajectory (with a sign), on which the Eilenberger equations are solved. The CdGM states give rise to the anomalous branch, which has a zero energy at $d = 0$. Such branch exists in all single flux quantum vortices, and it is the only branch that goes from negative to positive energies as $d$ changes from $-\infty$ to $+\infty$ \cite{Volovik93JETPL}. Also, other subgap branches may exist, which are less thoroughly studied. Some considerations of the upper branches can be found in a paper by Kopnin \cite{Kopnin98PRB}. 

Within the Eilenberger equations formalism, the spectrum of an Abrikosov vortex in an $s$-wave superconductor has been studied in a number of papers \cite{KramerPesch,Klein89PRB,Klein90PRB,Ullah+90PRB,Schopohl+95PRB,Rainer+96PRB,Hayashi+97PRB,Eschrig+99PRB,Miranovic+2004PRB}. For some model order parameter profiles in a vortex, even analytical solutions of the Eilenberger equations exist \cite{Schopohl98}.

Scanning tunnel spectroscopy (STS) provides a tool to measure directly the local spectrum on the surface of metals \cite{STM2007Review}. The first measurement of the local density of states in a vortex using STS has been reported by Hess \textit{et al.} \cite{Hess+89PRL}. This experiment was followed by many other studies \cite{STM2007Review}. The typical subgap structure (at energies smaller than $\Delta_{\infty}$) of the quasiparticle spectrum observed in $s$-wave superconductors is as follows: in the center of the vortex there is a peak in the density of states at $E = 0$ \cite{Hess+89PRL} (the so-called zero bias anomaly), which fans out when moving away from the vortex center, so that two position-dependent peaks appear \cite{Hess+90PRL}. These spectral features soon found a theoretical explanation in terms of the contribution to the density of states from the CdGM states, or from the anomalous spectral branch \cite{Shore_ZeroBias89,Klein90PRB,Gygi_ZeroBias90,Gygi+91PRB,Ullah+90PRB}. It is noteworthy that most STS data look as if the spectrum of the vortex is continuous, like the spectrum derived from the quasiclassical theory. Difficulties with resolving discrete CdGM states are partly connected with energy resolution limits of STS due to finite temperatures. Features resembling CdGM states have been found using STS only recently \cite{Chen+2018NatCom} in superconductors with a large ratio $\Delta_{\infty}/\mu \sim 1$ -- see Ref. \cite{Chen+2020PRL} and references therein. Thus, studies of the spectra of Abrikosov vortices even in conventional superconductors remain topical to date. Moreover, in recent years increased interest in vortex spectra has arisen in connection with observations of signatures of Majorana states in vortices in several superconducting compounds \cite{Wang+2018Science,Liu+2018PRX,Machida+2019NatMat,Yuan+2019NatPhys,Kong+2019NatPhys,Zhu+2020Science,Liu+2020NatCom}.

The effects of different degrees of disorder on the spectrum of vortices in $s$-wave superconductors have been studied in the experimental papers \cite{Renner+91,Ning+2010JPhys}. It has been found that with increasing disorder the sub-gap spectral features are blurred and eventually disappear. Such behavior can be explained in terms of the disorder-averaged Eilenberger equations \cite{Miranovic+2004PRB}. This formalism can be applied to a vortex only at relatively large impurity concentrations -- $c_{\mathrm{imp}} \gg \xi^{-2}$. In the superclean limit, when there are only few impurities per vortex area ($\sim \xi^2$), the averaged effect of impurities on measurable quantities might be smaller than the mesoscopic fluctuations of these quantities. 
In view of the availability of experimental techniques allowing precise manipulation of adatoms on metallic surfaces \cite{Eigler+90,Kim+2018Science}, a thorough theoretical examination of effects of individual impurities on spectra of Abrikosov vortices is relevant. 

%In this case, a solution of the equations of the theory of superconductivity (BdG or Gor'kov equations \cite{Kopnin-book}) with given positions of defects is required. To our knowledge, such solutions have been missing to date (except for considerations of some special cases \cite{Larkin+Ovchinnikov98PRB,Koulakov+Larkin99PRB,Koulakov+Larkin99PRB(2),Han+2000PRB}, mentioned above). 

The present paper provides a study of the whole subgap spectrum of a 2D Abrikosov vortex in an $s$-wave superconductor both in the presence and absence of a point impurity. In a sense, we extend the analysis of Larkin, Ovchinnikov and Koulakov \cite{Larkin+Ovchinnikov98PRB,Koulakov+Larkin99PRB,Koulakov+Larkin99PRB(2)} from the energy range $\abs{E} \ll \Delta_{\infty}$ to the range $\abs{E} < \Delta_{\infty}$. On the other hand, the mentioned authors considered fine spectral features on a scale of the order of $\Delta_{\infty}^2/\mu$, which is beyond the energy resolution limit of our partly quasiclassical approach. Thus, the present paper and Refs. \cite{Larkin+Ovchinnikov98PRB,Koulakov+Larkin99PRB,Koulakov+Larkin99PRB(2)} are, in fact, related to different aspects of the same problem. %and cannot be deduced from each other.

Let us outline the structure of the paper and our main results. We start by giving the basic equations in Sec. \ref{sec:basic}. Here, two models of the vortex are introduced: a simplistic coreless vortex, and a more realistic vortex with a self-consistent order parameter profile. The spectral properties obtained within both model are qualitatively similar.
%. Within the first model, the modulus of the order parameter profile in the vortex is assumed constant, so that there is no core. This approximately corresponds to the low-temperature limit \cite{KramerPesch}. The second model is applicable at high temperatures: here, the order parameter profile is determined from the Ginzburg-Landau equations. 

In Sec. \ref{sec:clean} we study a vortex without impurity within the Eilenberger equations formalism. First, some properties of the anomalous spectral branch are derived, which have not been previously reported. Next, upper spectral branches are considered. We prove analytically that if London screening can be neglected (the screening length is infinite), there is an \emph{infinite} number of upper branches. This statement holds for any monotonic order parameter profile $\abs{\Delta (r)}$ in a vortex, where $r$ is the distance from the vortex center. Physically, this somewhat surprising phenomenon is connected with the slow decay of the supervelocity $v_S$ with distance: $v_S \propto r^{-1}$. This results in sufficiently slow variations of the local Doppler shift of the gap edge on a straight quasiparticle trajectory, so that an infinite amount of bound Andreev states appears on such trajectory (provided that the superconductor has no boundaries). The upper branches appear very close to the gap edge -- at energies in the range $\abs{\Delta_{\infty} - E} < 0.03 \Delta_{\infty}$.

In Sec. \ref{sub:LDOS} the local density of states is calculated. We find that the quasiclassical spectrum of a vortex has a position-dependent gap (more precisely, a double-gap symmetric with respect to the Fermi energy) that is much larger that the CdGM minigap $\epsilon_0$. In fact, in the center of the vortex this gap appears at energies right above the zero-bias anomaly and has a width of approximately $\Delta_{\infty}$. The gap has a quite large spatial extent -- it disappears only at a distance of the order of $10\xi$ from the vortex center. Such pronounced spectral feature should be observable in conventional superconductors, given the energy resolution achieved in recent STS experiments \cite{Machida+2019NatMat,Chen+2020PRL}.

Section \ref{sec:ImpStates} is devoted to impurity-induced states. First, we consider a quite general gapped 2D superconducting system with a magnetic or nonmagnetic point impurity. We find that the impurity induces up to four (two per spin projection) discrete quasiparticle states, whose energies are confined to the local gap at the position of the impurity.
Next, for our vortex system we calculate the energies and wavefunctions of impurity states for different impurity positions and scattering phases. We claim that the impurity-induced states should be observable in STS spectra of vortices in $s$-wave superconductors.
Finally, the modification of the local spectral gap due to the impurity is discussed.

Our main results are summarized in the conclusion. Most of our calculations are given in detail in the appendices.

%Here, we want to make a remark concerning the relation of our results to the results of Koulakov, Larkin and Ovchinnikov \cite{Larkin+Ovchinnikov98PRB,Koulakov+Larkin99PRB,Koulakov+Larkin99PRB(2)}. These authors studied within the BdG equations formalism the low-energy part of the vortex spectrum ($\abs{E} \ll \Delta_{\infty}$) in the presence of point impurities. They found impurity-induced modifications of the discrete spectrum on an energy scale of the order of $\Delta_{\infty}^2/\mu$. Our quasiclassical approach does not resolve such small energy scales, however, it allows us to study impurity states within the whole subgap range of energies. Thus, our results and the results from Refs. \cite{Larkin+Ovchinnikov98PRB,Koulakov+Larkin99PRB,Koulakov+Larkin99PRB(2)} are related to different aspects of the same problem and cannot be deduced from each other.

%%%%%%%%%%%%%%%%%%%%%%%%%%%%%%%%%%%%%%%%%%%%%%%%%%%%%%%%%%%%%%%%%%%%%%%%%%%%%%%%%%%%%%%%%%%%%%%%%%%%%%%%%%%%%%%%%%%%%%%%%%%%%%%%%%%%%%%%%%%%%%%%%%%
\section{Basic equations}
\label{sec:basic}

Our analysis is based on the Gor'kov equation for the energy-dependent retarded Green functions $\hat{G}_E(\vec{r},\vec{r}')$ and $\hat{F}_E^{\dagger}(\vec{r},\vec{r}')$:

\begin{eqnarray}
	& \biggl\{  H_0(\vec{r}) + U(\vec{r}-\vec{r}_i) + \hat{\tau}_z [\vec{J}(\vec{r}-\vec{r}_i) \hat{\boldsymbol{\sigma}} - E - i\eps]  & \nonumber \\
	& + \left. \left(
	\begin{array}{cc}
	  0 &  \!\! -\Delta(\vec{r}) \\
		\Delta^*(\vec{r}) & \!\! 0
	\end{array} \right) \right\} \!\! \left(
	\begin{array}{c}
		\hat{G}_E(\vec{r},\vec{r}') \\
		-\hat{F}_E^{\dagger}(\vec{r},\vec{r}')
	\end{array} \right)
	\!\! = \!\! \left( \!\!
	\begin{array}{c}
		\delta(\vec{r} - \vec{r}') \\
		0
	\end{array}  \!\! \right) \!\! . &
	\label{eq:Gorkov}
\end{eqnarray}
Here, 
\begin{equation}
	H_0(\vec{r}) = -\frac{\hbar^2 \nabla^2}{2m}  - \mu = -\frac{\hbar^2 \nabla^2}{2m}  - \frac{\hbar^2 k_F^2}{2m},
	\label{eq:H0_S}
\end{equation}
%
%$\mu$ is the chemical potential, $k_F$ is the Fermi wavenumber, 
$m$ is the electron mass, $\hat{\tau}_z$ is a Pauli matrix in Nambu space, $U(\vec{r})$ is the electrical potential of the impurity positioned at $\vec{r} = \vec{r}_i$, and $\vec{J}(\vec{r})$ is its exchange field, $\hat{\boldsymbol{\sigma}} = \{ \sigma_x,\sigma_y,\sigma_z \}$ are the Pauli matrices, $\eps$ is an infinitely small positive quantity, $\Delta (\vec{r})$ is the superconducting order parameter, and $\delta(\vec{r})$ is the Dirac delta function. The Green functions $\hat{G}_E(\vec{r},\vec{r}')$ and $\hat{F}_E^{\dagger}(\vec{r},\vec{r}')$ are $2 \times 2$ matrices in spin space. For explicit definitions of these functions in terms of electron field operators the reader may refer to Ref. \cite{Bespalov2018}.

In Eq. \eqref{eq:Gorkov} we have not taken into account the magnetic field of the vortex. It is known that in superconductors with the magnetic field screening length much larger than the coherence length the vector potential can be neglected compared to the gradient of $\Delta$ in the vicinity of the vortex core (see Sec. 12.5 in Ref. \cite{Kopnin-book}). This statement is valid if we use the simplest gauge, such that the order parameter has the form
\begin{equation}
	\Delta(\vec{r}) = \abs{\Delta(r)} \frac{x-iy}{\sqrt{x^2 + y^2}},
	\label{eq:Delta}
\end{equation}
where we placed the origin in the center of the vortex. Then we can neglect the magnetic field at $r \lesssim \xi$ in extreme type-II superconductors or in a thin-film geometry.

We will perform detailed calculations of the subgap spectrum of the vortex within two models. In the first model we put $\abs{\Delta(r)} = \mathrm{const}$, to which we refer as the coreless vortex. This model may be relevant at near-zero temperatures, where the order parameter modulus is known to experience a sharp jump at $r \ll \xi$ \cite{KramerPesch}. Consideration from Refs. \cite{Gygi+91PRB} and \cite{Volovik93JETPL_core} show that $\abs{\Delta(r)}$ in fact jumps to a value that is smaller than $\Delta_{\infty}$, and then with increasing $r$ it approaches the asymptotic value $\Delta_{\infty}$ in a more smooth manner. In view of this, we admit that our coreless model is somewhat crude, however, it captures the main qualitative features of the spectrum of a realistic vortex and allows us to obtain some exact analytical results. Our second model of a vortex uses a function $\abs{\Delta(r)}$ obtained by solving the Ginzburg-Landau equation, and thus it is applicable at temperatures close to the superconducting transition temperature. We refer to this model as the vortex with core. Many qualitative results obtained in this paper are valid for a quite general order parameter profile, provided that $\abs{\Delta(r)}$ is a monotonically non-decreasing function, and it has a limit at $r \to \infty$:
\begin{equation}
	\lim_{r\to \infty} \abs{\Delta(r)} = \Delta_{\infty}.
	\label{eq:Delta_inf}
\end{equation}

We will use the Green functions to calculate the local density of states. In particular, the spin-up/spin-down densities of states, $\nu_{\uparrow}$ and $\nu_{\downarrow}$, are given by
\begin{equation}
	\nu_{\sigma}(E,\vec{r}) = \pi^{-1} \mathrm{Im} [ G_{E \sigma\sigma} (\vec{r},\vec{r})] = \pi^{-1} \mathrm{Im} [ G_{ER \sigma\sigma} (\vec{r},\vec{r})],
	\label{eq:nu_general}
\end{equation}
where $\sigma = \uparrow,\downarrow$, and we have defined the regular part of the Green function $\hat{G}_{ER}(\vec{r},\vec{r}')$ by subtracting the logarithmic pecularity from it:
\begin{equation}
	\hat{G}_{ER}(\vec{r},\vec{r}') = \hat{G}_E(\vec{r},\vec{r}') - \frac{m}{\pi \hbar^2} \ln \frac{2}{k_F \abs{\vec{r} - \vec{r}'} e^{\gamma}}.
	\label{eq:GER_2D}
\end{equation}
Here, $\gamma = 0.577...$ is the Euler-Mascheroni constant. The origin of the peculiarity can be understood as follows: for sufficiently small values of $\abs{\vec{r} - \vec{r}'}$, in the left-hand side of Eq. \eqref{eq:Gorkov} all terms can be neglected except for those containing differentiation. Then, the equation for $\hat{G}_E$ takes the form
\begin{equation}
	- \frac{\hbar^2 \nabla^2}{2m} \hat{G}_E(\vec{r},\vec{r}') = \delta (\vec{r} - \vec{r}').
	\label{eq:Poisson}
\end{equation}
This Poisson equation is formally equivalent to the equation for the electric potential of a point charge in 2D, which is known to have a logarithmic peculiarity at $\vec{r} = \vec{r}'$. It should be noted that the peculiarity appears in the real part of the Green function, and subtracting it does not affect the density of states, which is proportional to the imaginary part of $G_{E\sigma\sigma}$.  

%The subtraction procedure given by Eq. \eqref{eq:GER_2D} does not affect the imaginary part of the Green function, which is responsible for the density of states. This justifies Eq. \eqref{eq:nu_general}.

If we consider the system without impurity, the Green functions have no spin structure and hence are scalars, which we denote as $G_E^{(0)}(\vec{r},\vec{r}')$ and $F_E^{(0) \dagger}(\vec{r},\vec{r}')$. It is shown in Appendix \ref{app:quasiclassics} that the Green functions with coinciding coordinates can be written in terms of the quasiclassical Green functions $g_E(\vec{r},\vec{n})$ and $f_E^{\dagger}(\vec{r},\vec{n})$ as follows:
\begin{equation}
	G_{ER}^{(0)}(\vec{r},\vec{r}) = i \pi \nu_0 \int g_E(\vec{r},\vec{n}) \frac{d\vec{n}}{2\pi},
	\label{eq:GER(r,r)}
\end{equation}
\begin{equation}
	F_{E}^{\dagger (0)}(\vec{r},\vec{r}) = i \pi \nu_0 \int f_E^{\dagger}(\vec{r},\vec{n}) \frac{d\vec{n}}{2\pi},
	\label{eq:FE(r,r)}
\end{equation}
where $\nu_0 = m/(2 \pi \hbar^2)$ is the normal density of state per spin projection, $\vec{n}$ is a 2D unit vector, and integration is performed over the directions of $\vec{n}$. The functions $g_E(\vec{r},\vec{n})$ and $f_E^{\dagger}(\vec{r},\vec{n})$ (and $f_E(\vec{r},\vec{n})$) can be determined from the Eilenberger equations \cite{EilenbergerEquation,Kopnin-book}:
\begin{eqnarray}
	& -i\hbar v_F \vec{n} \nabla g_E + \Delta(\vec{r}) f_E^{\dagger} - f_E \Delta^*(\vec{r}) = 0, & \label{eq:Eilenberger_g} \\
	& -i\hbar v_F \vec{n} \nabla f_E - 2(E + i\eps) f_E + 2\Delta(\vec{r}) g_E = 0, & \label{eq:Eilenberger_f} \\
	& i\hbar v_F \vec{n} \nabla f_E^{\dagger} - 2(E + i\eps) f_E^{\dagger} + 2 \Delta^*(\vec{r}) g_E = 0, & \label{eq:Eilenberger_f+}
\end{eqnarray}
where $v_F = \hbar k_F/m$ is the Fermi velocity.

\section{Subgap spectrum in the clean case}
\label{sec:clean}

\subsection{Spectral branches}
\label{sub:branches}

\begin{figure}[ht]
	\centering
		\includegraphics[width = 0.9\linewidth]{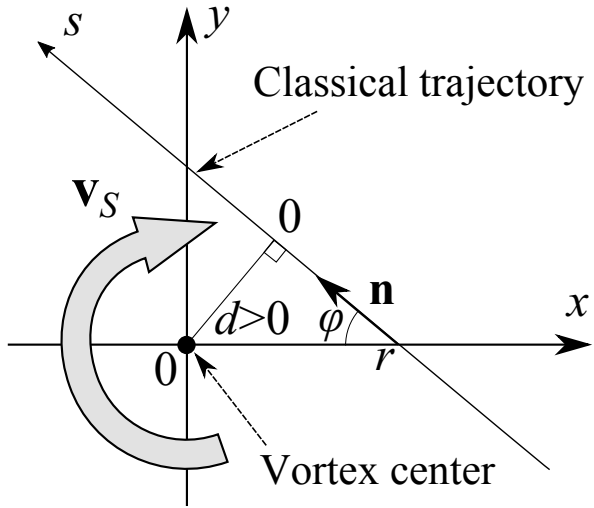}
	\caption{The coordinate system.}
	\label{fig:2DFrame}
\end{figure}

This section is devoted to the spectral properties of a vortex without impurities. First, we will take a closer look at subgap spectral branches -- dependencies of the quasiparticle energy $E$ vs. impact parameter $d$. By definition, the pair $(d,E)$ belongs to a spectral branch, if for these parameters the quasiclassical Green functions have a pole.
We introduce the impact parameter $d$ as in Fig. \ref{fig:2DFrame}, such that $d>0$ for classical trajectories directed towards the supervelocity $\vec{v}_S$, and $d<0$ for trajectories directed along $\vec{v}_S$. Such definition of $d$ provides that the anomalous spectral branch has a positive energy for $d>0$. We introduce a coordinate $s$ on the classical trajectories, such that $s=0$ corresponds to the point that is closest to the vortex center.

To calculate the energies at which the Green functions have poles, a convenient parametrization of these functions is desirable. Schopohl \cite{Schopohl98} found that a parametrization in terms of two complex Riccati amplitudes exists. In our case, the symmetry of the system allows for an even simpler parametrization in terms of one real function $\psi_d(E,s)$ (see Appendix \ref{app:psi}):
\begin{equation}
	g_E(\vec{r},\vec{n}) = i\cot \left( \frac{\psi_d(s)+\psi_d(-s)}{2} + i\eps \right),
	\label{eq:g_psi}
\end{equation}
\begin{equation}
	f_E(\vec{r},\vec{n}) = \frac{i \exp \left( i \frac{\psi_d(s) - \psi_d(-s)}{2} + i \theta(s) \right)}{\sin \left( \frac{\psi_d(s)+\psi_d(-s)}{2} + i\eps \right)},
	\label{eq:f_psi}
\end{equation}
\begin{equation}
	f_E^{\dagger}(\vec{r},\vec{n}) = \frac{i \exp \left( i \frac{\psi_d(-s) - \psi_d(s)}{2} - i \theta(s) \right)}{\sin \left( \frac{\psi_d(s)+\psi_d(-s)}{2} + i\eps \right)},
	\label{eq:f+_psi}
\end{equation}
where $\theta(s)$ is the order parameter phase on the classical trajectory. The function $\psi_d$ satisfies the differential equation
\begin{equation}
	\frac{d\psi_d}{ds} = 2E + \frac{d}{d^2 + s^2} -2 \abs{\Delta\left( \sqrt{s^2 + d^2} \right)} \cos(\psi_d)
	\label{eq:psi}
\end{equation}
with the boundary condition
\begin{equation}
	\psi_d(-\infty) = - \arccos(E).
	\label{eq:bound_psi}
\end{equation}
For brevity here and further we omit $E$ in the list of arguments of $\psi_d$ and use dimensionless energies and lengths: energy is measured in units of $\Delta_{\infty}$, and lengths are measured in units of $\hbar v_F/\Delta_{\infty}$, which is of the order of $\xi$.

It can be seen from Eqs. \eqref{eq:g_psi} - \eqref{eq:f+_psi} that the Green functions has a pole when
\begin{equation}
	\sin\left( \frac{\psi_d(s) + \psi_d(-s)}{2} \right) = 0.
	\label{eq:pole1}
\end{equation}
It follows form Eq. \eqref{eq:psi} that 
\begin{eqnarray}
	& \frac{d}{ds} \left[ \frac{\psi_d(s) + \psi_d(-s)}{2} \right] = 2 \abs{\Delta\left( \sqrt{s^2 + d^2} \right)} & \nonumber \\
	& \times \sin \left( \frac{\psi_d(s) + \psi_d(-s)}{2} \right) \sin \left( \frac{\psi_d(s) - \psi_d(-s)}{2} \right). &
	\label{eq:pole_all_s}
\end{eqnarray}
Hence, if Eq. \eqref{eq:pole1} is satisfied for some $s$, then it holds for all $s$, and Eq. \eqref{eq:pole1} is equivalent to
\begin{equation}
	\psi_d(0) =  \pi n,
	\label{eq:psi(0)}
\end{equation}
where $n$ is an integer. For $d>0$ we denote as $E^{(n)}(d)$ the energies for which Eq. \eqref{eq:psi(0)} is satisfied -- thus, $E^{(n)}(d)$ yields the $n$-th spectral branch. Note that the right-hand side of Eq. \eqref{eq:psi} is discontinuous at $d=0$. To ensure continuity of the functions $E^{(n)}(d)$ at $d=0$, the numbering of the branches has to be shifted for $d<0$, so that the $n$-th branch should be defined by
\begin{equation}
	\psi_d(0) =  \pi (n-1)
	\label{eq:psi(0)_d<0}
\end{equation}
for negative $d$.

Next, we note that $\psi_d(0)$ is a monotonically increasing function of $E$ (see Appendix \ref{app:psi}), which means that $E^{(n+1)}(d)>E^{(n)}(d)$. 

For $E = 0$ and very small positive $d$ one can see that $\psi_d(s) \approx -\pi/2$ for $-s \gg d$. In the vicinity of $s = 0$ one has
\begin{equation}
	\psi_d(s) \approx -\frac{\pi}{2} + \int_{-\infty}^s \frac{d}{d^2 + s^{\prime2}} ds' = \arctan \left( \frac{s}{d} \right),
	\label{eq:psi_small_d}
\end{equation}
so that $\psi_d(0) \approx 0$ and hence $E^{(0)}(0) = 0$. This means that the $0$-th branch should be identified as the anomalous branch.

We will derive two properties of this branch. First, for $d>0$ we have the estimate
\begin{equation}
	E^{(0)}(d) \geq \abs{\Delta(d)} - \frac{1}{2d}.
	\label{eq:E(0)_estimate}
\end{equation}
To prove this, let us assume the opposite. Then we find  that for $E = E^{(0)}(d)$ the function $\psi_d(s)$ cannot cross zero at any $s \leq 0$, because at $\psi_d = 0$
\begin{eqnarray}
	& \frac{d \psi_d}{ds}(s) = 2E^{(0)}(d) + \frac{d}{d^2 + s^2} - 2 \abs{\Delta \left( \sqrt{s^2 + d^2} \right)} & \nonumber \\
	& \leq 2E^{(0)}(d) + \frac{1}{d} - 2\abs{\Delta(d)} <0. &
	\label{eq:wrong2}
\end{eqnarray}
This means that $\psi_d(0) < 0$, so our assumption is wrong. The condition \eqref{eq:E(0)_estimate} has a simple physical interpretation: the energy of an Andreev state in the vortex cannot be lower than the local gap edge estimated by taking into account the local lowering of the gap due to the Doppler shift. 

Another property of the anomalous branch is that $E^{(0)}(d)$ is a monotonically increasing function of $d$, which is proven in Appendix \ref{app:poles}.

The analysis of higher spectral branches is more complicated, however, in some quite general cases their qualitative behavior can be even derived analytically. A general property of the spectrum is that if the pair $(E,d)$ belongs to some spectral branch, then the pair $(-E,-d)$ also belongs to some branch, which is a consequence of the particle-hole symmetry of the quasiclassical approximation. In other words, we have $E^{(-n)}(-d) = - E^{(n)}(d)$. Hence, to obtain a picture of the whole spectrum it is sufficient to calculate its positive part, $E>0$. 

Now, let us consider an order parameter profile with the following asymptotic behavior at $r \to \infty$:
\begin{equation}
	\abs{\Delta(r)} = 1 - h/r^2 + o(r^{-2})
	\label{eq:Delta=1-b/r^2}
\end{equation}
with $h \geq 0$. For such profiles a critical value of the impact parameter exists, which equals
\begin{equation}
	d_{\mathrm{cr}} = 1/4 - 2h,
	\label{eq:d_crit}
\end{equation}
such that for $d<d_{\mathrm{cr}}$ the number of spectral branches with $E>0$ is finite, and for $d>d_{\mathrm{cr}}$ there is an infinite number of branches -- see Appendix \ref{app:qualitative} for a proof. If the difference $1 - \abs{\Delta(r)}$ decays slower than $r^{-2}$ when $r \to \infty$, this can be interpreted as $h \to \infty$ in Eq. \eqref{eq:Delta=1-b/r^2}. Hence, for any monotonically non-decreasing function $\abs{\Delta(r)}$ we have an infinite number of spectral branches for impact parameters $d>1/4$.

Concerning our two vortex models, for the coreless vortex $d_{\mathrm{cr}} = 1/4$, and for the vortex obtained by solving the Ginzburg-Landau equation $d_{\mathrm{cr}} = 0$. Analytical considerations (see Appendix \ref{app:qualitative}) show that for a coreless vortex at $d<1/4$ there are no branches with positive energies, except for the anomalous branch, and numerical calculations for the vortex with core revealed no branches with $E>0$ for $d<0$. The absence of spectral branches with $E>0$ for $d<0$ should not be surprising, because the Doppler shift of the gap edge for trajectories with $d<0$ is positive, and hence the effective local gap can be even larger than $\Delta_{\infty}$, which hampers the formation of subgap states. Thus, the whole picture of the spectrum is strongly asymmetric with respect to the change of sign of $d$, which contradicts the conclusion of Kopnin \cite{Kopnin98PRB}, who found that the upper spectral branches are represented by even functions of $d$. The latter conclusion is a consequence of a mistake in calculations in Ref. \cite{Kopnin98PRB}, consisting in taking the Doppler shift of the quasiparticle energy with the same sign for quasiparticles moving along and towards the supervelocity.

Formally, the infinite number of spectral branches appears due to the slow decay of the supervelocity ($v_s \sim r^{-1}$) and due to the slow asymptotic of the order parameter ($\Delta_{\infty} - \abs{\Delta(r)} \sim r^{-2}$). In turn, these features are connected with the fact that we neglected the screening of the magnetic field. If it is taken into account, at distances form the vortex center that are larger than the magnetic field screening length (being either the London or the Pearl length) the supervelocity decays faster than $r^{-1}$, which also results in a faster decay of the difference $\Delta_{\infty} - \abs{\Delta(r)}$. Then, we will have a finite number of spectral branches, although their number can be arbitrarily large provided that the magnetic field screening length is sufficiently large.

Now we briefly describe the numerical procedure to calculate the spectral branches. To determine $E^{(n)}(d)$ for a given positive $d$ one needs to find such a value of $E$ that Eqs. \eqref{eq:psi}, \eqref{eq:bound_psi} and \eqref{eq:psi(0)} are satisfied. We solve this problem by integrating Eq. \eqref{eq:psi} with the initial condition \eqref{eq:psi(0)} towards negative $s$. When reaching sufficiently large $\abs{s}$, we can determine $\psi_d(-\infty)$. The value $\psi_d(-\infty)$ is a monotonic function of $E$, because the right-hand side of Eq. \eqref{eq:psi} is monotonous in $E$. This allows us to search the value of $E$ which satisfied Eq. \eqref{eq:bound_psi} using a simple bisection method.

\begin{figure}[htb]
	\centering
		\includegraphics[width = \linewidth]{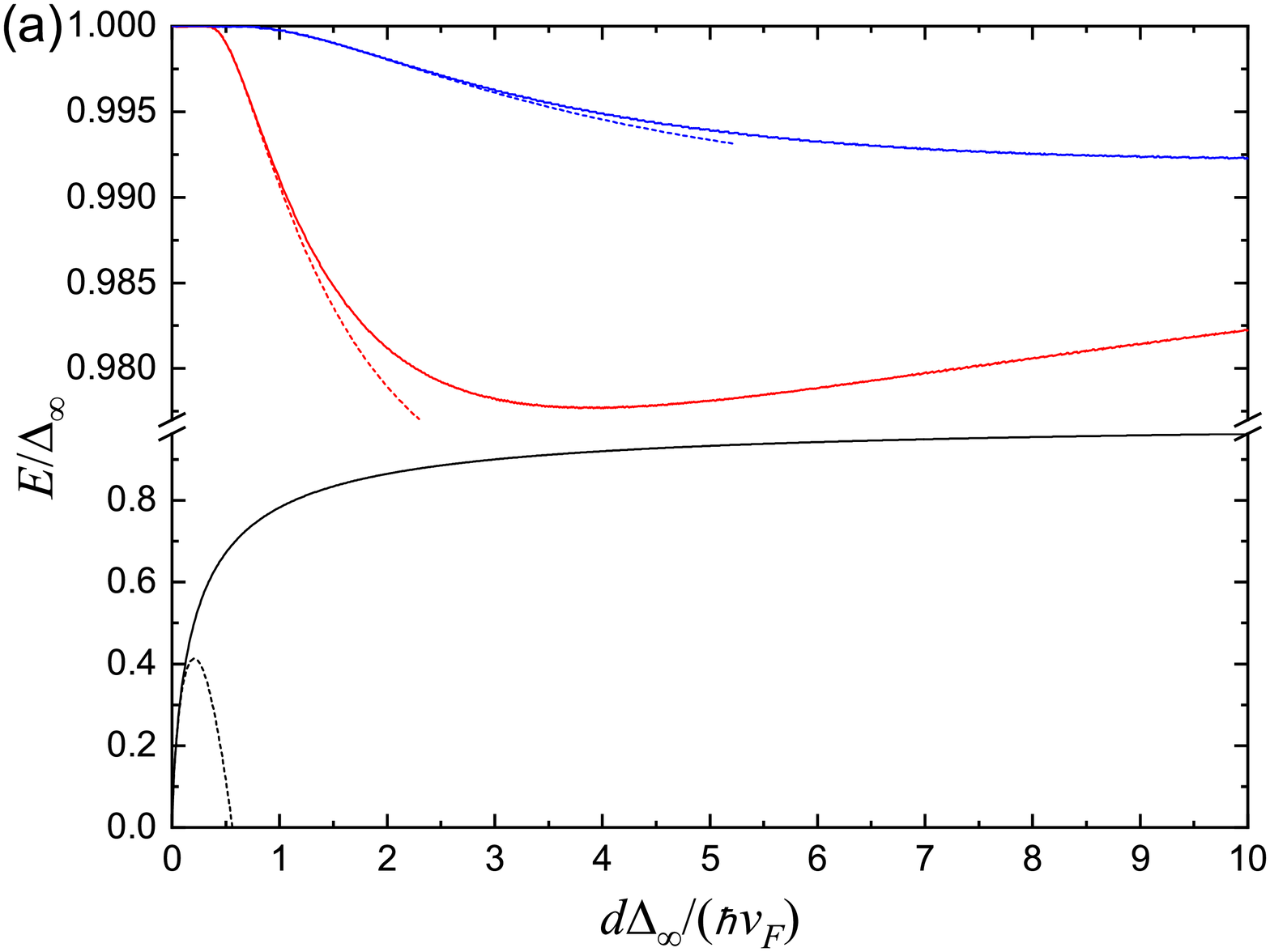}
		\includegraphics[width = \linewidth]{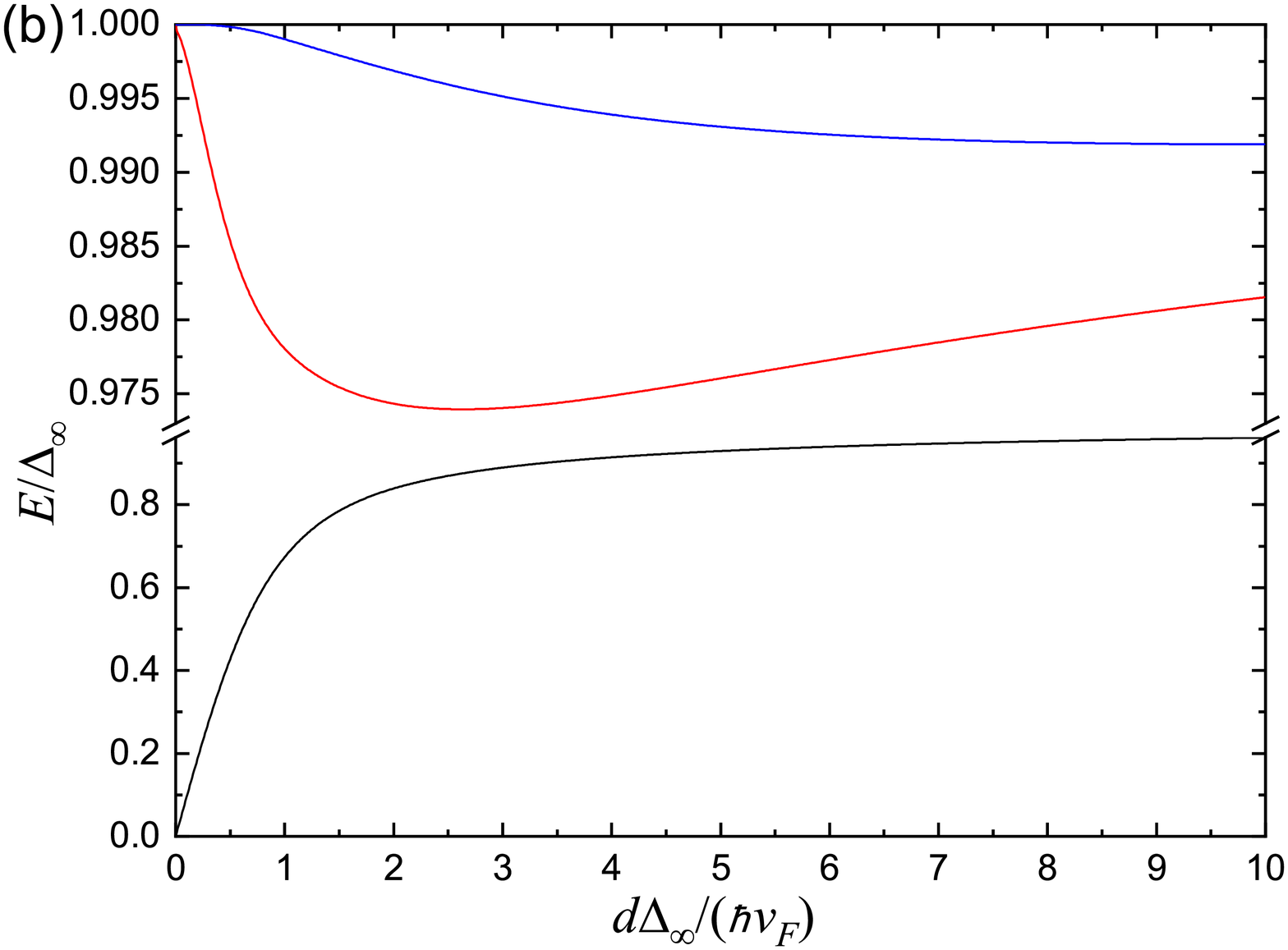}
	\caption{The anomalous spectral branch and two upper branches ($n=1,2$) for a coreless vortex (a) and for a vortex with core (b). The dashed lines correspond to asymptotic expressions \eqref{eq:E0(d)} and \eqref{eq:E(n)_approx}.}
	\label{fig:Branches}
\end{figure}

The calculated spectral braches with numbers $n=0$,1 and 2 are shown in Fig. \ref{fig:Branches}. For the coreless vortex we obtained some asymptotic expressions for $E^{(n)}(d)$, which we write down here without derivation. For the anomalous branch in the limit $\abs{d} \ll 1$ we have
\begin{equation}
	E^{(0)}(d) \approx 2d (\ln \abs{d}^{-1} - \gamma).
	\label{eq:E0(d)}
\end{equation}
For branches with $n>0$ in the limit $d - 1/4 \ll 1$ we found
\begin{eqnarray}
	& E^{(n)}(d) = 1 - \frac{8}{d^2} \exp \left( \frac{2}{\sqrt{4d - 1}} \left[ 2 \arg \Gamma \left( 1 + \frac{i}{2} \sqrt{4d-1} \right) \right. \right. & \nonumber\\
	& \left. \left. + \arccos \left( 1 - \frac{1}{2d} \right) - \pi(n+1) \right] \right), &
	\label{eq:E(n)_approx}
\end{eqnarray}
where $\Gamma(z)$ is the gamma function. Graphs of Eqs. \eqref{eq:E0(d)} and \eqref{eq:E(n)_approx} are shown in Fig. \ref{fig:Branches}a.

Finally, we note that the approach to the calculation of the spectral branches using Eqs. \eqref{eq:psi}, \eqref{eq:bound_psi} and \eqref{eq:psi(0)} is equivalent to the method used in Ref. \cite{Bardeen+69PR}. However, in this paper only the anomalous branch has been studied and other order parameter profiles have been used (step-like and hyperbolic tangent-like), so a quantitative comparison with our results is not possible.

\subsection{Green functions and local density of states}
\label{sub:LDOS}

In this Section we will calculate the Green functions with coinciding arguments and the local density of states.

Let us start with the Green functions. We take a point on the $x$-axis: $\vec{r} = (r,0)$. We parametrize the vector $\vec{n}$ in Eqs. \eqref{eq:GER(r,r)} and \eqref{eq:FE(r,r)} by an angle $\varphi$, so that $\vec{n} = (-\cos \varphi, \sin \varphi)$ -- see Fig. \ref{fig:2DFrame}. For our fixed position $\vec{r}$ the parameters $d$ and $s$ are given by
\begin{equation}
	d(\varphi) = r \sin \varphi, \qquad s(\varphi) = -r \cos \varphi.
	\label{eq:sd_2D}
\end{equation}
Using these relations and Eqs. \eqref{eq:GER(r,r)}, \eqref{eq:FE(r,r)}, \eqref{eq:g_psi} and \eqref{eq:f+_psi}, taking into account that $\theta(s) = 0$ on the positive $x$-axis, we obtain
\begin{equation}
	G_{ER}^{(0)}(\vec{r},\vec{r}) =  -\nu_0 \int_{-\pi/2}^{\pi/2} \cot \left( \frac{\psi_{d}(s) + \psi_{d}(-s)}{2} + i\eps \right) d\varphi,
	\label{eq:GER(r,r)_psi}
\end{equation}
\begin{equation}
	F_{E}^{\dagger (0)}(\vec{r},\vec{r}) = -\nu_0 \int_{-\pi/2}^{\pi/2} \frac{\cos \left(\frac{\psi_d(s) - \psi_d(-s)}{2} \right)} {\sin \left( \frac{\psi_d(s) + \psi_d(-s)}{2} + i\eps \right)} d \varphi.
	\label{eq:FE(r,r)_psi}
\end{equation}
We assume here that $d = d(\varphi)$ and $s = s(\varphi)$.

\begin{figure}[ht]
	\centering
		\includegraphics[width=0.9\linewidth]{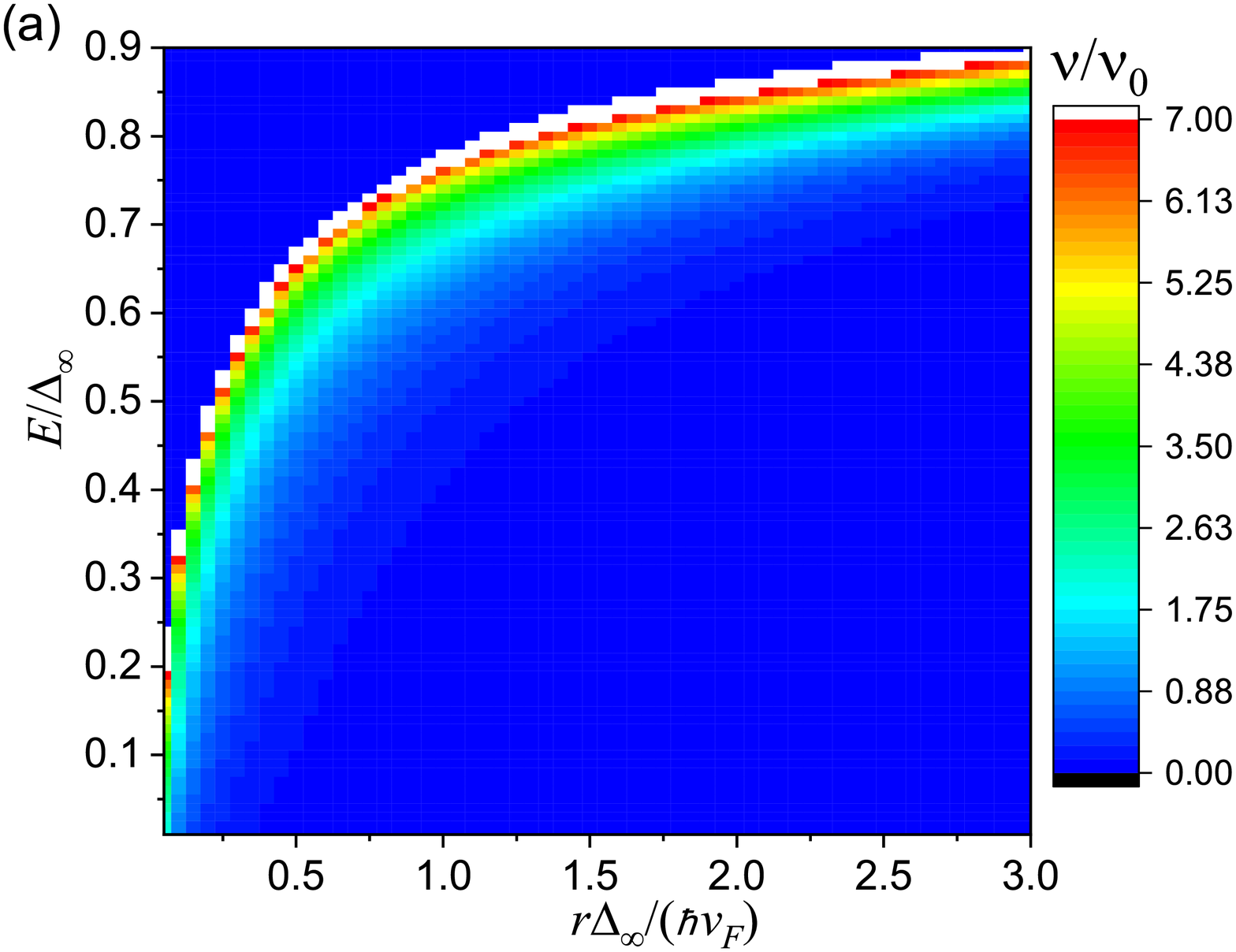}
		\includegraphics[width=0.9\linewidth]{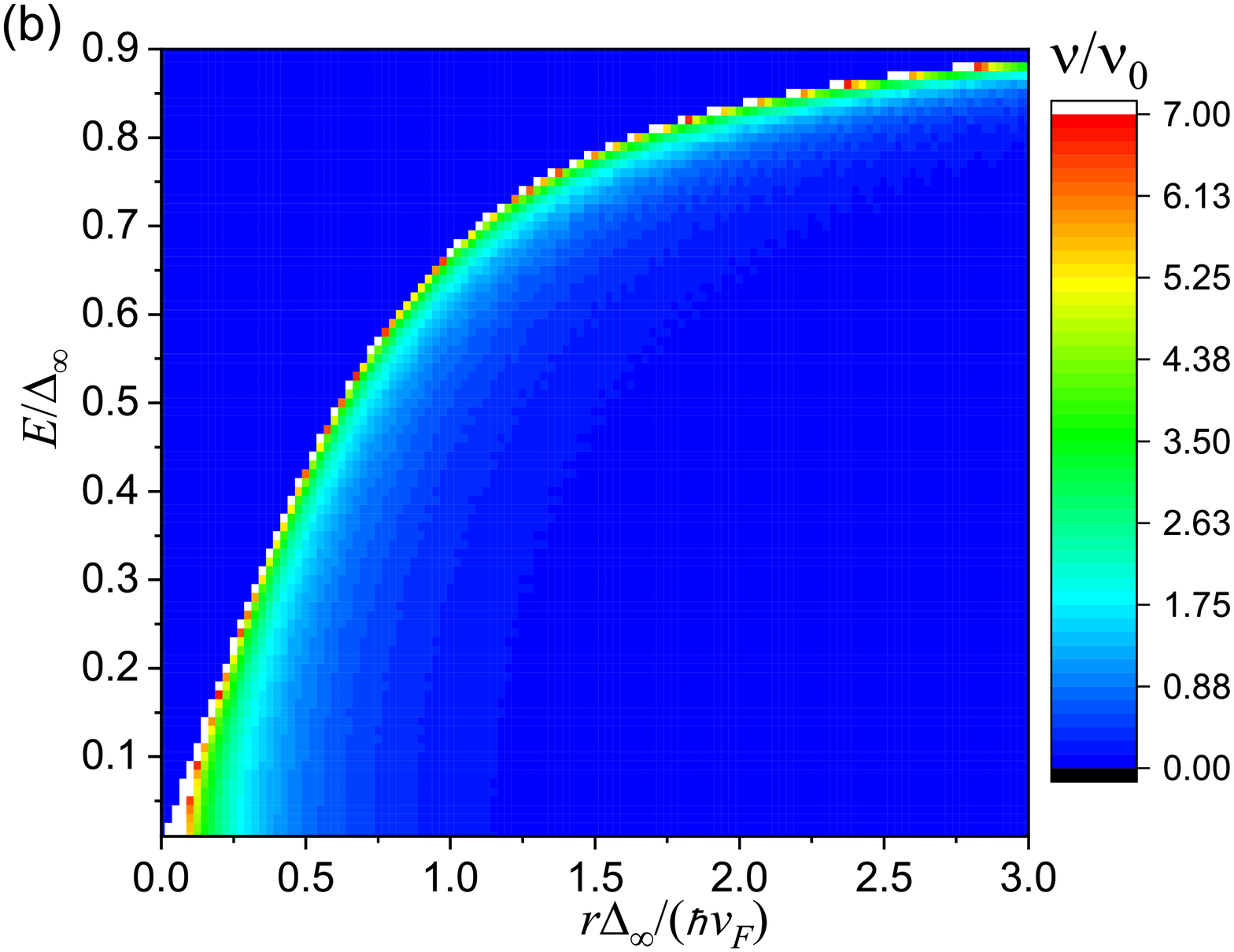}
	\caption{Density of states in a coreless vortex (a) and in a vortex with a core (b).} \label{fig:DOS2D}
\end{figure}

Now we concentrate on the density of states per spin projection, which is given by
\begin{equation}
	\nu(E,\vec{r}) = \frac{1}{\pi} \mathrm{Im} \left[ G_{ER}^{(0)}(\vec{r},\vec{r}) \right].
	\label{eq:nu}
\end{equation}
Using Eq. \eqref{eq:GER(r,r)_psi}, for $E>0$ we obtain
\begin{equation}
  \nu(E,\vec{r}) = \nu_0 \!\! \sum_{n=0}^{+\infty}  \int_0^{\pi/2} \!\!\! \delta \left( \frac{\psi_d(s) + \psi_d(-s)}{2} - \pi n\right) \! d\varphi.
	\label{eq:nu2D}
\end{equation}
In this sum the contributions to the density of states from all spectral branches are explicitly separated. After integration we obtain
\begin{equation}
   \nu(E,\vec{r})  = \nu_0 \sum_{n=0}^{+\infty}
	 \abs{ \frac{\partial}{\partial \varphi} \left( \frac{\psi_d(s) + \psi_d(-s)}{2}\right) \biggl|_{E^{(n)}(d)=E}}^{-1}  \!\!\!\!.
	\label{eq:nu2D_numerical}
\end{equation}
We used this expression to calculate numerically the local density of states: first, the value of $\varphi$ for which the relation $\psi_d(s) + \psi_d(-s) = 2\pi n$ holds was calculated, and then the derivative of $(\psi_d(s) + \psi_d(-s))/2$ with respect to $\varphi$ was determined. We limited ourselves to the range of energies $E<0.9<\min_d E^{(1)}(d)$, so that only the anomalous branch ($n=0$) contributed to the density of states. The resulting profiles of $\nu(E,r)$ are shown in Fig. \ref{fig:DOS2D}. It can be seen that $\nu(E,r) >0$ for $E<E^{(0)}(r)$, but for $E>E^{(0)}(r)$ the density of states completely vanishes. Thus, the quasiclassical spectrum has a position-dependent gap. 
%(see Fig. \ref{fig:nu0} for a picture of the whole gap in the $r$-$E$ plane). 
In the vicinity of $E = E^{(0)}(r)$ the density of states is proportional to $[E^{(0)}(r) - E]^{-1/2}$. Such behavior has been found in the low-energy limit in Ref. \cite{Waxman93AnnPhys}. Remarkably, the described above spectral features have not been mentioned in preceding papers where the density of states has been calculated using the Eilenberger equations \cite{Ullah+90PRB,Schopohl+95PRB,Rainer+96PRB,Eschrig+99PRB}. This is most likely due to typical quasiclassical calculations of $\nu(E,\vec{r})$ relying on solving the Eilenberger equations \eqref{eq:Eilenberger_g} - \eqref{eq:Eilenberger_f+} with a small, but finite $\eps$. This results in the smoothing of the peaks in the density of states and in $\nu(E,\vec{r})$ being positive everywhere. Another drawback of this approximation is that the contributions to the density of states from spectral branches with $n>0$ cannot be resolved, unless $\eps$ is taken extremely small. Indeed, the profiles of $\nu(E,\vec{r})$ obtained in Refs. \cite{Ullah+90PRB,Schopohl+95PRB,Rainer+96PRB} have only one peak around $E = 1$. On the other hand, our approach based on Eq. \eqref{eq:nu2D_numerical} assumes an infinitesimal $\eps$ and thus allows to resolve the contributions to $\nu(E,\vec{r})$ from any spectral branch, if desired.

\begin{figure}[htb]
	\centering
		\includegraphics[width=\linewidth]{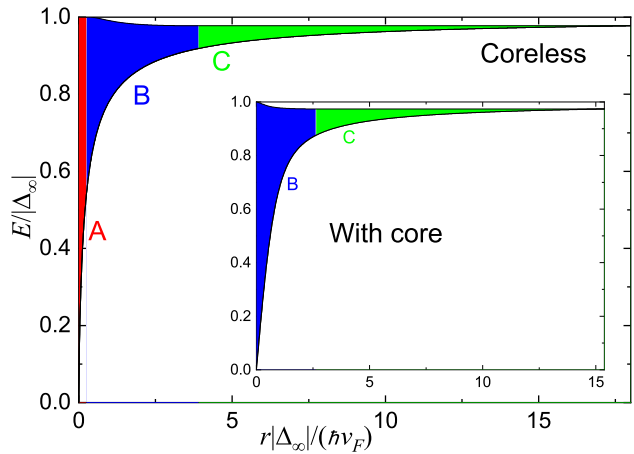}
	\caption{The gap in the local density of states of a coreless vortex and of a vortex with core (inset). The meaning of different colors is explained in Sec. \ref{sub:LDOS}.} \label{fig:nu0}
\end{figure}

Now we depict the whole spectral gap -- the region in the $r$-$E$ plane, where $\nu(E,r) = 0$. This can be done using Eq. \eqref{eq:nu2D_numerical}: it can be seen that the density of states vanishes if for all $d \in (0,r)$ and for all $n = 0,1,2...$ the inequality $E \neq E^{(n)}(d)$ holds. The spectral gaps for a coreless vortex and a vortex with core are shown in Fig. \ref{fig:nu0}. The boundaries of the gap are determined by the branches with numbers $n=0$ and $n=1$ only. In fact, the lower boundary is given by $E = E^{(0)}(r)$ for all $r$. The energy corresponding to the upper boundary, which we denote as $E_{\mathrm{max}}(r)$, behaves differently in the tree parts depicted in different colors in Fig. \ref{fig:nu0}.
For a coreless vortex, region (A) corresponds to $r<1/4$, region (B) -- to $1/4<r<d_{\mathrm{min}}^{(1)}$, and region (C) -- to $d_{\mathrm{min}}^{(1)}<r<r_c$, where $d_{\mathrm{min}}^{(1)}$ is the value of the impact parameter at which the spectral branch $E^{(1)}(d)$ has a minimum, and $r_c$ is the distance from the vortex core at which the spectral gap closes. The latter quantity is determined by the equation
\begin{equation}
	E^{(0)}(r_c) = E^{(1)}_{\mathrm{min}},
	\label{eq:r_c}
\end{equation}
where $E^{(1)}_{\mathrm{min}} = E^{(1)}(d_{\mathrm{min}}^{(1)})$ is the minimum energy of the branch with $n=1$. The upper boundary of the spectral gap is given by
\begin{equation}
	E_{\mathrm{max}}(r) = \left\{
	\begin{array}{ll}
		1 & \mbox{in (A),} \\
		E^{(1)}(r) & \mbox{in (B),} \\
		E_{\mathrm{min}}^{(1)} & \mbox{in (C).}
	\end{array} \right.
	\label{eq:Emax}
\end{equation}

The qualitative difference of the spectral gap of the vortex with core from the one of the coreless vortex consists in the absence of region (A), so that region (B) extends from $r=0$ to $r = d_{\mathrm{min}}^{(1)}$. Numerical constants that characterize the spectral gap for the two models of a vortex are given in Table \ref{tab:parameters}.

\begin{table}
\begin{ruledtabular}
	\centering
		\begin{tabular}{c|ccc}
			& $d_{\mathrm{min}}^{(1)}$ & $E_{\mathrm{min}}^{(1)}$ & $r_c$ \\		
			\hline
			Coreless & 3.9 & 0.9777 & 17.9  \\		
			With core & 2.6 & 0.974 & 15.4  \\
		\end{tabular}
	\caption{Spectral parameters of a vortex within two models.}	\label{tab:parameters}
\end{ruledtabular}
\end{table}

Concerning experimental implications of the obtained results, we think that observations of traces of the upper spectral branches using STS should be problematic because of the close proximity of their energies to the bulk gap $\Delta_{\infty}$. Spectral features due to the upper branches may be hard to distinguish from superconducting pairing anisotropy effects, which are present in any real superconductor. We suppose that angle-resolved measurements of the density of states are required to find the upper branches. Pairing anisotropy, as well as Fermi surface anisotropy, which is present in all materials, also lead to a smearing of the inverse-square root singularity in the in the local density of states $\nu(E,\vec{r})$ close to the local gap. However, the gap itself is not that much affected by this anisotropy as long as the superconducting order parameter is nodeless. As such, the local vanishing of the density of states near the center of vortices should be detectable in STS experiments on conventional superconductors.

\section{Impurity states in a 2D vortex}
\label{sec:ImpStates}

\subsection{General considerations}
\label{sub:Imp_general}

In this section we will analyze Eq. \eqref{eq:Gorkov} in the presence of the impurity potentials $V(\vec{r})$ and $\vec{J}(\vec{r})$ in the case of a fairly general superconducting system to obtain an equation for the energies of discrete impurity-induced states. Our considerations will be based on the theory developed in Ref. \cite{Bespalov2018} for a 3D system. 

We assume that we are dealing with a point impurity, so that $V(\vec{r})$ and $\vec{J}(\vec{r})$ are localized on a scale that is much smaller than $k_F^{-1}$. In this case one may choose a spin quantization axis, such that the electron spin is not rotated upon scattering if it is directed along this axis. In addition, in the spin-up and spin-down channels the point impurity acts as an $s$-wave scatterer, so that it is completely characterized by two scattering phases $\alpha_{\uparrow}$ and $\alpha_{\downarrow}$ for spin-up and spin-down electrons, respectively. These phases depend on energy, however in the narrow energy interval of interest, $E \sim \Delta_{\infty}$, they can be considered almost constant.

For our choice of the spin quantization axis one can see that the components of the Green functions with spin indices  $\uparrow \downarrow$ and $\downarrow \uparrow$ vanish, and the equations for the components with indices $\uparrow \uparrow$ and $\downarrow \downarrow$ decouple. Acting like in Ref. \cite{Bespalov2018}, one can express the solutions of Eq. \eqref{eq:Gorkov} in terms of the Green functions $G_E^{(0)}(\vec{r},\vec{r}')$ and $F_E^{\dagger(0)}(\vec{r},\vec{r}')$ without impurity -- see Appendix \ref{app:Imp_2D}. Then, to determine the energies of discrete impurity states, we need to find the impurity-induced poles of the Green functions. According to Appendix \ref{app:Imp_2D}, such poles appear only at energies for which the local density of states without impurity $\nu(E,\vec{r}_i)$ vanishes. This is quite natural: the appearance of discrete states localized by the impurity at energies lying inside the local continuous spectrum, corresponding to $\nu(E,\vec{r}_i) \neq 0$, is very unlikely. Inside the local spectral gap at position $\vec{r}_i$, the energies of impurity states with spin up are the solutions of the equation
\begin{equation}
	{\cal D}_{\uparrow}(E) = 0,
	\label{eq:Dup=0}
\end{equation}
where
\begin{eqnarray}
	& \hspace{-4cm} {\cal D}_{\uparrow}(E) = \abs{F_E^{\dagger (0)} (\vec{r}_i,\vec{r}_i)}^2 & \nonumber \\
	& + \!\! \left[ \frac{m \cot \alpha_{\uparrow}}{2 \hbar^2} \! - \! G_{ER}^{(0)}(\vec{r}_i,\vec{r}_i) \right] \!\!\! \left[ \frac{m \cot \alpha_{\downarrow}}{2\hbar^2}
	\! + \! G_{ER}^{(0)}(\vec{r}_i,\vec{r}_i) \right] \!\! . &
	\label{eq:D_up_2D'}
\end{eqnarray}
Note that $G_{ER}^{(0)}(\vec{r}_i,\vec{r}_i)$ is real here. To obtain an equation for spin-down impurity states, one should swap $\uparrow$ and $\downarrow$ in Eqs. \eqref{eq:Dup=0} and \eqref{eq:D_up_2D'}.

Now we will analyze Eqs. \eqref{eq:Dup=0} and \eqref{eq:D_up_2D'}. Let the function $G_{ER}^{(0)}(\vec{r}_i,\vec{r}_i)$ be real in some energy interval $E \in (E_{\mathrm{min}},E_{\mathrm{max}})$. The function ${\cal D}_{\uparrow}(E)$ can be written in the form
\begin{equation}
	{\cal D}_{\uparrow}(E) = - {\cal D}_{\uparrow+}(E) {\cal D}_{\uparrow-}(E),
	\label{eq:D(2D)_product}
\end{equation}
where
\begin{eqnarray}
	& {\cal D}_{\uparrow \pm}(E) = G_{ER}^{(0)}(\vec{r}_i,\vec{r}_i) - \frac{m}{4\hbar^2} (\cot \alpha_{\uparrow} - \cot \alpha_{\downarrow}) & \nonumber \\
	& \pm \sqrt{\left( \frac{m}{4\hbar^2} \right)^2 \left( \cot \alpha_{\uparrow} + \cot \alpha_{\downarrow} \right)^2 + \abs{F_E^{\dagger (0)} (\vec{r}_i,\vec{r}_i)}^2}. &
	\label{eq:D(2D)_pm}
\end{eqnarray}
By direct differentiation and using the relation
\begin{equation}
	\abs{\frac{\partial F_E^{\dagger(0)}(\vec{r}, \vec{r})}{\partial E}} < \frac{\partial G_{ER}^{(0)}(\vec{r},\vec{r})}{\partial E},
	\label{eq:F'<G'}
\end{equation}
derived in Appendix \ref{app:quasiclassics}, we can prove that the functions ${\cal D}_{\uparrow \pm}(E)$ increase with increasing energy, hence on the interval $E \in (E_{\mathrm{min}},E_{\mathrm{max}})$ they vanish no more than once. Thus, on this interval Eq. \eqref{eq:Dup=0} has no more than two roots. Taking into account that ${\cal D}_{\uparrow+}(E) \geq {\cal D}_{\uparrow-}(E)$, we can have the following four qualitatively different situations (we do not consider the cases when ${\cal D}_{\uparrow\pm}(E)$ vanish at the boundaries of the interval $(E_{\mathrm{min}},E_{\mathrm{max}})$):

\begin{itemize}
	\item{(i)} $\lim\limits_{E \to E_{\mathrm{min}}} \!\!\! {\cal D}_{\uparrow+}(E)>0$ and $\lim\limits_{E \to E_{\mathrm{max}}} \!\!\! {\cal D}_{\uparrow-}(E)<0$. Equation \eqref{eq:Dup=0} has no roots.
	\item{(ii)} $\lim\limits_{E \to E_{\mathrm{min}}} \!\!\! {\cal D}_{\uparrow-}(E)>0$ or $\lim\limits_{E \to E_{\mathrm{max}}} \!\!\! {\cal D}_{\uparrow+}(E)<0$. Equation \eqref{eq:Dup=0} has no roots.
	\item{(iii)} $\lim\limits_{E \to E_{\mathrm{min}}} \!\!\! {\cal D}_{\uparrow+}(E)<0$ and $\lim\limits_{E \to E_{\mathrm{max}}} \!\!\! {\cal D}_{\uparrow-}(E)>0$. Equation \eqref{eq:Dup=0} has two roots.
	\item{(iv)} In all other cases, there is one root.
\end{itemize}

We want to mention that Eqs. \eqref{eq:Dup=0} and \eqref{eq:D_up_2D'} can be used to express the energies of Yu-Shiba-Rusinov states \cite{Yu1965,Shiba68PTP,Rusinov1969JETP} induced by a magnetic impurity in a uniform superconductor in terms of the scattering phases $\alpha_{\uparrow}$ and $\alpha_{\downarrow}$, as has been done by Rusinov \cite{Rusinov1969JETP}.

\subsection{Impurity states in a vortex}
\label{sub:Imp_Vortex}

In this section we will apply the developed above approach to find the impurity states in our system with a vortex.

To determine the number of spin-up impurity states for each combination of parameters $r_i$, $\alpha_{\uparrow}$ and $\alpha_{\downarrow}$, according to Sec. \ref{sub:Imp_general}, one needs to analyze the behavior of the functions $G_{ER}^{(0)}(\vec{r}_i,\vec{r}_i)$ and $F_E^{\dagger (0)}(\vec{r}_i,\vec{r}_i)$ in the vicinity of the energies corresponding to the boundaries of the local spectral gap at position $\vec{r}_i$. Such analysis is given in Appendix \ref{app:poles}. We found that for $r_i<d^{(1)}_{\mathrm{min}}$ there may be from 0 to 2 impurity states per electron spin projection. For $r_i \in (d^{(1)}_{\mathrm{min}},r_c)$ there are 1 or 2 impurity states per spin projection. We want to stress that even nonmagnetic impurities induce bound states (this does not contradict Anderson's theorem \cite{Anderson1959JPCS}, because the order parameter is inhomogeneous in space). For $r_i>r_c$ no bound impurity states appear, however, there may be quasibound states of the Yu-Shiba-Rusinov type.

Technically, the calculation of the energies of impurity states consists of two steps. First, the signs of ${\cal D}_{\uparrow+}(E)$ and ${\cal D}_{\uparrow-}(E)$ at energies lying close to the boundaries of the local spectral gap are determined, which can be done using the relations derived in Appendix \ref{app:poles}. This is required to find out whether the monotonic functions ${\cal D}_{\uparrow+}(E)$ and ${\cal D}_{\uparrow-}(E)$ have a root or not. Second, the roots are calculated using a simple bisection procedure.

\begin{figure*}[ht]
	\centering
		\includegraphics[width = 0.325\linewidth]{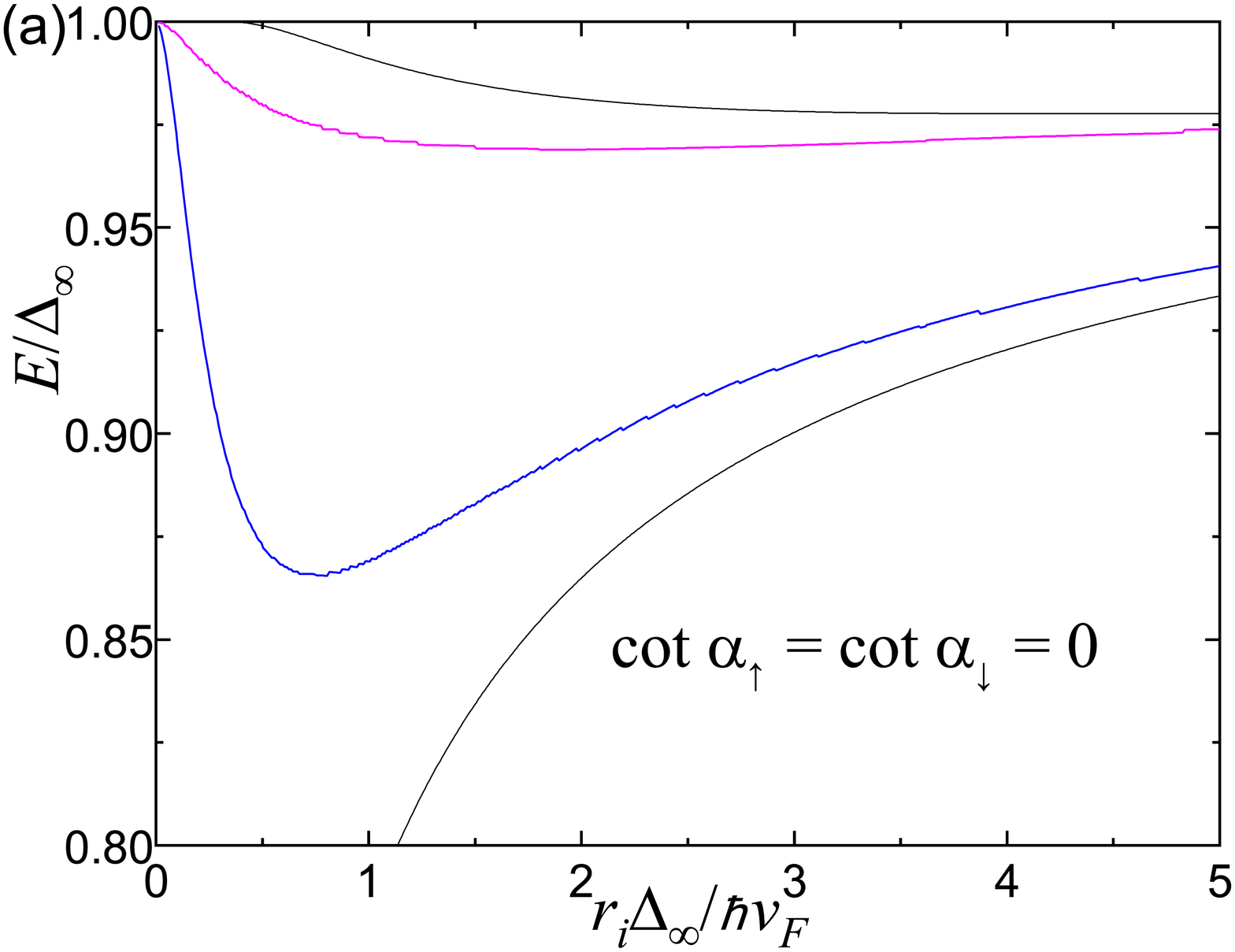}
		\includegraphics[width = 0.325\linewidth]{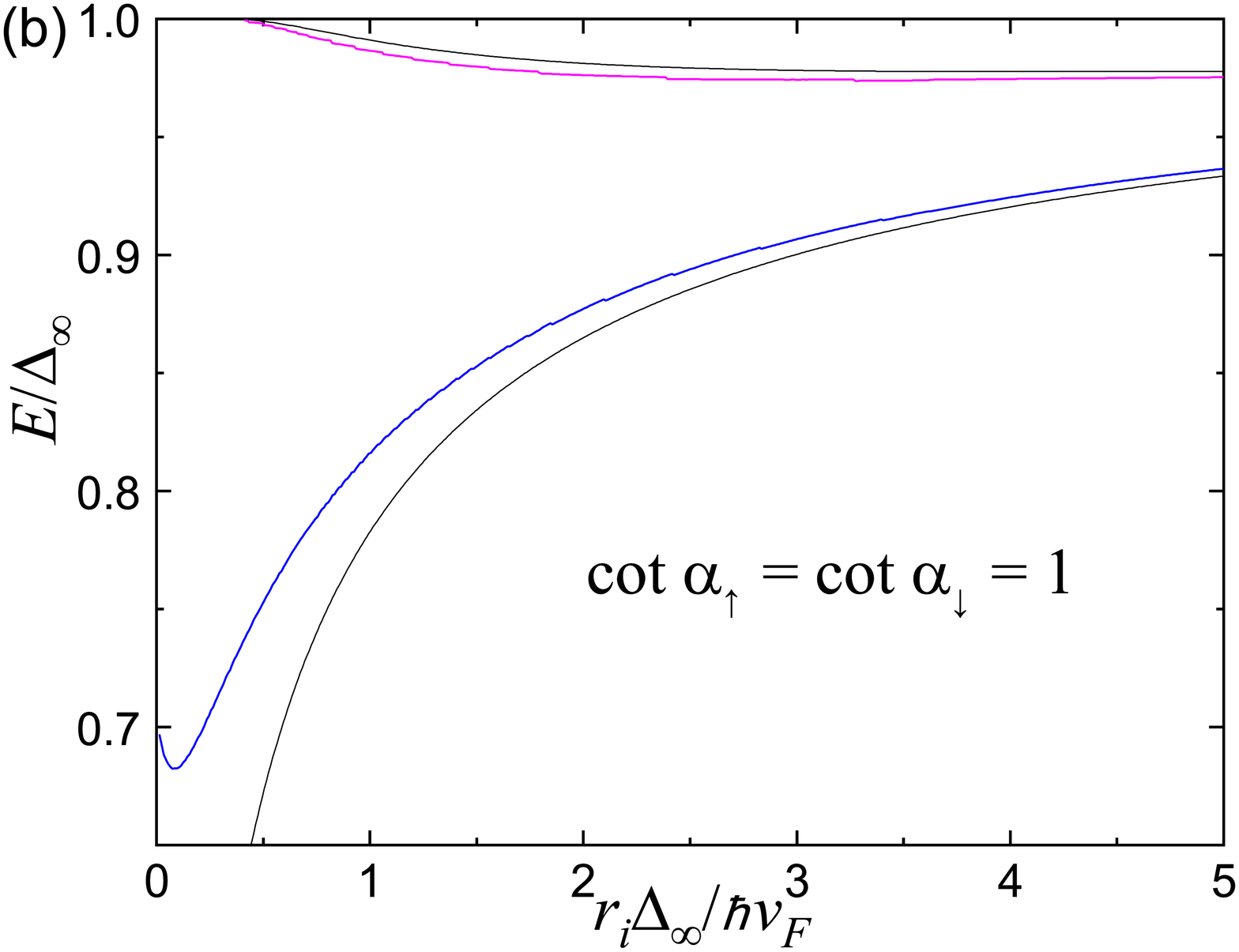}
		\includegraphics[width = 0.325\linewidth]{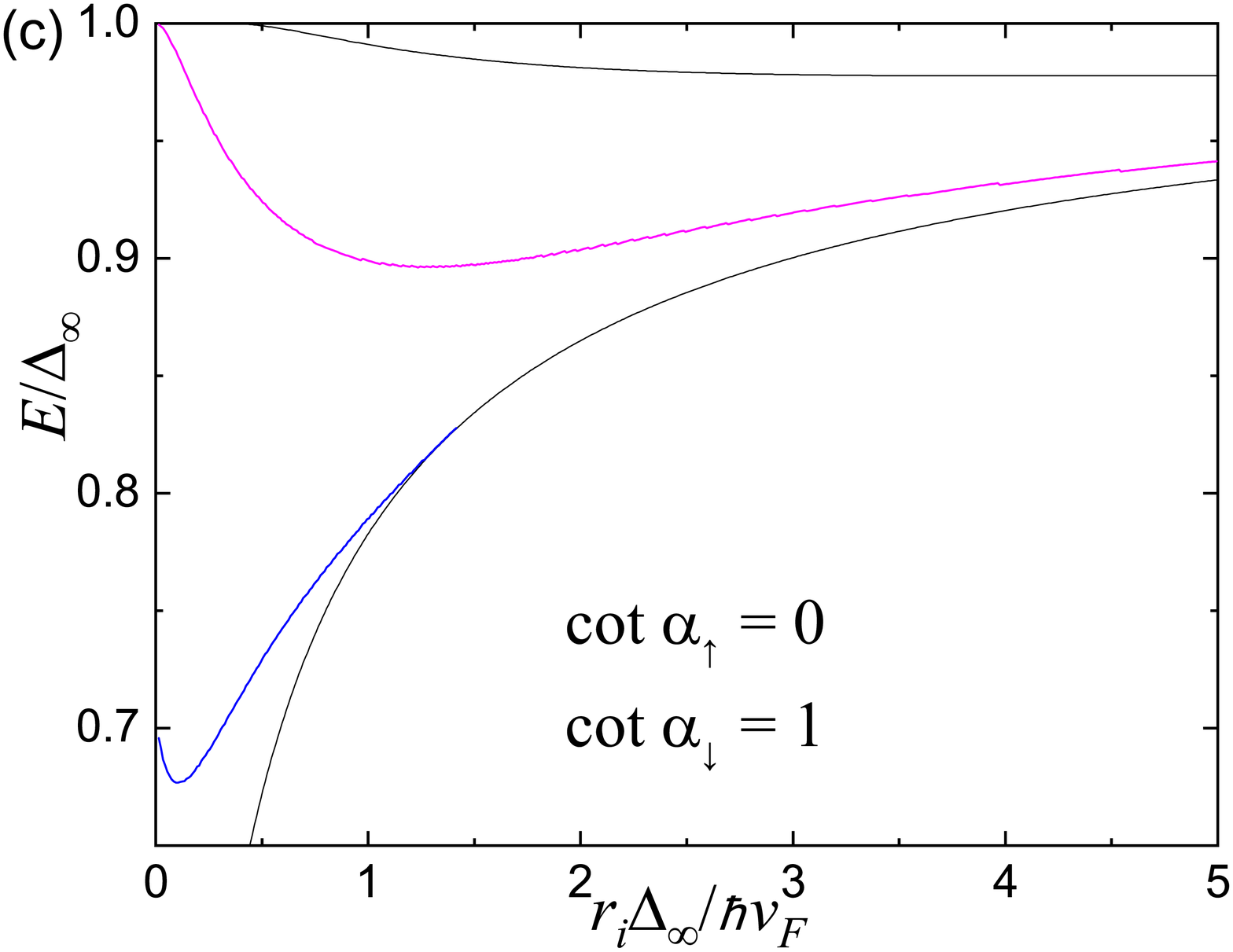}
		\includegraphics[width = 0.325\linewidth]{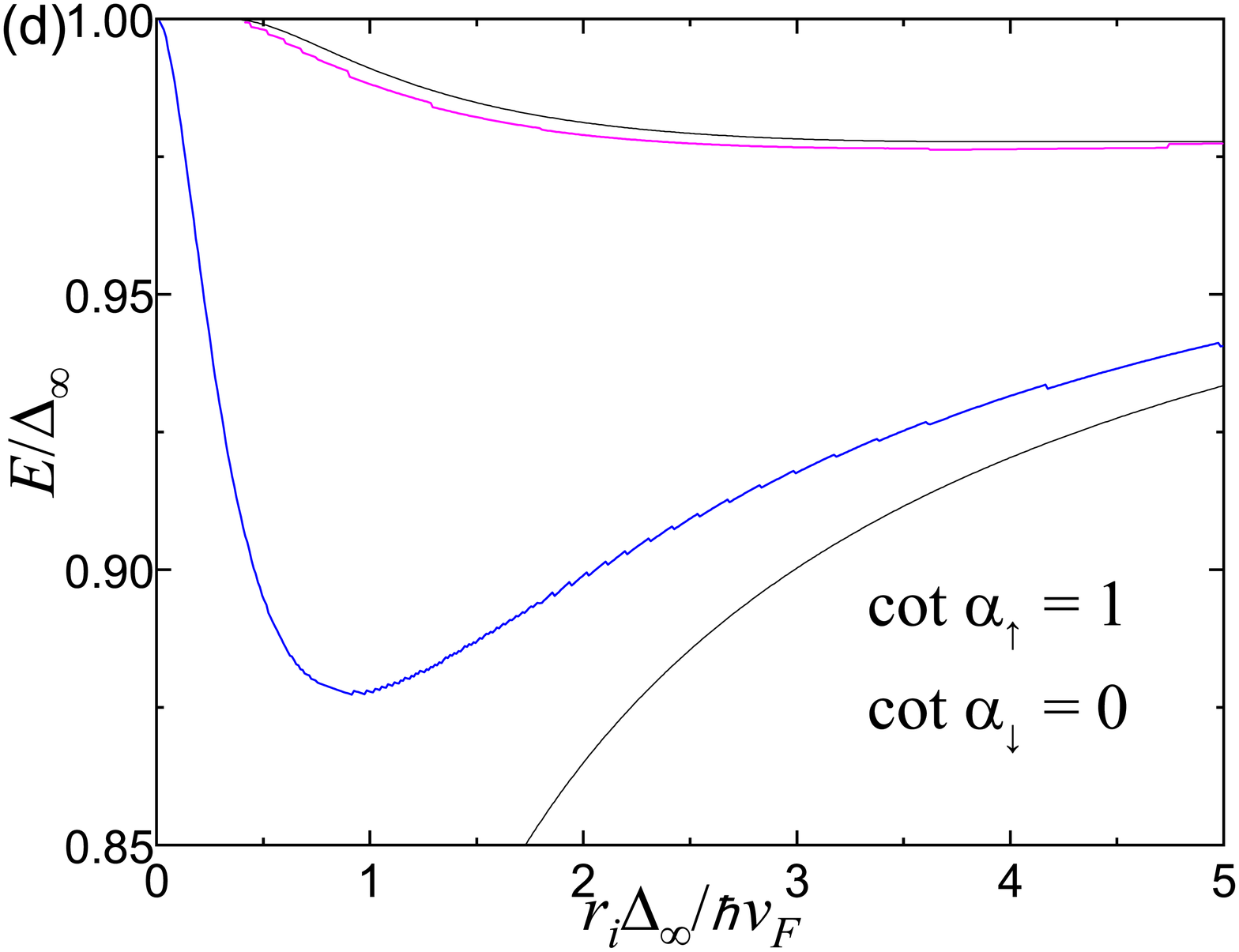}
		\includegraphics[width = 0.325\linewidth]{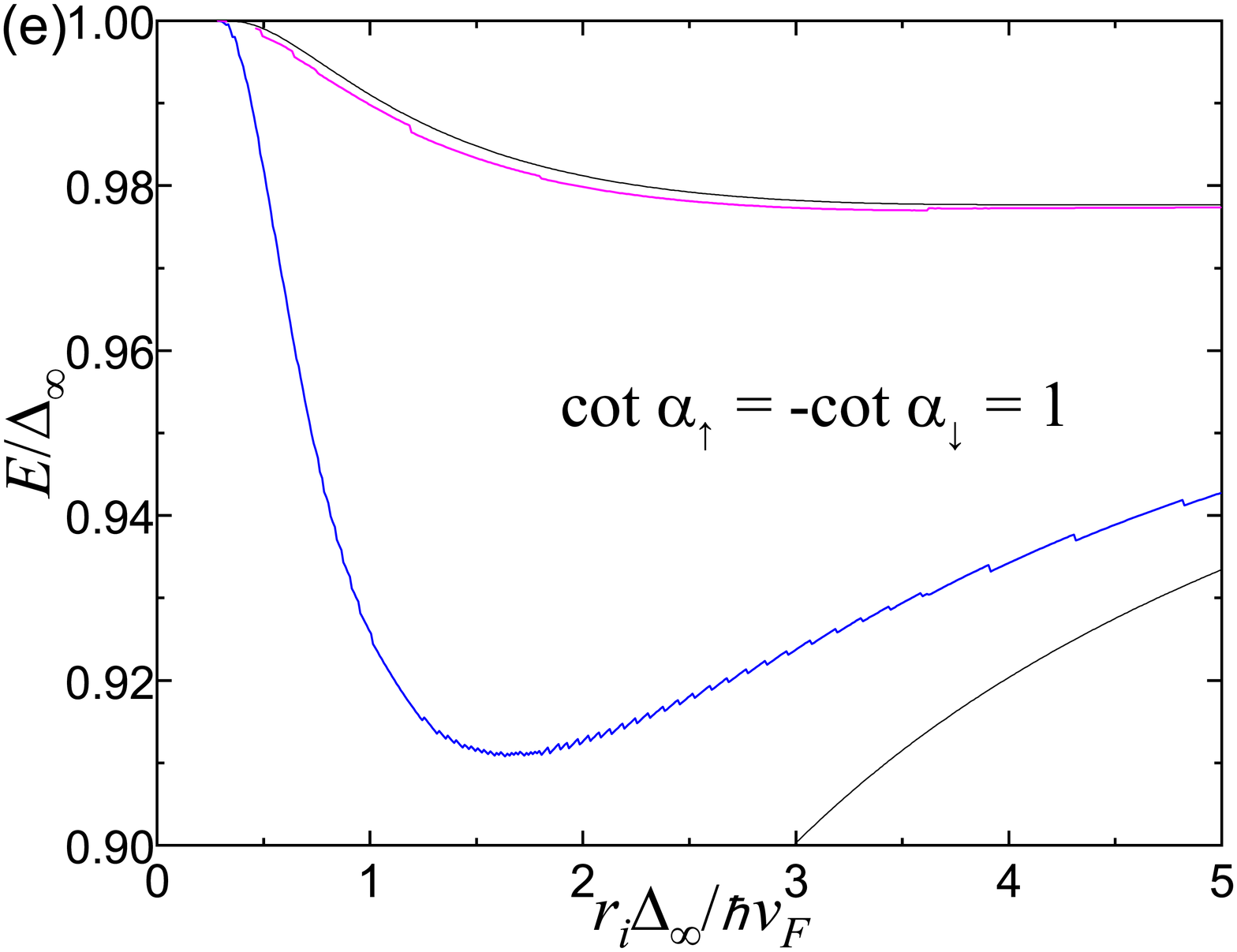}
		\includegraphics[width = 0.325\linewidth]{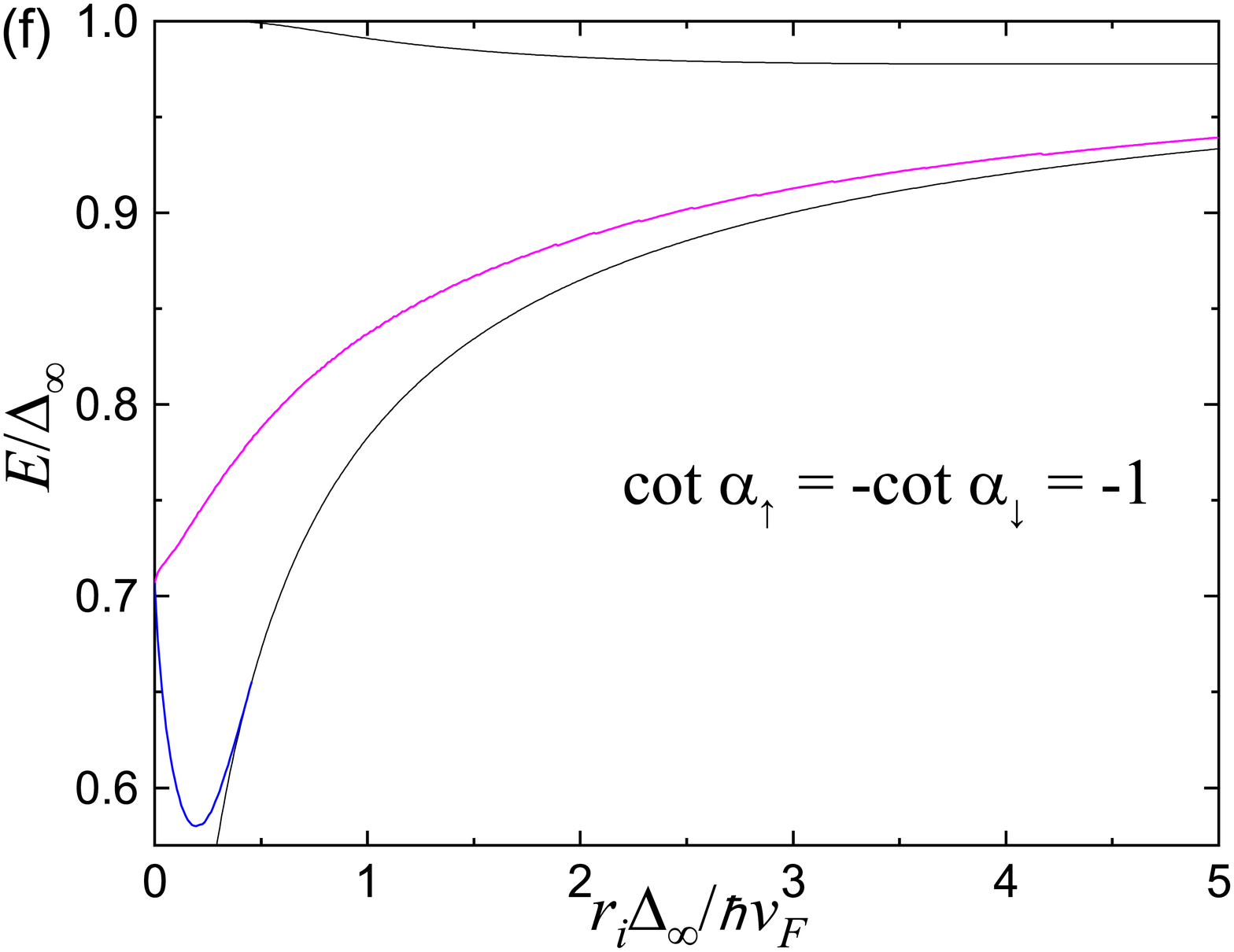}
	\caption{Energies of spin-up impurity states vs. $r_i$ for a coreless vortex for impurities with different scattering phases (shown in the graphs). (a),(b) -- nonmagnetic impurity, (c)-(f) -- magnetic impurity. The thin black lines show $E_{\mathrm{min}}(r_i)$ and $E_{\mathrm{max}}(r_i)$.} \label{fig:Eimp_2D}
\end{figure*}

\begin{figure*}[ht]
	\centering
		\includegraphics[width = 0.325\linewidth]{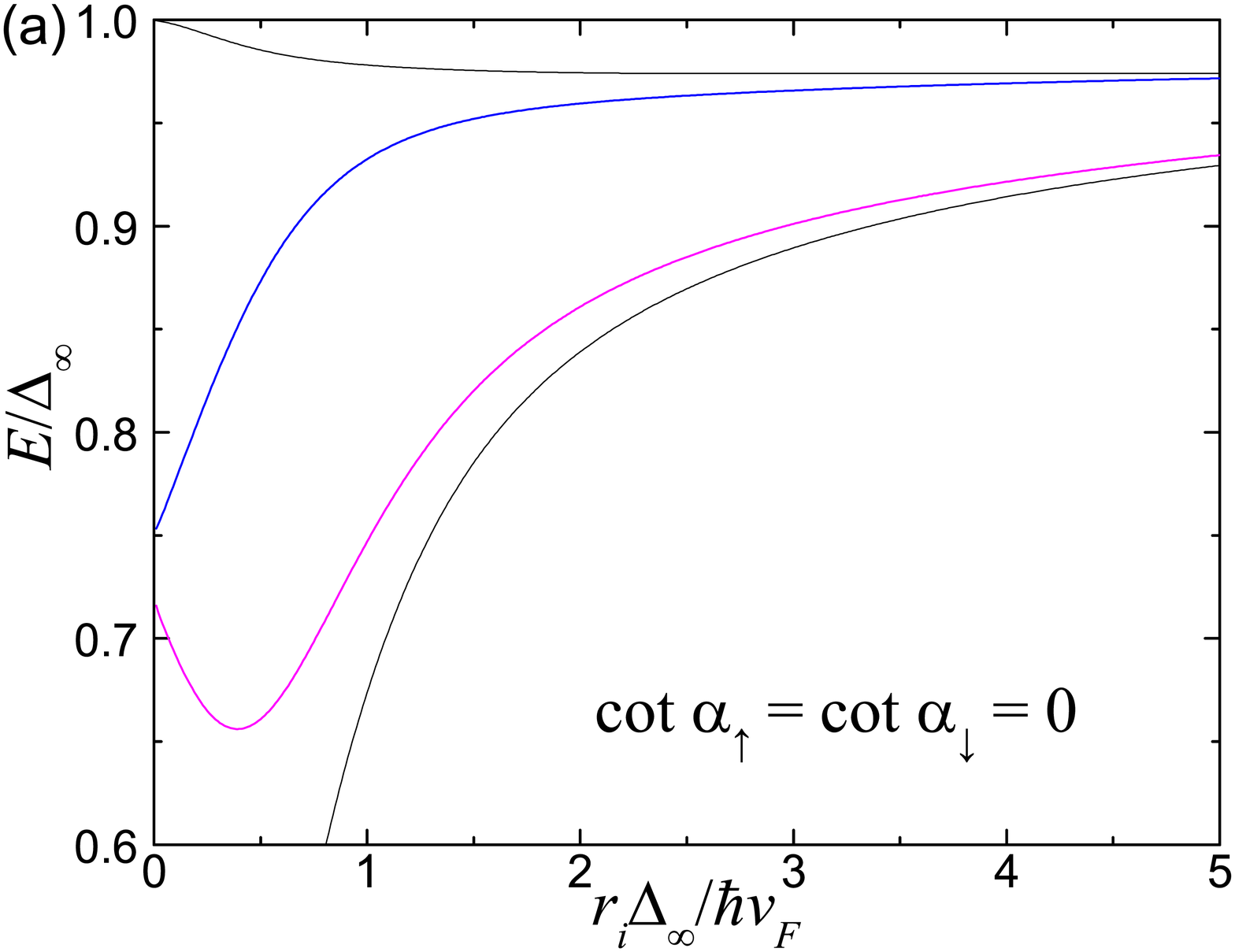}
		\includegraphics[width = 0.325\linewidth]{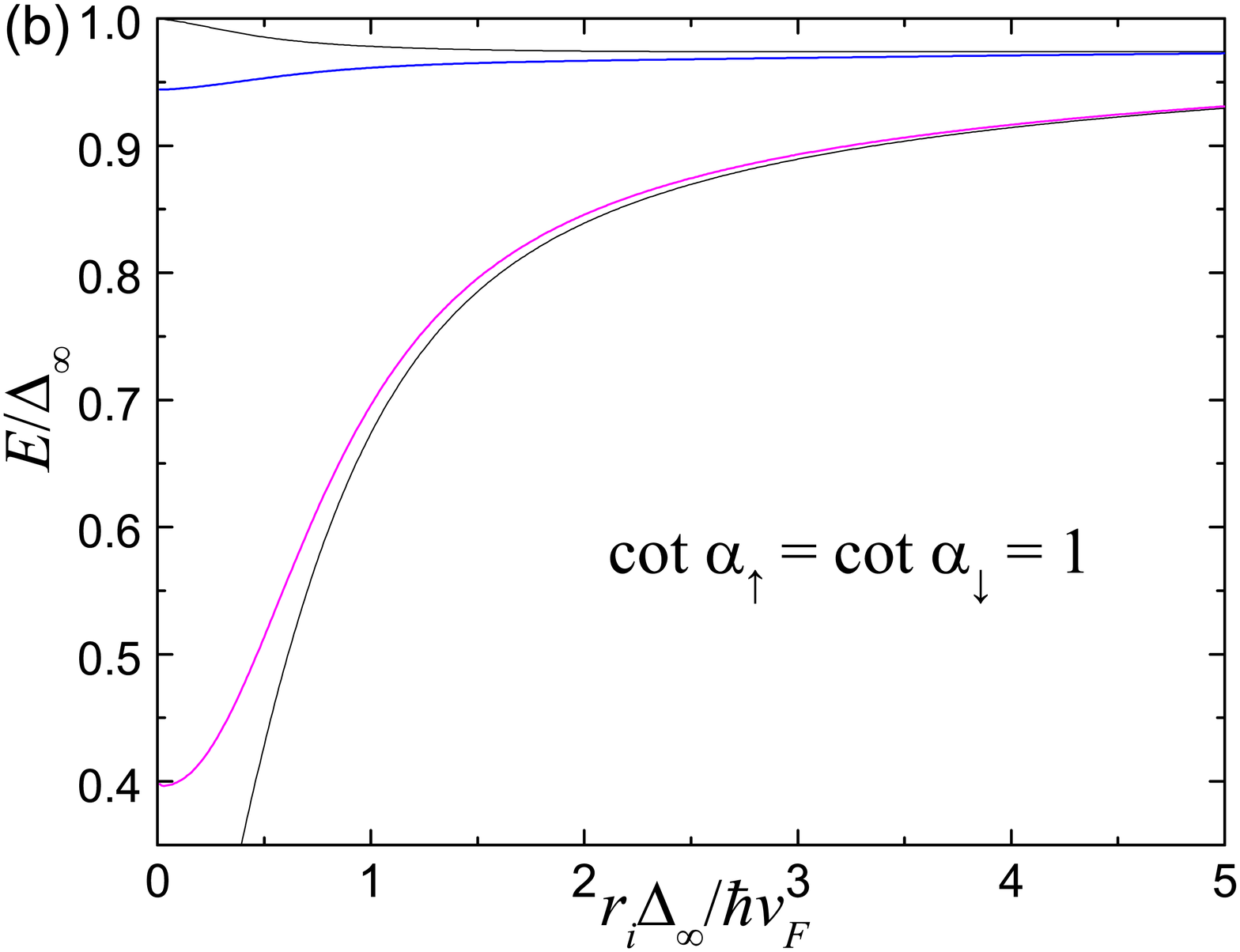}
		\includegraphics[width = 0.325\linewidth]{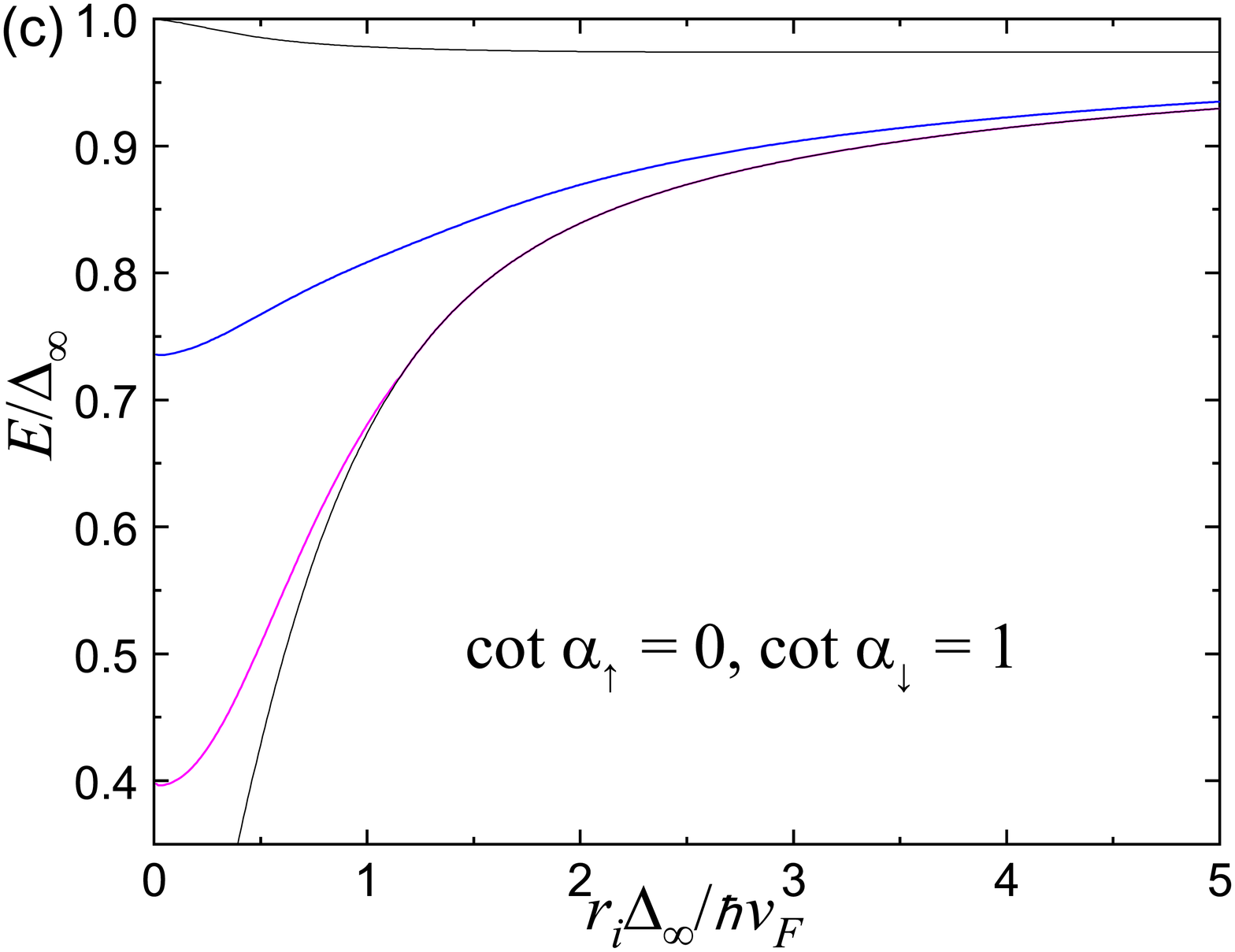}
		\includegraphics[width = 0.325\linewidth]{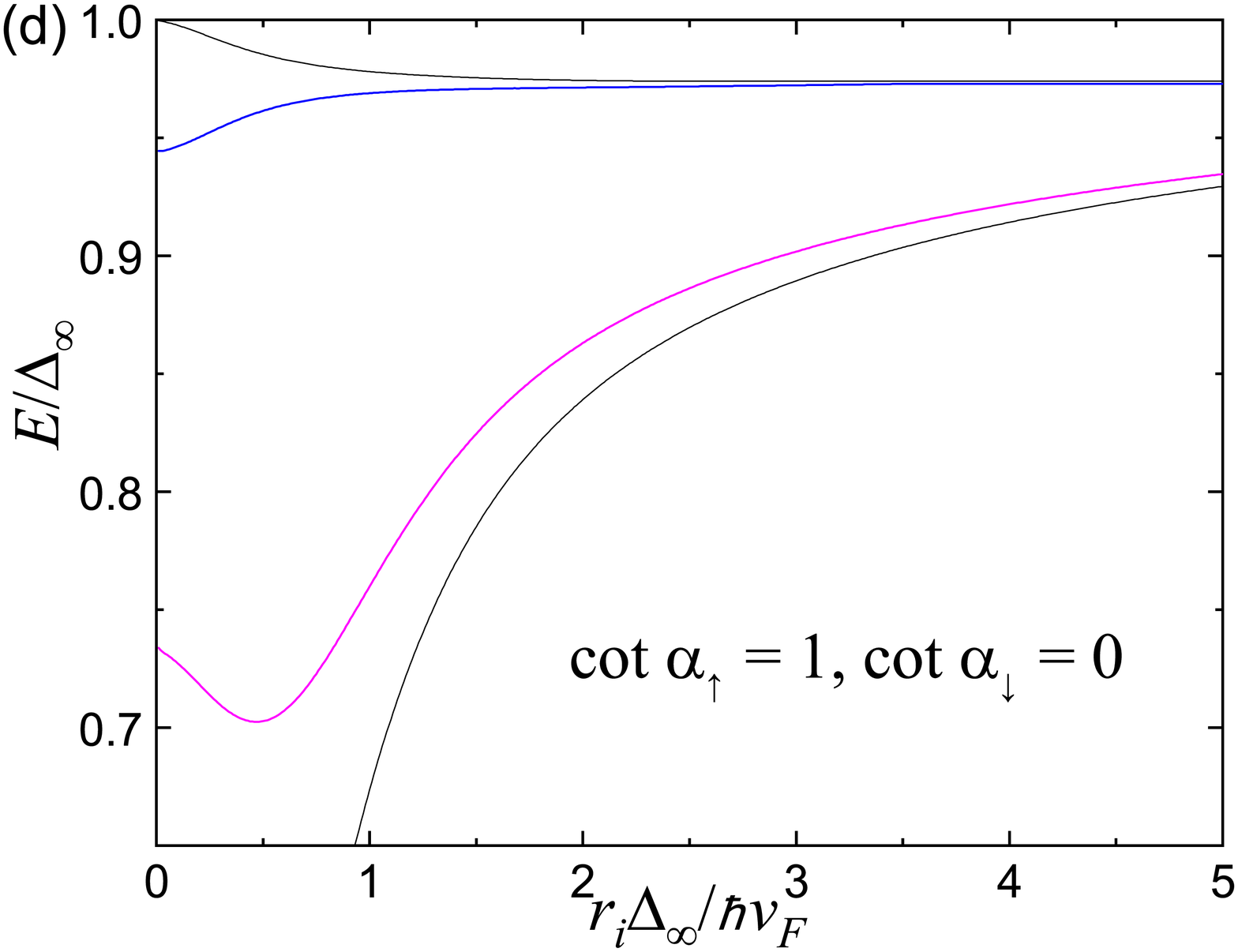}
		\includegraphics[width = 0.325\linewidth]{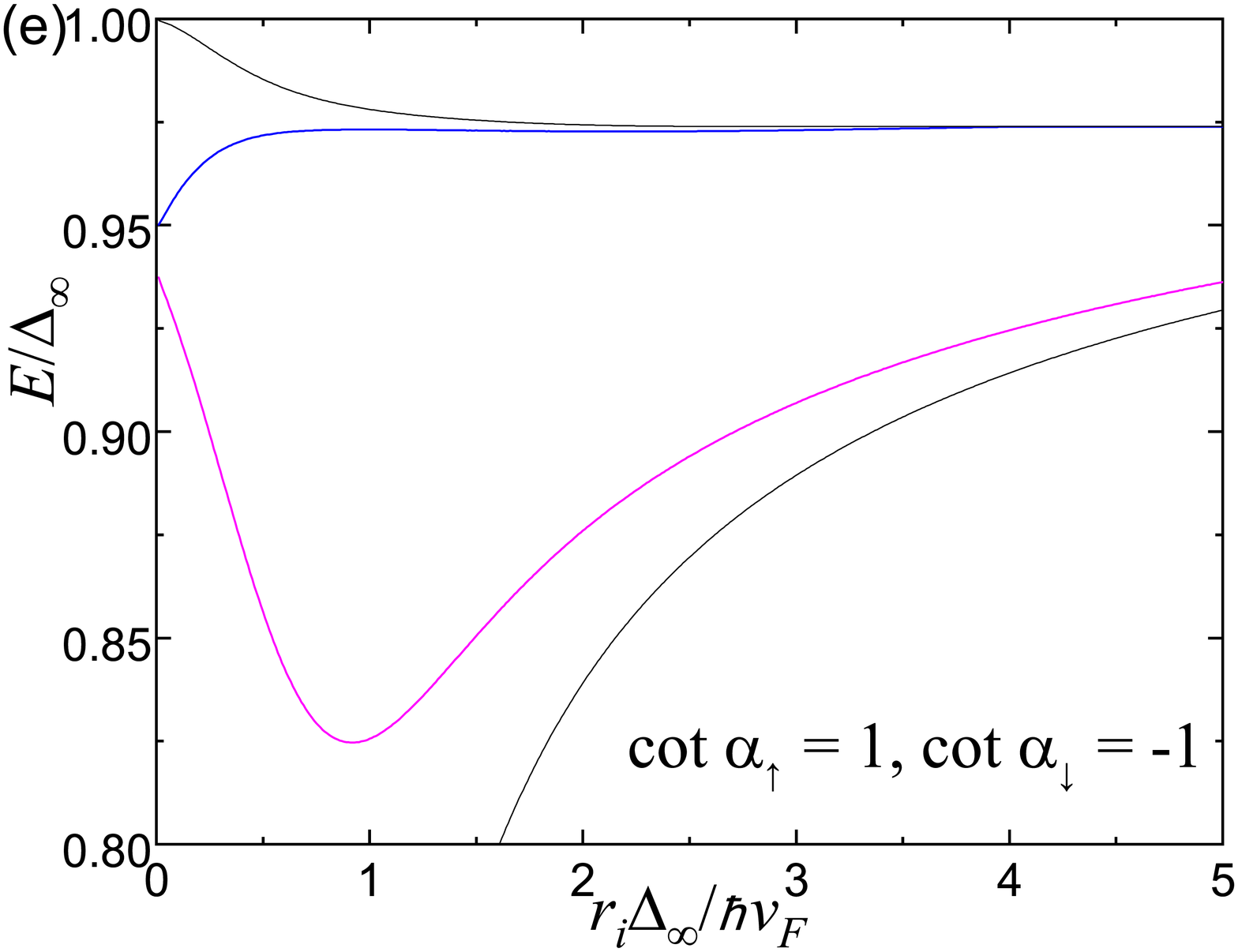}
		\includegraphics[width = 0.325\linewidth]{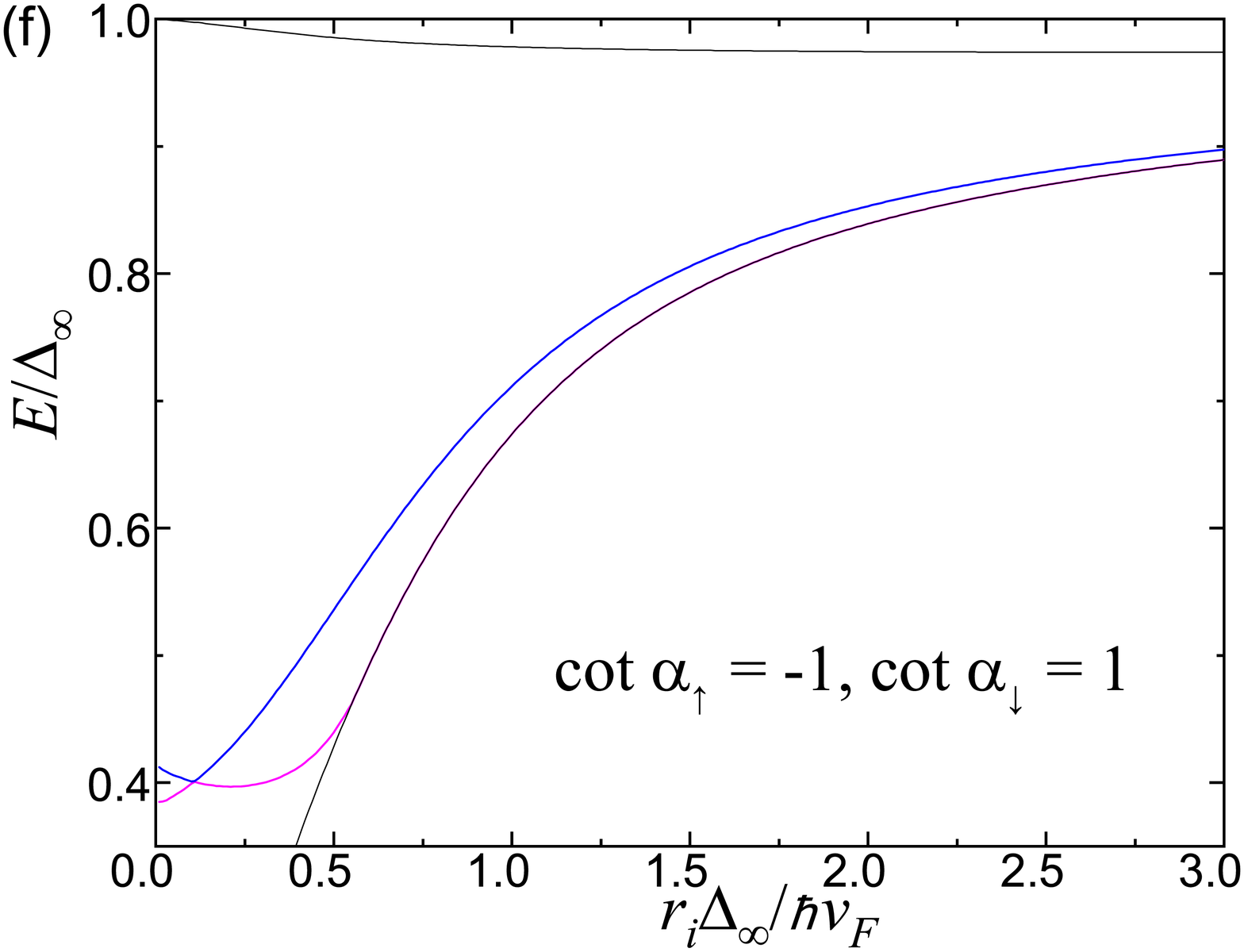}
	\caption{Energies of spin-up impurity states vs. $r_i$ for a vortex with core for impurities with different scattering phases. Notations are the same as in Fig. \ref{fig:Eimp_2D}.} \label{fig:Eimpc_2D}
\end{figure*}

The calculated dependencies of the energies of spin-up impurity states vs $r_i$ for a coreless vortex and for a vortex with core are shown in Figs. \ref{fig:Eimp_2D} and \ref{fig:Eimpc_2D}, respectively. An interesting feature can be seen in Fig. \ref{fig:Eimpc_2D}f: two graphs of energy vs $r_i$ have a point of intersection. At this point ${\cal D}_{\uparrow+}(E)$ and ${\cal D}_{\uparrow-}(E)$ vanish simultaneously, which becomes possible when $\alpha_{\uparrow} + \alpha_{\downarrow} = 0$, according to Eq. \eqref{eq:D(2D)_pm}.

%Here, we want to make a remark concerning the relation of our results to the results of Koulakov, Larkin and Ovchinnikov \cite{Larkin+Ovchinnikov98PRB,Koulakov+Larkin99PRB,Koulakov+Larkin99PRB(2)}. These authors studied within the BdG equations formalism the low-energy part of the vortex spectrum ($\abs{E} \ll \Delta_{\infty}$) in the presence of point impurities. They found impurity-induced modifications of the discrete spectrum on an energy scale of the order of $\Delta_{\infty}^2/\mu$. Our quasiclassical approach does not resolve such small energy scales, however, it allows us to study impurity states within the whole subgap range of energies. Thus, our results and the results from Refs. \cite{Larkin+Ovchinnikov98PRB,Koulakov+Larkin99PRB,Koulakov+Larkin99PRB(2)} are related to different aspects of the same problem and cannot be deduced from each other.

\begin{figure*}[htb]
	\centering
		\includegraphics[width = 0.325\linewidth]{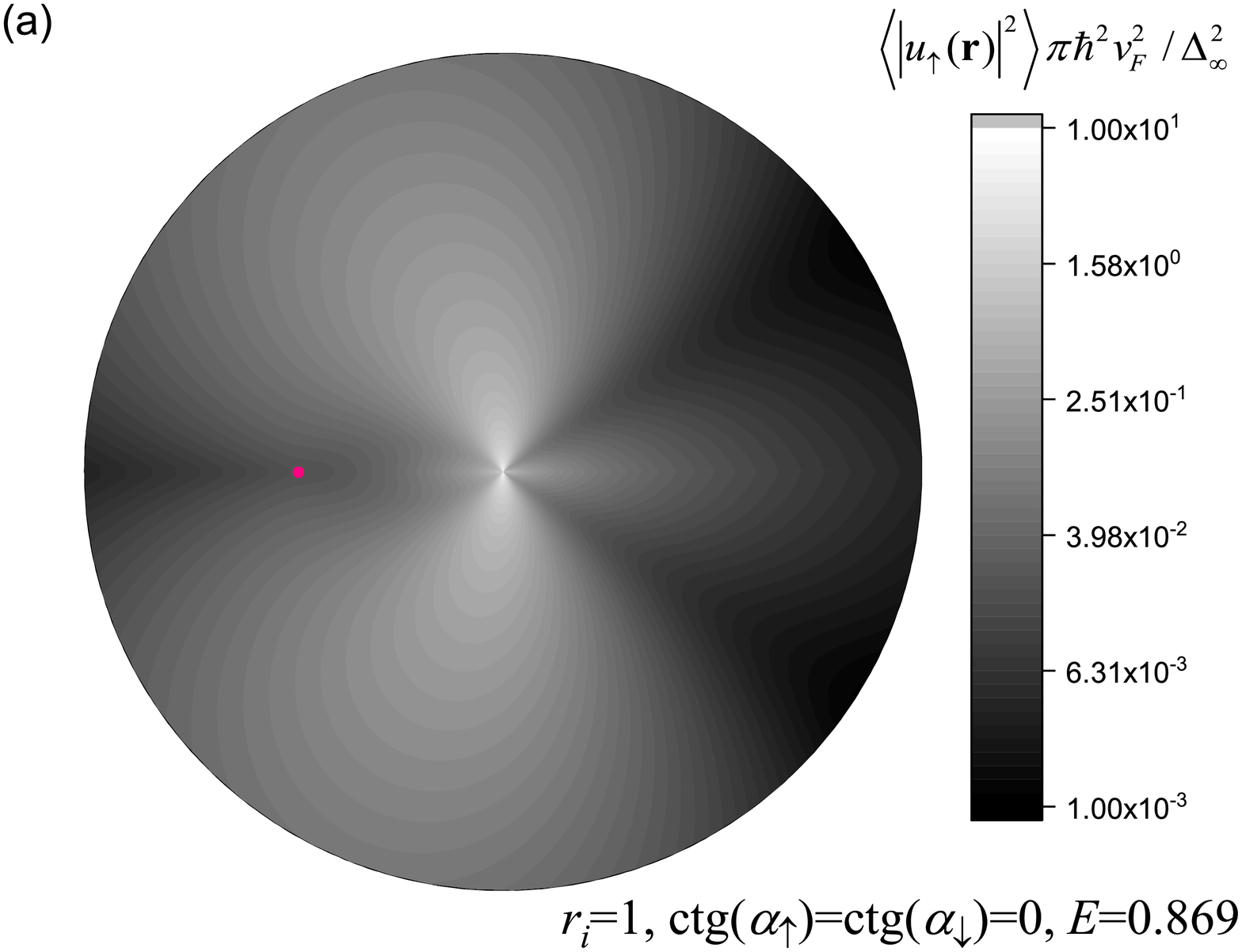}
		\includegraphics[width = 0.325\linewidth]{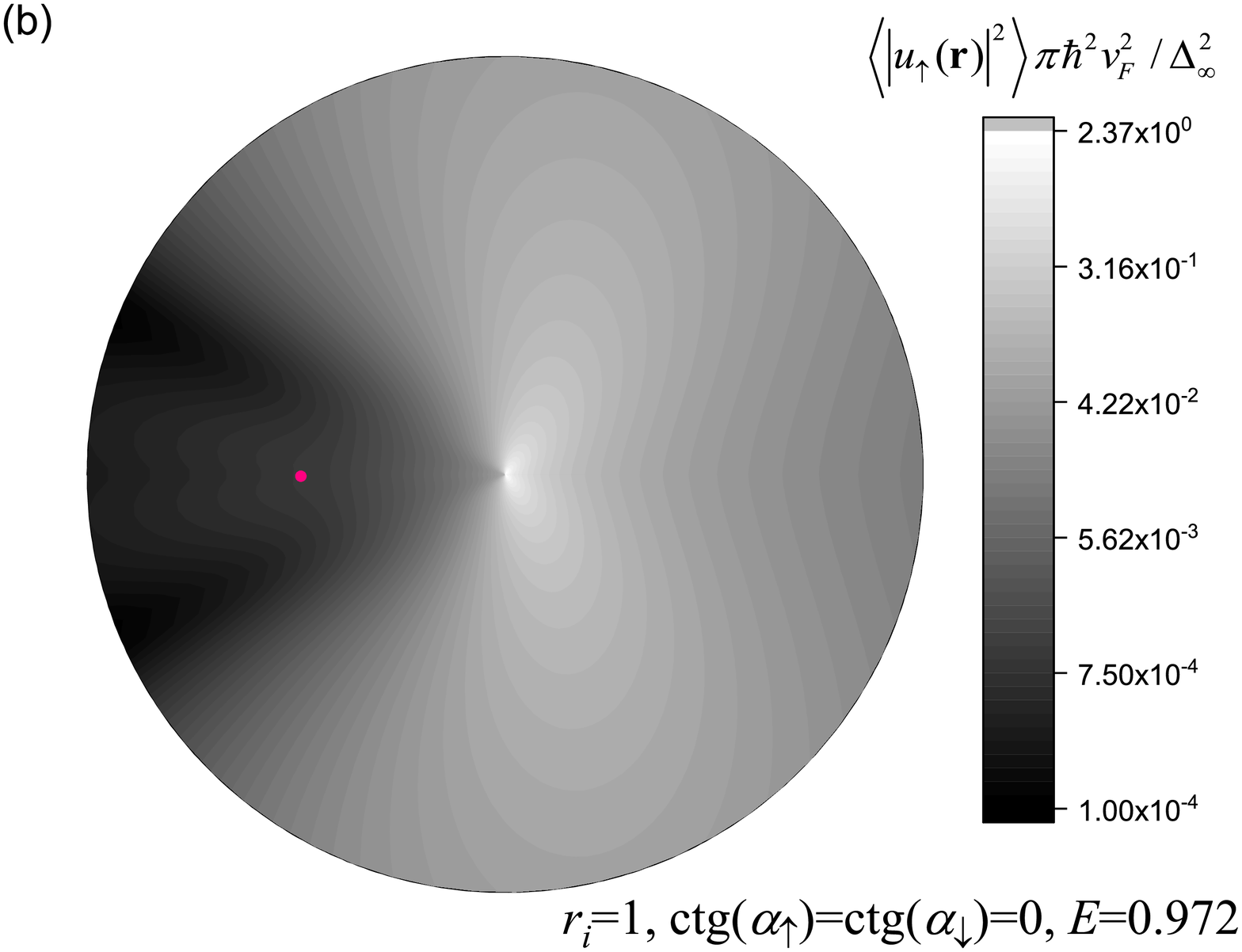}
		\includegraphics[width = 0.325\linewidth]{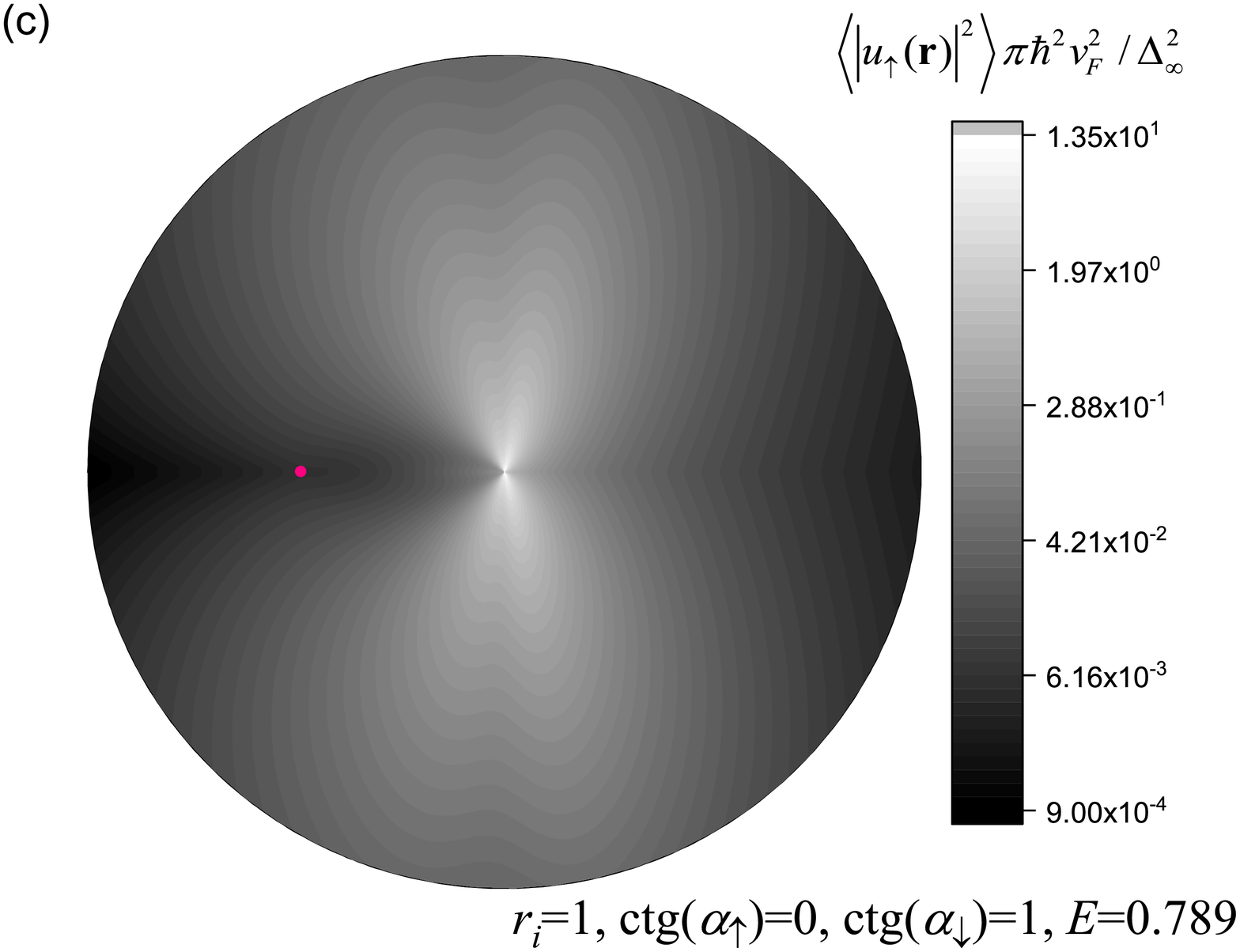}
		\includegraphics[width = 0.325\linewidth]{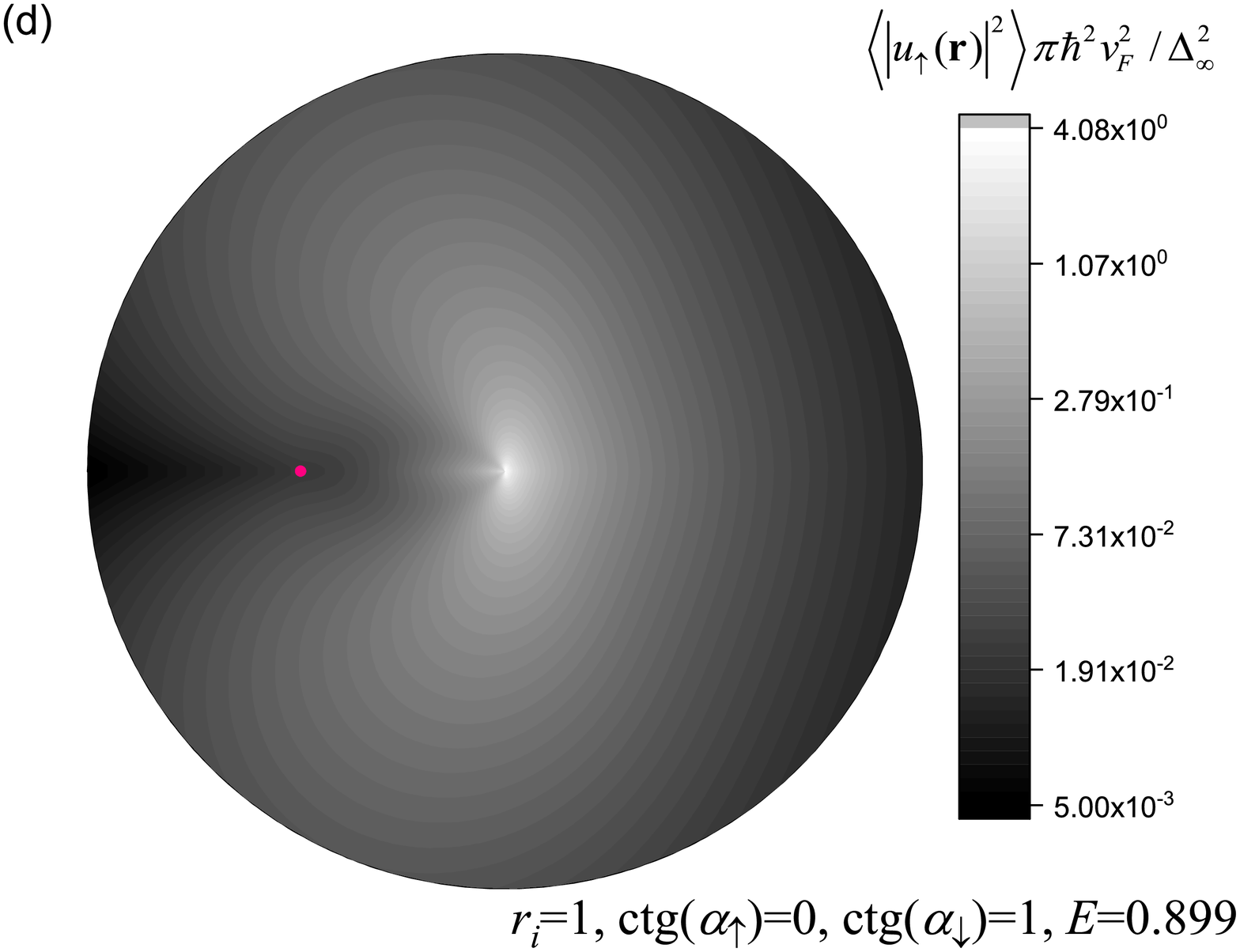}
		\includegraphics[width = 0.325\linewidth]{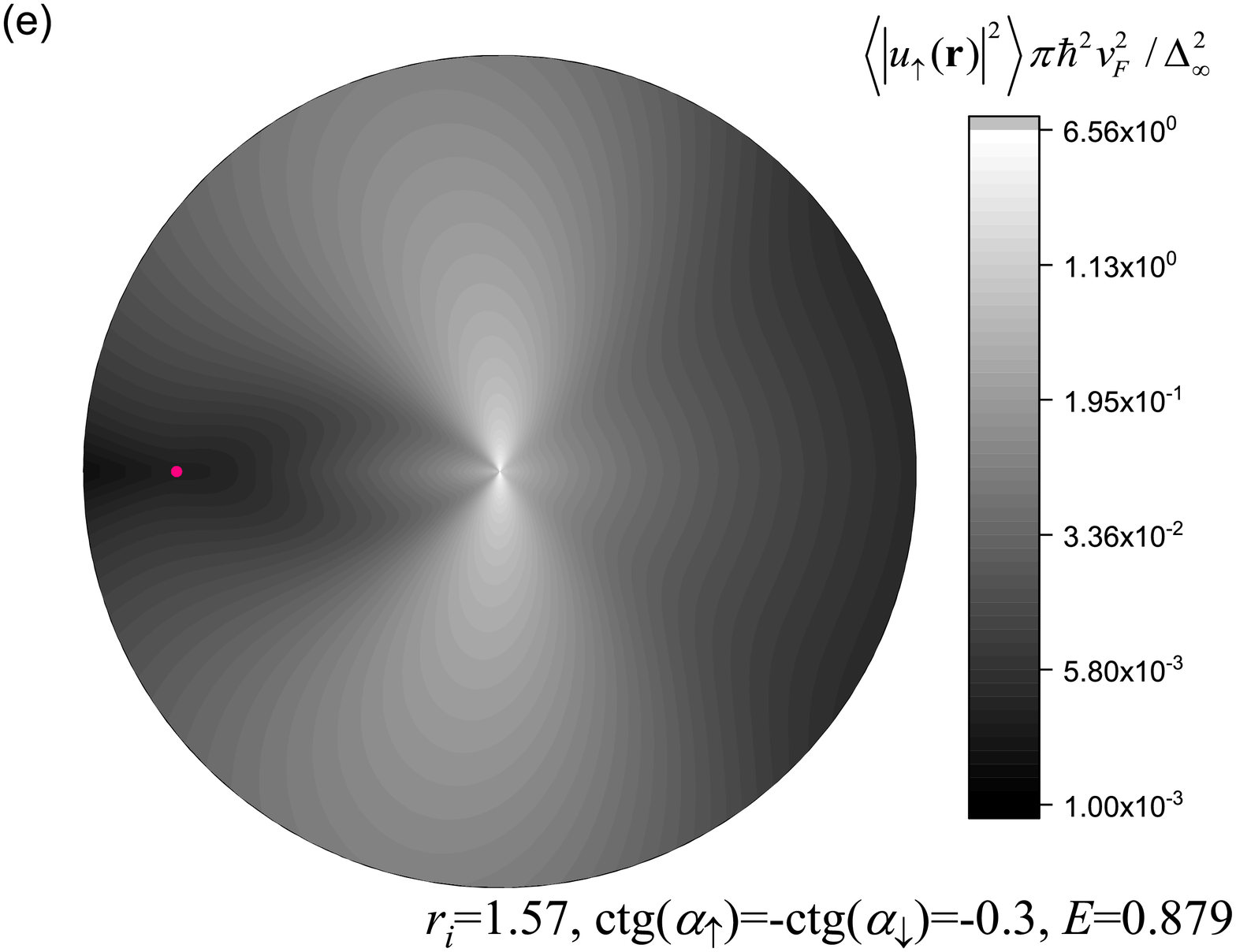}
		\includegraphics[width = 0.325\linewidth]{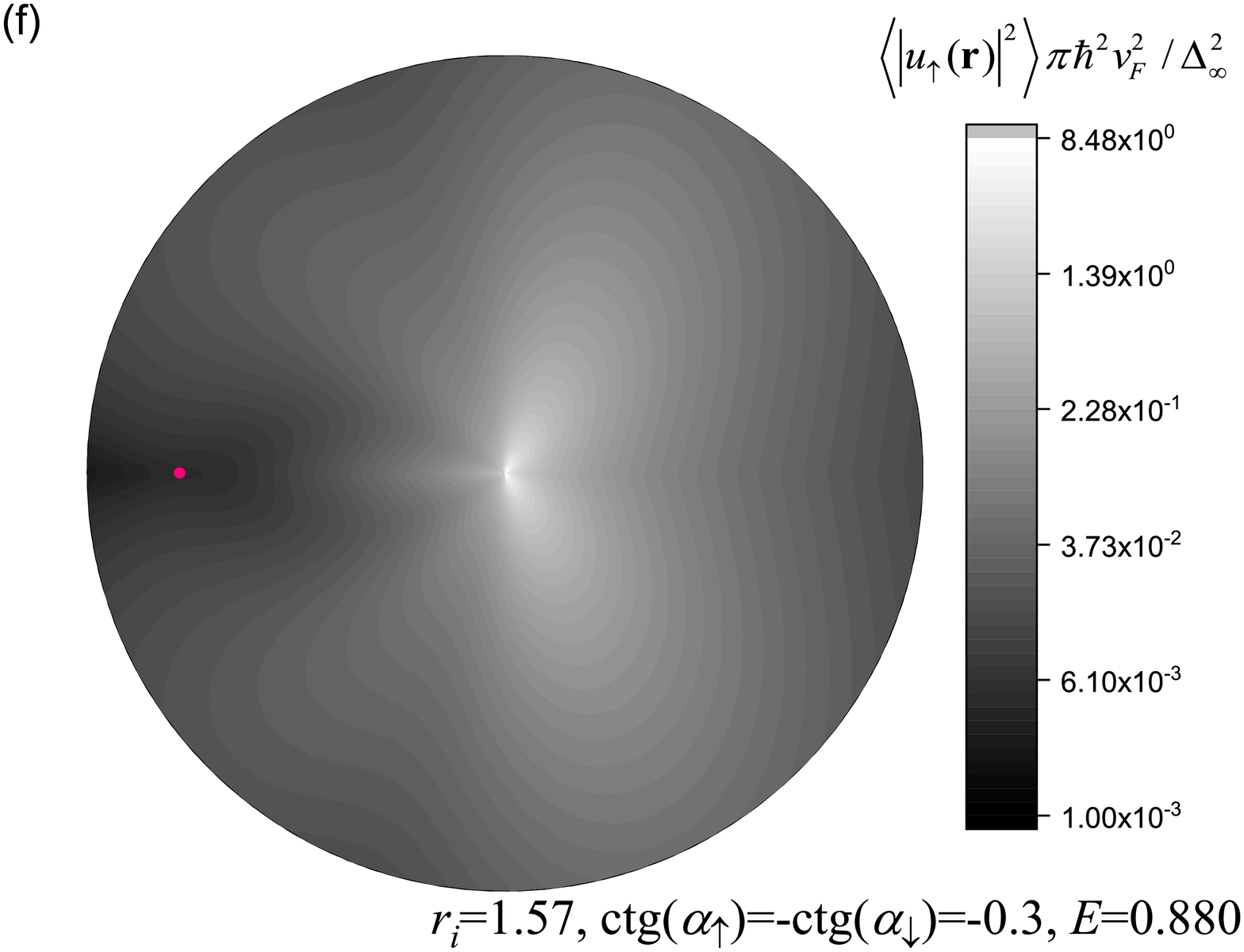}		
	\caption{Wavefunctions of spin-up impurity states in a coreless vortex (the impurity is in the center of the graphs). The radius of the area shown is $2\hbar v_F/\Delta_{\infty}$. The pink point marks the center of the vortex. Graphs in the pairs (a) and (b), (c) and (d), (e) and (f) correspond to two impurity states with the same parameters of the impurity (shown at the bottom of the graphs). Graphs (f) and (e) correspond to parameters, that are very close to a crossing of two curves of the impurity state energy vs $r_i$ dependencies, like in Fig. \ref{fig:Eimpc_2D}f.}
	\label{fig:|u|^2}
\end{figure*}

Now let us discuss the contribution of impurity states to the local density of states. Each spin-up impurity state with energy $E_{\uparrow i}$ corresponds to a normalized solution of the BdG equations $(u_{\uparrow i}(\vec{r}),v_{\uparrow i}(\vec{r}))$ (generally, the wavefunctions have also two spin-down components, however, in our case they vanish due to the special choice of the spin quantization axis). The contribution of one such state to the spin-up density of states is
\begin{equation}
	\delta \nu_{\uparrow i}(E,\vec{r}) = \abs{u_{\uparrow i}(\vec{r})}^2 \delta(E - E_{\uparrow i}).
	\label{eq:nu_up_i}
\end{equation}
Each solution of the BdG equations with spin up corresponds to a solution of these equations with spin down with a wavefunction $(u_{\downarrow i}(\vec{r}),v_{\downarrow i}(\vec{r}))$ and with energy $E_{\downarrow i} = -E_{\uparrow i}$. The components of the wavefunction can be taken in the form $u_{\downarrow i}(\vec{r}) = - v_{\uparrow i}^*(\vec{r})$, $v_{\downarrow i}(\vec{r}) = u_{\uparrow i}^*(\vec{r})$. The functions $\abs{u_{\uparrow i}(\vec{r})}^2$ and $\abs{v_{\uparrow i}(\vec{r})}^2$ oscillate in space on a scale of the order of the Fermi wavelength. It is shown in Appendix \ref{app:wavefunctions} that after averaging over these oscillations in the quasiclassical approximation one obtains
\begin{equation}
	\mean{\abs{u_{\uparrow i}(\vec{r})}^2} = \mean{\abs{v_{\uparrow i}(\vec{r})}^2},
	\label{eq:mean=mean}
\end{equation}
where $\mean{...}$ means spatial averaging. Hence, to determine the spatially averaged wavefunctions of all impurity states it is sufficient to calculate only the functions  $\mean{\abs{u_{\uparrow i}(\vec{r})}^2}$ and $\mean{\abs{u_{\downarrow i}(\vec{r})}^2}$, corresponding to positive energies.
%It follows from this that the averaged conyributions from discrete impurity states to the spin-up and spin-down densities of states, $\delta_{\uparrow}$ and $\delta_{\downarrow}$, are related via
%
%\begin{equation}
%	\mean{\delta \nu_{\uparrow}(E,\vec{r})} = \mean{\delta \nu_{\downarrow}(-E,\vec{r})}.
%	\label{eq:nu_up_down}
%\end{equation}
%
%Hence, to determine the density of states at all energies it is sufficient to determine it at positive energies only, which means that we need to calculate only the functions  $\mean{\abs{u_{\uparrow i}(\vec{r})}^2}$ and $\mean{\abs{u_{\downarrow i}(\vec{r})}^2}$, corresponding to positive energies.
The method for calculating $\mean{\abs{u_{\uparrow i}(\vec{r})}^2}$ is described in Appendix \ref{app:wavefunctions}. Characteristic profiles of these functions for a coreless vortex are shown in Fig. \ref{fig:|u|^2}.

The continuous part of the vortex spectrum is also affected by the impurity. As demonstrated in Appendix \ref{app:Imp_2D}, the impurity influences the continuous spectrum in the range of energies, lying outside the local spectral gap at position $\vec{r}_i$ without impurity, $(E^{(0)}(r_i),E_{\mathrm{max}}(r_i))$. This means that for all positions $\vec{r}$ the local gap is reduced to this energy range. Certainly, the local gap at $r>r_i$ remains unchanged. Technically, if the impurity is located sufficiently far from the vortex center, so that $r_i>r_c$, the spectral gap should completely disappear. However, the larger the distance $r_i$, the smaller the contribution of the impurity to the density of states in the vicinity of the vortex center.

To end this section, we briefly consider the influence of anisotropy effects on our results. As we have mentioned in Sec. \ref{sub:LDOS}, in real $s$-wave superconductors the order parameter and Fermi surface are always somewhat anisotropic. This anisotropy generally does not eliminate the local gap, the existence of which is the main prerequisite for the appearance of discrete impurity states. Given this, we expect that anisotropy effects will not strongly affect these states.

\section{Conclusion}
\label{sec:Conclusion}

To sum up, we have analyzed the subgap spectrum of a 2D Abrikosov vortex in an $s$-wave superconductor in the absence and presence of a point impurity. We worked in the limit $\Delta_{\infty} \ll \mu$, so that the quasiclassical approximation could be used. We considered two models of a vortex: a vortex with constant modulus of the order parameter and a vortex with an order parameter profile determined from the Ginzburg-Landau equations. The results obtained within both models are qualitatively the same.

First, we calculated the spectral branches -- Andreev state energy vs. impact parameter dependencies -- for a clean vortex, assuming an infinite magnetic field screening length. In addition to the well-known anomalous branch, we found an infinite number of upper branches. The existence of these branches becomes possible because of the Doppler effect connected with spontaneous currents in the vortex. These currents lower the effective gap edge, creating a potential well that is large enough to accommodate an infinite number of Andreev states. If screening of the magnetic field is taken into account, the number of spectral branches becomes finite, but it can be arbitrary large provided that the screening length is large enough.

Second, we calculated the local density of states of a clean vortex. We found a large position-dependent gap in the spectrum with a width of the order of $\Delta_{\infty}$ and spatial extent of the order of ten coherence lengths. The existence of such gap is a necessary condition for the appearance of discrete impurity states.

Finally, we studied the influence of an impurity on the vortex spectrum. We proved that a point impurity induces up to 4 discrete quasiparticle states and reduces the width of spectral gap mentioned above. The energies and wavefunctions of the impurity states were calculated for different positions and scattering phases of the defect. We claim that the local gap of a clean vortex as well as the impurity-induced effects can be detected in STS experiments on conventional superconductors.

We are grateful to A. S. Mel’nikov for stimulating discussions and for thorough reading of this paper. The work has been supported by Russian Foundation for Basic Research grants No. 18-42-520037 and No. 19-31-51019, and the Russian State Contract No. 0035-2019-0021.

\appendix

\section{Quasiclassical approximation for Green functions in 2D}
\label{app:quasiclassics}

In this Appendix, within the quasiclassical approximation we will derive some useful expressions for the Green function in a clean two-dimensional superconductor.

We start by considering the solution of the Gor'kov equation in vacuum at $E = 0$:
\begin{equation}
	G_{0R}(\vec{r},\vec{r}') = \frac{mi}{2\hbar^2} \mathrm{H}_0^{(1)}(k_F \abs{\vec{r} - \vec{r}'}),
	\label{eq:G_vac_2D}
\end{equation}
where $\mathrm{H}_0^{(1)}$ is the Hankel function of the first kind. For small arguments, $z \ll 1$, it has the following asymptotic behavior:
\begin{equation}
	\mathrm{H}_0^{(1)}(z) \approx 1 + \frac{2i}{\pi} \left( \ln \frac{z}{2} + \gamma \right).
	\label{eq:Hankel_smallz}
\end{equation}
Thus, there is a logarithmic peculiarity at $\vec{r} = \vec{r}'$.

In a superconductor, for $\abs{\vec{r} - \vec{r}'} \ll \xi$ in the left-hand side of Eq. \eqref{eq:Gorkov} one can neglect all terms except for the one containing $H_0$. Then, the local solution of the Gor'kov equation has the form
\begin{widetext}
\begin{equation}
	G_E^{(0)}(\vec{r},\vec{r}') = \frac{mi}{2\hbar^2} \mathrm{H}_0^{(1)}(k_F \abs{\vec{r} - \vec{r}'}) + \int g_E'(\vec{r}',\vec{n}) e^{ik_F (\vec{r} - \vec{r}') \vec{n}} \frac{d\vec{n}}{2\pi}.
	\label{eq:GE_close}
\end{equation}
Assuming that the spatial scale for $g_E'(\vec{r'},\vec{n})$ is of the order of $\xi$ (which is proven by its relation to the quasiclassical function $g_E(\vec{r'},\vec{n})$, see below), we may substitute in Eq. \eqref{eq:GE_close} $g_E'(\vec{r}',\vec{n}) \approx g_E'((\vec{r}'+\vec{r})/2,\vec{n})$. Now we obtain the quasiclassical Green function $g_E(\vec{R},\vec{n})$ from Eq. \eqref{eq:GE_close} according to the definition \cite{Kopnin-book}
\begin{equation}
	g_E(\vec{R},\vec{n}) = \int d^3 \vec{r} \int d^3 \vec{r}' \int_{-\xi_m}^{\xi_m} \frac{d \xi_p}{\pi i} G_E (\vec{r},\vec{r}') e^{i\vec{n} ( \vec{r}' - \vec{r})(k_F + \xi_p/(\hbar v_F))} \delta \left( \vec{R} - \frac{\vec{r} + \vec{r}'}{2} \right),
	\label{eq:gE_def}
\end{equation}
\end{widetext}
where $\xi_m$ is an energy such that $\abs{E} \ll \xi_m \ll \mu$. When calculating $g_E(\vec{R},\vec{n})$ we use the relation
\begin{equation}
	\frac{mi}{2\hbar^2} \mathrm{H}_0^{(1)}(k_F r) = \int \frac{e^{i\vec{k} \vec{r}}}{\frac{\hbar^2 k^2}{2m} - \frac{\hbar^2 k_F^2}{2m} - i \eps} \frac{d^2 \vec{k}}{(2\pi)^2}.
	\label{eq:H0_int}
\end{equation}
After some integration we obtain
\begin{equation}
	g_E(\vec{r},\vec{n}) = 1 - \frac{2i \hbar^2}{m} g'(\vec{r},\vec{n}).
	\label{eq:g'g}
\end{equation}
Expressing $g'$ through $g_E$ in Eq. \eqref{eq:GE_close}, we have
\begin{eqnarray}
	& \hspace{-3cm} G_E(\vec{r},\vec{r}') \approx - \frac{m}{2\hbar^2} \mathrm{Y}_0(k_F \abs{\vec{r} - \vec{r}'}) & \nonumber \\
	& + \frac{mi}{2\hbar^2} \int g_E \left( \frac{\vec{r} + \vec{r}'}{2},\vec{n'} \right) e^{ik_F (\vec{r} - \vec{r}') \vec{n}} \frac{d\vec{n}}{2\pi},&
	\label{eq:GE_close'}
\end{eqnarray}
where $\mathrm{Y}_0 (z)$ is the Neumann function. When deriving Eq. \eqref{eq:GE_close'} we have used that $\mathrm{H}_0^{(1)}(z) = \mathrm{J}_0(z) + i\mathrm{Y}_0(z)$, and
\begin{equation}
	\int e^{ik_F \vec{n} \vec{r}} \frac{d \vec{n}}{2\pi} = \mathrm{J}_0 (k_F r),
	\label{eq:J0}
\end{equation}
where $\mathrm{J}_0(z)$ is the Bessel function. Similarly to Eq. \eqref{eq:GE_close'} one can derive
\begin{equation}
	F_E^{\dagger}(\vec{r},\vec{r}') \approx \frac{mi}{2\hbar^2} \int f_E^{\dagger} \left( \frac{\vec{r} + \vec{r}'}{2},\vec{n'} \right) e^{ik_F (\vec{r} - \vec{r}') \vec{n}} \frac{d\vec{n}}{2\pi}.
	\label{eq:FE_close}
\end{equation}
Equations \eqref{eq:GER(r,r)} and \eqref{eq:FE(r,r)} follow from Eqs. \eqref{eq:GE_close'} and \eqref{eq:FE_close}.

Now we will derive an important property of the Green functions with coinciding arguments that is used in Sec. \ref{sec:ImpStates}. We start with the known expansions
\begin{equation}
	G_E^{(0)}(\vec{r},\vec{r}') = \sum_n \frac{u_n^{(0)}(\vec{r}) u^{(0)*}_n(\vec{r}')}{E_n^{(0)} - E - i\eps},
	\label{eq:G_BdG}
\end{equation}
\begin{equation}
	F_E^{\dagger (0)}(\vec{r},\vec{r}') = \sum_n \frac{v^{(0)}_n(\vec{r}) u^{(0)*}_n(\vec{r}')}{E_n^{(0)} - E - i\eps},
	\label{eq:F_BdG}
\end{equation}
where $(u_n^{(0)}(\vec{r}),v_n^{(0)}(\vec{r}))$ are the quasiparticle wavefunctions of the system without impurities, and $E_n^{(0)}$ are the corresponding energies of the quasiparticles. Let us differentiate Eqs. \eqref{eq:G_BdG} and \eqref{eq:F_BdG} by energy at $E \neq E_n^{(0)}$ and then substitute $\vec{r} = \vec{r}'$:
\begin{equation}
	\frac{\partial G_{ER}^{(0)}(\vec{r},\vec{r})}{\partial E} = \sum_{n>0} \left[ \frac{\abs{u_n^{(0)}(\vec{r})}^2}{(E_n^{(0)} - E)^2} + \frac{\abs{v_n^{(0)}(\vec{r})}^2}{(E_n^{(0)} + E)^2} \right],
	\label{eq:G'1}
\end{equation}
\begin{eqnarray}
	& \frac{\partial F_E^{\dagger (0)}(\vec{r},\vec{r})}{\partial E}  =  \sum\limits_{n>0} v^{(0)}_n(\vec{r}) u^{(0)*}_n(\vec{r}) & \nonumber \\
	& \times \left[\frac{1}{(E_n^{(0)} - E)^2} -  \frac{1}{(E_n^{(0)} + E)^2} \right], &
	\label{eq:F'1}
\end{eqnarray}
where summation is over positive energies, and we used the fact that states with negative energies $-E_n^{(0)}$ have wavefunctions $(v_n^{(0)*}(\vec{r}),-u_n^{(0)*}(\vec{r}))$. Within the quasiclassical approximation  the relation 
\begin{equation}
	G_{ER}^{(0)}(\vec{r},\vec{r}) = - G_{-ER}^{(0)}(\vec{r},\vec{r})
	\label{eq:E_vs_-E:G}
\end{equation}
holds, which follows from Eq. \eqref{eq:GER(r,r)} and the property $g_{-E}(\vec{r},\vec{n}) = -g_E(\vec{r},-\vec{n})$, which is valid for such energies that the term $i\eps$ can be discarded in the Eilenberger equations. It follows from the above that
\begin{eqnarray}
	& \frac{\partial G_{ER}^{(0)}(\vec{r},\vec{r})}{\partial E} \approx \frac{1}{2} \left[ \frac{ \partial G_{ER}^{(0)}(\vec{r},\vec{r})}{\partial E} - \frac{\partial G_{-ER}^{(0)}(\vec{r},\vec{r})}{\partial E} \right]  & \nonumber \\
	& \hspace{-0.5cm} = \!\! \sum\limits_{n>0} \frac{\abs{u_n^{(0)}(\vec{r})}^2 + \abs{v_n^{(0)}(\vec{r})}^2}{2} \left[ \frac{1}{(E_n^{(0)} - E)^2} + \frac{1}{(E_n^{(0)} + E)^2} \right] \!\!. &
	\label{eq:G'2}
\end{eqnarray}
Since
\[ \abs{v^{(0)}_n(\vec{r}) u^{(0)*}_n(\vec{r})} \leq \frac{\abs{u_n^{(0)}(\vec{r})}^2 + \abs{v_n^{(0)}(\vec{r})}^2}{2} \]
and 
\[ \frac{1}{(E_n^{(0)} \!\! - E)^2} + \frac{1}{(E_n^{(0)} \!\! + E)^2} > \! \abs{ \frac{1}{(E_n^{(0)} \!\! - E)^2} - \frac{1}{(E_n^{(0)} \!\! + E)^2}}, \]
one can see that Eqs. \eqref{eq:F'1} and \eqref{eq:G'2} yield Eq. \eqref{eq:F'<G'}.

Now consider the Green functions with non-coincident arguments in the limiting case $k_F \abs{\vec{r} - \vec{r}'} \gg 1$. Following Gor'kov and Kopnin \cite{Gorkov+72JETP_eng}, we write the Green functions in the form
\begin{widetext}
\begin{equation}
	G_E(\vec{r},\vec{r}') = \frac{m}{\hbar^2} \sqrt{\frac{i}{2\pi k_F \abs{\vec{r} - \vec{r}'}}} \left[ \tilde{g}_{E+} (\vec{r}',\vec{n},\abs{\vec{r}- \vec{r}'}) e^{ik_F \abs{\vec{r} - \vec{r}'}} + \tilde{g}_{E-} (\vec{r}',\vec{n},\abs{\vec{r}- \vec{r}'}) e^{-ik_F \abs{\vec{r} - \vec{r}'}} \right],
	\label{eq:GE_far}
\end{equation}
\begin{equation}
	F_E^{\dagger}(\vec{r},\vec{r}') = \frac{m}{\hbar^2} \sqrt{\frac{i}{2\pi k_F \abs{\vec{r} - \vec{r}'}}} \left[ \tilde{f}^{\dagger}_{E+} (\vec{r}',\vec{n},\abs{\vec{r}- \vec{r}'}) e^{ik_F \abs{\vec{r} - \vec{r}'}} + \tilde{f}^{\dagger}_{E-} (\vec{r}',\vec{n},\abs{\vec{r}- \vec{r}'}) e^{-ik_F \abs{\vec{r} - \vec{r}'}} \right],
	\label{eq:FE_far}
\end{equation}
where $\vec{n} = (\vec{r} - \vec{r}')/\abs{\vec{r}-\vec{r}'}$. Like in the 3D case \cite{Gorkov+72JETP_eng,Bespalov2018}, the Andreev equations for $\tilde{g}_{E \pm}(\vec{r}',\vec{n},s)$ and $\tilde{f}^{\dagger}_{E \pm}(\vec{r}',\vec{n},s)$ can be derived:
\begin{equation}
	\mp i\hbar v_F \frac{\partial \tilde{g}_{E \pm}}{\partial s} - E \tilde{g}_{E \pm} + \Delta(\vec{r}' + s\vec{n}) \tilde{f}^{\dagger}_{E \pm} = 0,
	\label{eq:Andreev_gpm}
\end{equation}
\begin{equation}
	\pm i\hbar v_F \frac{\partial \tilde{f}_{E\pm}^\dagger}{\partial s} - E \tilde{f}_{E \pm}^\dagger + \Delta^*(\vec{r}' + s\vec{n}) \tilde{g}_{E \pm} = 0.
	\label{eq:Andreev_fpm}
\end{equation}
Let us derive the boundary conditions for these functions. For this we transform Eq. \eqref{eq:GE_close'} in the limit $k_F\abs{\vec{r} - \vec{r}'} \gg 1$ (but $\abs{\vec{r} - \vec{r}'} \ll \xi$). In this limit the integral in Eq. \eqref{eq:GE_close'} can be calculated using the stationary phase approximation. Using also the asymptotic expression for the Neumann function, we obtain
\begin{equation}
	G_E(\vec{r},\vec{r}') \approx \frac{mi}{2\hbar^2} \sqrt{\frac{1}{2\pi k_F \abs{\vec{r} - \vec{r}'}}} \left\{ [1 + g_E(\vec{r}',\vec{n})] e^{ik_F \abs{\vec{r} - \vec{r}'} -i\pi/4} + [-1 + g_E(\vec{r}',-\vec{n})] e^{-ik_F \abs{\vec{r} - \vec{r}'} + i \pi/4} \right\}.
	\label{eq:GE_intermediate}
\end{equation}
\end{widetext}
Comparing this with Eq. \eqref{eq:GE_far}, we see that
\begin{equation}
	\tilde{g}_{E+}(\vec{r}',\vec{n},0) = \frac{1}{2} [ 1 + g_E(\vec{r}',\vec{n})],
	\label{eq:tg+_bound}
\end{equation}
\begin{equation}
	\tilde{g}_{E-}(\vec{r}',\vec{n},0) = \frac{i}{2} [ -1 + g_E(\vec{r}',-\vec{n})].
	\label{eq:tg-_bound}
\end{equation}
Similarly one obtains
\begin{equation}
	\tilde{f}^{\dagger}_{E+}(\vec{r}',\vec{n},0) = \frac{1}{2} f^{\dagger}_E(\vec{r}',\vec{n}),
	\label{eq:tf+_bound}
\end{equation}
\begin{equation}
	\tilde{f}_{E-}(\vec{r}',\vec{n},0) = \frac{i}{2} f^{\dagger}_E(\vec{r}',-\vec{n}).
	\label{eq:tf-_bound}
\end{equation}
Let us define the following two functions:
\begin{equation}
	\tilde{g}_E(\vec{r}',\vec{n},s) = \left\{
	\begin{array}{ll}
	  \tilde{g}_{E+}(\vec{r}',\vec{n},s) & \mbox{when } s>0, \\
		-i\tilde{g}_{E-}(\vec{r}',-\vec{n},-s) & \mbox{when } s<0, \\
	\end{array} \right.
	\label{eq:tg_def}
\end{equation}
\begin{equation}
	\tilde{f}^{\dagger}_E(\vec{r}',\vec{n},s) = \left\{
	\begin{array}{ll}
	  \tilde{f}^{\dagger}_{E+}(\vec{r}',\vec{n},s) & \mbox{when } s>0, \\
		-i\tilde{f}^{\dagger}_{E-}(\vec{r}',-\vec{n},-s) & \mbox{when } s<0. \\
	\end{array} \right.
	\label{eq:tf_def}
\end{equation}
Equations \eqref{eq:Andreev_gpm} and \eqref{eq:Andreev_fpm} together with the boundary conditions \eqref{eq:tg+_bound}-\eqref{eq:tf-_bound} yield
\begin{equation}
	- i\hbar v_F \frac{\partial \tilde{g}_E}{\partial s} - E \tilde{g}_E + \Delta(\vec{r}' + s\vec{n}) \tilde{f}^{\dagger}_E = -i\hbar v_F \delta(s),
	\label{eq:Andreev_tg}
\end{equation}
\begin{equation}
	i\hbar v_F \frac{\partial \tilde{f}_E^\dagger}{\partial s} - E \tilde{f}_E^\dagger + \Delta^*(\vec{r}' + s\vec{n}) \tilde{g}_E = 0.
	\label{eq:Andreev_tf}
\end{equation}
Finally, Eqs. \eqref{eq:GE_far} and \eqref{eq:FE_far} can be written in the form
\begin{widetext}
\begin{equation}
	G_E(\vec{r},\vec{r}') = \frac{mi}{\hbar^2} \sqrt{\frac{1}{2\pi k_F \abs{\vec{r} - \vec{r}'}}} \left[ \tilde{g}_E (\vec{r}',\vec{n},\abs{\vec{r}- \vec{r}'}) e^{ik_F \abs{\vec{r} - \vec{r}'}-i\pi/4} + \tilde{g}_E (\vec{r}',-\vec{n},-\abs{\vec{r}- \vec{r}'}) e^{-ik_F \abs{\vec{r} - \vec{r}'} +i\pi/4} \right],
	\label{eq:GE_far'}
\end{equation}
\begin{equation}
	F_E^{\dagger}(\vec{r},\vec{r}') = \frac{mi}{\hbar^2} \sqrt{\frac{1}{2\pi k_F \abs{\vec{r} - \vec{r}'}}} \left[ \tilde{f}^{\dagger}_E (\vec{r}',\vec{n},\abs{\vec{r}- \vec{r}'}) e^{ik_F \abs{\vec{r} - \vec{r}'}-i\pi/4} + \tilde{f}^{\dagger}_E (\vec{r}',-\vec{n},-\abs{\vec{r}- \vec{r}'}) e^{-ik_F \abs{\vec{r} - \vec{r}'}+i\pi/4} \right].
	\label{eq:FE_far'}
\end{equation}
\end{widetext}

\section{Parametriazation of quasiclassical Green functions in terms of $\psi_d(s)$}
\label{app:psi}
In this Appendix we will derive Eqs. \eqref{eq:g_psi} - \eqref{eq:bound_psi}. We start with the Riccati parametrization of the Green functions:
\begin{equation}
	g_E = \frac{1-ab}{1+ab}, \qquad f_E = \frac{-2ia}{1+ab}, \qquad f_E^{\dagger} = \frac{-2ib}{1+ab},
	\label{eq:Riccati}
\end{equation}
where the Riccati amplitudes $a(s)$ and $b(s)$ satisfy the following equations on a classical trajectory (see Fig. \ref{fig:2DFrame}):
\begin{equation}
	\frac{da}{ds} + [-2i(E + i\eps) + \Delta^* a]a - \Delta = 0,
	\label{eq:a}
\end{equation}
\begin{equation}
	\frac{db}{ds} - [-2i(E + i\eps) + \Delta b]b + \Delta^* = 0,
	\label{eq:b}
\end{equation}
Here we use the dimensionless units introduced in Sec. \ref{sub:branches}. The boundary conditions for $a$ and $b$ read
\begin{equation}
	a(-\infty) = ie^{i\theta(-\infty) - i \arccos(E + i\eps)},
	\label{eq:bound_a}
\end{equation}
\begin{equation}
	b(+\infty) = ie^{-i\theta(+\infty) - i \arccos(E + i\eps)}.
	\label{eq:bound_b}
\end{equation}
If one substitutes
\begin{equation}
	a(s) = i e^{i\psi_d(s) + i \theta(s)}
	\label{eq:a_psi}
\end{equation}
into Eqs. \eqref{eq:a} and \eqref{eq:bound_a} and takes into account that on a classical trajectory with impact parameter $d$
\begin{equation}
	\frac{d \theta}{ds} = -\frac{d}{d^2 +s^2},
	\label{eq:dtheta/ds}
\end{equation}
one obtains Eqs. \eqref{eq:psi} and \eqref{eq:bound_psi} with $E + i\eps$ instead of $E$. The same equations are obtained if one substitutes
\begin{equation}
	b(s) = i e^{i\psi_d(-s) - i \theta(s)}
	\label{eq:b_psi}
\end{equation}
into Eqs. \eqref{eq:b} and \eqref{eq:bound_b}. The imaginary contribution $i\eps$ can be easily taken into account, if one notes that at real energies the right-hand sides of Eqs. \eqref{eq:psi} and \eqref{eq:bound_psi} are monotonically increasing functions of $E$, and hence $\partial \psi_d(s)/\partial E > 0$. This means that to obtain $\psi_d$ at a complex energy $E + i\eps$ one should simply add $(\partial \psi_d(s)/\partial E) i\eps$ to $\psi_d$ determined at a real energy $E$, which is equivalent to adding $i \eps$, because $\eps$ is infinitely small, and $\partial \psi_d(s)/\partial E > 0$. Using this fact, we can obtain Eqs. \eqref{eq:g_psi} - \eqref{eq:f+_psi} from Eqs. \eqref{eq:Riccati}, \eqref{eq:a_psi} and \eqref{eq:b_psi}.

\section{Behavior of the Green functions in the clean case in the vicinity of their singularities}
\label{app:poles}

This Appendix is mainly devoted to the properties of the functions $G_{ER}^{(0)}(\vec{r},\vec{r})$, $F_E^{\dagger (0)}(\vec{r},\vec{r})$ and ${\cal D}_{\uparrow \pm}(E)$ in the vicinity of their singularities. Considerations of the functions ${\cal D}_{\uparrow \pm}(E)$ are necessary to determine the number of bound impurity states at a given position of the impurity, according to Sec. \ref{sub:Imp_general}.

For a start, let us calculate the Green functions of a coreless vortex at $E = 1$ and $r<1/4$. It turns out that these functions are finite at the gap edge. According to Schopohl \cite{Schopohl98}, the Riccati amplitude $a(s)$ [see Appendix \ref{app:psi}] at $E=1$ on a classical trajectory parallel to the $x$-axis with impact parameter $d<1/4$ equals $a_d^-(s)$, given by Eq. \eqref{eq:ad+-}. From Eq. \eqref{eq:a_psi} we then obtain
\begin{equation}
	\psi_d(s) = - \ln a_d^-(s) - i \ln \left( \frac{-s + id}{\sqrt{s^2 + d^2}} \right) - \frac{\pi}{2}.
	\label{eq:ad-psi}
\end{equation}
To determine the Green at a point $\vec{r} = (r,0)$ we substitute this into Eqs. \eqref{eq:GER(r,r)_psi} and \eqref{eq:FE(r,r)_psi}:
\begin{equation}
	G_{\Delta_{\infty} R}^{(0)} (\vec{r},\vec{r}) = \nu_0 \int_0^{2\pi} \frac{r}{\sqrt{1 - 4r \cos \varphi}} d\varphi,
	\label{eq:GER(1)}
\end{equation}
\begin{equation}
	F_{\Delta_{\infty}}^{\dagger (0)} (\vec{r},\vec{r}) = \frac{\nu_0}{2} \int_0^{2\pi} \frac{2r - \cos \varphi}{\sqrt{1 - 4r \cos \varphi}} d\varphi.
	\label{eq:FE(1)}
\end{equation}
These relations are useful for calculations of ${\cal D}_{\uparrow \pm}(\Delta_{\infty})$ for $r_i<1/4$. 

For $E<1$ the Green functions may have singularities when for some angle $\varphi$ [see Fig. \ref{fig:2DFrame}] and some integer $n$ $E^{(n)}(d(\varphi)) = E$. Then, the integrands in Eqs. \eqref{eq:GER(r,r)_psi} and \eqref{eq:FE(r,r)_psi} become infinite, because then Eq. \eqref{eq:pole1} is satisfied, where one assumes $d = d(\varphi)$ and $s = s(\varphi)$ [see Eq. \eqref{eq:sd_2D}]. Let us consider the function $\psi_d(s)$ at parameters $d>0$ and $E>0$ that are close to some number $d_0$ and the corresponding energy $E^{(n)}(d_0)$, respectively, so that Eq. \eqref{eq:pole1} is not exactly satisfied:
\begin{equation}
	d = d_0 + d_1, \qquad E = E^{(n)}(d_0) + E_1,
	\label{eq:Ed_perturbed}
\end{equation}
where $d_1$ and $E_1$ are small perturbations. The function $\psi_d(s)$ then can be written in the form $\psi_d(s) = \psi_{d_0}(s) + \tilde{\psi}(s)$, where $\psi_{d_0}(s)$ corresponds to the energy $E^{(n)}(d_0)$, so that $\psi_{d_0}(s) + \psi_{d_0}(-s) = 2\pi n$, and $\tilde{\psi}(s)$ is small. By linearizing Eqs. \eqref{eq:psi} and \eqref{eq:bound_psi} we obtain the following equations for $\tilde{\psi}(s)$:
\begin{widetext}
\begin{equation}
	\frac{d\tilde{\psi}}{ds} = 2E_1 + \frac{s^2 - d_0^2}{(d_0^2 + s^2)^2} d_1 + 2 \abs{\Delta \left( \sqrt{s^2 + d_0^2} \right)} \sin(\psi_{d_0}(s)) \tilde{\psi} - 2 \frac{\partial\abs{\Delta \left( \sqrt{s^2 + d_0^2} \right)}}{\partial d_0} \cos(\psi_{d_0}(s)) d_1,
	\label{eq:tpsi}
\end{equation}
\begin{equation}
	\tilde{\psi}(-\infty) = \frac{E_1}{\sqrt{1 - E^{(n)}(d_0)^2}}.
	\label{eq:bound_tpsi}
\end{equation}
The solution of these equations is
\begin{equation}
	 \tilde{\psi}(s) = \!\! \int_{-\infty}^s \left[2 E_1 + \frac{s^{\prime 2} - d_0^2}{(d_0^2 + s^{\prime 2})^2} d_1 - 2 \frac{\partial\abs{\Delta \left( \sqrt{s^{\prime 2} + d_0^2} \right)}}{\partial d_0} \cos(\psi_{d_0}(s')) d_1\right] \! \exp \! \left( 2 \int_{s'}^s \abs{\Delta \left(\sqrt{\tilde{s}^2 + d_0^2} \right)}\sin \psi_{d_0}(\tilde{s}) d\tilde{s} \right) \!\! ds'.
	\label{eq:tpsi_solution}
\end{equation}
The Green function $g_E$ [Eq. \eqref{eq:g_psi}] at the energy $E$ and impact parameter $d$ is then
\begin{equation}
	g_E \approx \frac{2i}{\tilde{\psi}(s) + \tilde{\psi}(-s) + i\eps},
	\label{eq:g_tpsi}
\end{equation}
and the integrand in Eq. \eqref{eq:FE(r,r)_psi} is
\begin{equation}
  \frac{\cos \left(\frac{\psi_d(s) - \psi_d(-s)}{2} \right)} {\sin \left( \frac{\psi_d(s) + \psi_d(-s)}{2} + i\eps \right)} \approx  (-1)^n \frac{2 \cos \left(\frac{\psi_{d_0}(s) - \psi_{d_0}(-s)}{2} \right)} {\tilde{\psi}(s) + \tilde{\psi}(-s) + i\eps} 
	\label{eq:F_integrand}
\end{equation}
Equation \eqref{eq:tpsi_solution} yields
\begin{equation}
	\frac{\tilde{\psi}(s) + \tilde{\psi}(-s)}{2} = \int_{-\infty}^0 \left[2 E_1 + \frac{s^{\prime 2} - d_0^2}{(d_0^2 + s^{\prime 2})^2} d_1 - 2\frac{\partial \abs{\Delta}}{\partial d_0} d_1 \cos \psi_{d_0}(s') \right] \exp\left( 2 \int_{s'}^s \abs{\Delta \left(\sqrt{\tilde{s}^2 + d_0^2} \right)}\sin \psi_{d_0}(\tilde{s}) d\tilde{s} \right) ds'.
	\label{eq:tpsi(s)(-s)}
\end{equation}
Note that here the right-hand side vanishes, and hence $g_E$ becomes infinite when $E_1/d_1  =\mathrm{d} E^{(n)}(d)/ \mathrm{d}d$, where
\begin{equation}
	\frac{\mathrm{d} E^{(n)}}{\mathrm{d} d} (d_0) = - \frac{1}{2 {\cal N}(E^{(n)}(d_0),d_0,0)} \int_{-\infty}^0 \left[\frac{s^{\prime 2} - d_0^2}{(d_0^2 + s^{\prime 2})^2} - 2\frac{\partial \abs{\Delta}}{\partial d_0} \cos \psi_{d_0}(s') \right] \exp\left( 2 \int_{s'}^0 \abs{\Delta} \sin \psi_{d_0}(\tilde{s}) d\tilde{s} \right) ds' ,
	\label{eq:dEA/dd}
\end{equation}
\begin{equation}
	{\cal N}(E,d,s) = \int_{-\infty}^0 \exp\left( 2 \int_{s'}^s \abs{\Delta \left(\sqrt{\tilde{s}^2 + d^2} \right)} \sin \psi_d(\tilde{s}) d\tilde{s} \right) ds'.
	\label{eq:N}
\end{equation}
At this point we will make a small digression to prove that the energy of the anomalous spectral branch monotonically increases as a function of the impact parameter. In Eq. \eqref{eq:dEA/dd} we integrate the first term by parts:
\begin{equation}
	\frac{\mathrm{d} E^{(n)}}{\mathrm{d} d} (d_0) = {\cal N}^{-1}(E^{(n)}(d_0),d_0,0) \int_{-\infty}^0 \left[ \frac{s'}{d_0^2 + s^{\prime 2}} \abs{\Delta} \sin \psi_{d_0}(s') + \frac{\partial \abs{\Delta}}{\partial d_0} \cos \psi_{d_0}(s') \right]\exp\left( 2 \int_{s'}^0 \abs{\Delta} \sin \psi_{d_0}(\tilde{s}) d\tilde{s} \right) ds' .
	\label{eq:dEA/dd'}
\end{equation}
For the anomalous branch one can prove that 
\begin{equation}
	-\pi/2 < \psi_{d_0}(s) \leq 0 \qquad \mbox{for} \qquad s<0.
	\label{eq:psi_anomalous}
\end{equation}
 Indeed, $-\pi/2 < \psi_{d_0}(s)$ because of the boundary condition \eqref{eq:bound_psi} and because $d \psi_d/ds>0$ at $\psi_d = -\pi/2$, according to Eq. \eqref{eq:psi}. Now, let us assume that at $s = s_1<0$ the function $\psi_{d_0}(s)$ crosses zero for the first time, so that $\psi_{d_0}(s_1) = 0$, and
\begin{equation}
	\frac{d\psi_{d_0}}{ds}(s_1) = 2E + \frac{d}{d^2 + s_1^2} -2 \abs{\Delta\left( \sqrt{s_1^2 + d^2} \right)} \geq 0.
	\label{eq:anomalous_s0}
\end{equation}
Then
\[
	\psi_{d_0}(0) = \psi_{d_0}(s_1) + \int_{s_1}^0 \frac{d\psi_{d_0}}{ds} ds \geq \int_{s_1}^0 \left[ 2E + \frac{d}{d^2 + s^2} -2 \abs{\Delta\left( \sqrt{s^2 + d^2} \right)} \right] ds > \int_{s_1}^0 \frac{d\psi_{d_0}}{ds}(s_1) ds \geq 0,
\]
\end{widetext}
so that $\psi_{d_0}(0) > 0$, which contradicts the condition $\psi_{d_0}(0) = 0$ for the anomalous branch. This proves Eq. \eqref{eq:psi_anomalous}. This equation, in turn, means that the integrand in Eq. \eqref{eq:dEA/dd'} is positive, and hence the whole right-hand side is positive, what was to be shown.

For $d_0 = +0$, using Eq. \eqref{eq:psi_small_d}, one may obtain from Eq. \eqref{eq:dEA/dd'} the known result \cite{CdGM64} for the slope of the anomalous branch at $d = 0$:
\begin{equation}
	\frac{\mathrm{d} E^{(0)}}{\mathrm{d} d} (0) = \frac{\int_0^{\infty} \frac{\abs{\Delta(s)}}{s} \exp\left( -2 \int_0^s \abs{\Delta(s')} ds' \right) ds}{\int_0^{\infty} \exp\left( -2 \int_0^s \abs{\Delta(s')} ds' \right) ds}.
	\label{eq:dEA0/dd(0)}
\end{equation}

In the following we will need only Eq. \eqref{eq:tpsi(s)(-s)} with $d_1 = 0$ ($d=d_0$):
\begin{equation}
	\frac{\tilde{\psi}(s) + \tilde{\psi}(-s)}{2} = 2 [E - E^{(n)}(d)] {\cal N}(E^{(n)}(d),d,s),
	%\int_{-\infty}^0 \exp\left( 2 \int_{s'}^s \abs{\Delta \left(\sqrt{\tilde{s}^2 + d^2} \right)}\sin \psi_{d}(\tilde{s}) d\tilde{s} \right) ds',
	\label{eq:tpsi(s)(-s)'}
\end{equation}
which is valid for any $d$ and for $E \approx E^{(n)}(d)$. We will use Eq. \eqref{eq:tpsi(s)(-s)'} first to estimate $\nu(E,r)$ at $E \approx E^{(0)}(r)$. Using Eqs. \eqref{eq:g_tpsi} and \eqref{eq:tpsi(s)(-s)'}, we can write the real part of $g_E$ in the form
\begin{equation}
	\mathrm{Re}[g_E(d,s)] = \frac{\pi}{2 {\cal N}(E,d,s)}  \sum_n \delta(E - E^{(n)}(d)).
	\label{eq:Re_gE}
\end{equation}
Here, all spectral branches have been taken into account, and for convenience we use $d$ and $s$ as the arguments of $g_E$ instead of $\vec{r}$ and $\vec{n}$ (due to the roatational symmetry of the system, the value of $g_E$ is defined by two coordinates). Using Eqs. \eqref{eq:GER(r,r)} and \eqref{eq:nu}, we can write the density of states in the form
\begin{eqnarray}
	& \nu(E,\vec{r}) = \nu_0 \int_{-\pi}^{\pi} \mathrm{Re}[g_E(d(r,\varphi),s(r,\varphi))] \frac{d\varphi}{2 \pi} & \nonumber \\
	& = \nu_0 \int_{-\pi/2}^{\pi/2} \mathrm{Re}[g_E(d(r,\varphi),s(r,\varphi))] \frac{d\varphi}{\pi}. &
	\label{eq:nu2D_phi}
\end{eqnarray}
Here, we used that $g_E(d,s) = g_E(d,-s)$ -- see Eq. \eqref{eq:g_psi}.

Let us take $E \approx E^{(0)}(r) < \min_d E^{(1)}(d)$, so that in the sum in Eq. \eqref{eq:Re_gE} only the term with $n=0$ is relevant. One can see then that for $E > E^{(0)}(r)$ the density of states vanishes. For $E < E^{(0)}(r)$ the integrand in Eq. \eqref{eq:nu2D_phi} does not vanish only for $\varphi \approx \pi/2$, so that we can put $s(\varphi) = 0$ and $\psi_d \approx \psi_r$:
\begin{equation}
	\nu(E,\vec{r}) = \frac{\nu_0}{2 {\cal N}(E^{(0)}(r),r,0)} \int_0^{\pi/2} \!\!\!\! \delta \left( E - E^{(0)}(r \sin \varphi) \right) d\varphi.
	\label{eq:nu_near_EA0}
\end{equation}
In the vicinity of $\varphi = \pi/2$
\begin{equation}
	E^{(0)}(r \sin \varphi) \approx E^{(0)}(r) - \frac{r}{2} \frac{\mathrm{d}E^{(0)}}{\mathrm{d}d}(r) \left( \frac{\pi}{2} - \varphi \right)^2.
	\label{eq:EA0_expand}
\end{equation}
Now we can integrate over $\varphi$ in Eq. \eqref{eq:nu_near_EA0}:
\begin{equation}
	\nu(E,\vec{r}) = \frac{\nu_0 {\cal N}^{-1}(E^{(0)}(r),r,0)}{2 \sqrt{E^{(0)}(r) - E}} \left(2r \frac{\mathrm{d}E^{(0)}}{\mathrm{d}d}(r) \right)^{-1/2} \!\!\!\!\!\!\!\!\!.
	\label{eq:nu_near_EA0'}
\end{equation}
One can see that the density of states has an inverse square root singularity.

Now we will calculate the Green functions for $E \to E^{(0)}(r) +0$. Then, the imaginary term $i\eps$ in Eqs. \eqref{eq:g_tpsi} and \eqref{eq:F_integrand} can be discarded, and the main contribution to the integral in Eq. \eqref{eq:GER(r,r)_psi} comes from $\varphi \approx \pi/2$. We may use Eqs. \eqref{eq:g_tpsi} and \eqref{eq:tpsi(s)(-s)'} and put $s(\varphi) = 0$. After integrating $[E - E^{(0)}(d(\varphi))]^{-1}$ over $\varphi$ with the help of Eq. \eqref{eq:EA0_expand} we obtain
\begin{eqnarray}
  & \hspace{-4cm} G_{ER}^{(0)}(\vec{r},\vec{r}) \approx -\frac{\pi \nu_0}{2 {\cal N}(E^{(0)}(r),r,0)} & \nonumber \\
	& \times \left[ 2 r \frac{\mathrm{d} E^{(0)}}{\mathrm{d} d}(r) (E - E^{(0)}(r)) \right]^{-1/2}. &
	\label{eq:GER(Emin)}
\end{eqnarray}
Hence, $G_{ER}^{(0)}(\vec{r},\vec{r}) \to - \infty$ when $E \to E^{(0)}(r) + 0$. After doing similar transformations in Eq. \eqref{eq:FE(r,r)_psi}, using Eq. \eqref{eq:F_integrand} we find that $F_E^{\dagger(0)}(\vec{r},\vec{r}) \approx G_{ER}^{(0)}(\vec{r},\vec{r})$. Taking the difference of Eqs. \eqref{eq:GER(r,r)_psi} and \eqref{eq:FE(r,r)_psi} one can also prove that the difference  $F_E^{\dagger(0)}(\vec{r},\vec{r}) - G_{ER}^{(0)}(\vec{r},\vec{r})$ is finite at $E = E^{(0)}(r) + 0$. This means that for all $r_i<r_c$ we have
\begin{equation}	
	\lim_{E \to E^{(0)}(r_i)} {\cal D}_{\uparrow-}(E) = -\infty,
	\label{eq:D-(Emin)}
\end{equation}
\begin{eqnarray}
	 & \lim\limits_{E \to E^{(0)}(r_i)} \!\!\!\! {\cal D}_{\uparrow+}(E)  = \!\!\!\! \lim\limits_{E \to E^{(0)}(r_i)} \!\! \left[G_{ER}^{(0)}(\vec{r}_i,\vec{r}_i) \! - \! F_E^{\dagger (0)}(\vec{r}_i,\vec{r}_i) \right] & \nonumber\\
	& - \frac{m}{4\hbar^2} ( \cot \alpha_{\uparrow} - \cot \alpha_{\downarrow} ). &
	\label{eq:D(2d)+(Emin)}
\end{eqnarray}

Similar calculations can be done for the range of parameters $r < d_{\mathrm{min}}^{(1)}$ and $E \to E^{(1)}(r)-0$. %where $d_{\mathrm{min}}^{(1)}$ is the value of the impact parameter at which $E^{(1)}(d)$ has a minimum. 
We find then $G_{ER}^{(0)}(\vec{r},\vec{r}) \approx - F_E^{\dagger(0)}(\vec{r},\vec{r}) \propto [E^{(1)}(r) - E]^{-1/2}$. For a coreless vortex this means that if $r_i$ is in the range $1/4<r_i<d_{\mathrm{min}}^{(1)}$ [region B in Fig. \ref{fig:nu0}], then ${\cal D}_{\uparrow +}(E_{\mathrm{max}}(r_i)) = +\infty$, and
\begin{eqnarray}
  & \lim\limits_{E \to E_{\mathrm{max}}(r_i)} \!\!\! {\cal D}_{\uparrow-}(E) = \!\!\!\! \lim\limits_{E \to E_{\mathrm{max}}(r_i)} \!\! \left[G_{ER}^{(0)}(\vec{r}_i,\vec{r}_i) \! + \! F_E^{\dagger (0)}(\vec{r}_i,\vec{r}_i) \right] & \nonumber \\
	& - \frac{m}{4\hbar^2} ( \cot \alpha_{\uparrow} - \cot \alpha_{\downarrow} ). &
	\label{eq:D(2D)-(Emax)}
\end{eqnarray}
It follows from the considerations above that the case (ii) from Sec. \ref{sub:Imp_general} is impossible for $1/4<r_i<d_{\mathrm{min}}^{(1)}$ for a coreless vortex (or for $r_i<d_{\mathrm{min}}^{(1)}$ for a vortex with core).

Finally, consider the range of parameters $d_{\mathrm{min}}^{(1)} < r < r_c$ and $E \approx E^{(1)}_{\mathrm{min}}$. %, where $E^{(1)}_{\mathrm{min}} = E^{(1)}(d_{\mathrm{min}}^{(1)})$. 
Here, the main contribution to the integrals in Eqs. \eqref{eq:GER(r,r)_psi} and \eqref{eq:FE(r,r)_psi} comes from $\varphi \approx \varphi_0 = \arcsin(d_{\mathrm{min}}^{(1)}/r)$. For such $\varphi$ Eq. \eqref{eq:tpsi(s)(-s)'} yields
\begin{widetext}
\begin{equation}
	\frac{\tilde{\psi}(s) + \tilde{\psi}(-s)}{2} \approx 2 \left[ E - E_{\mathrm{min}}^{(1)} - \frac{1}{2} \left(d - d_{\mathrm{min}}^{(1)} \right)^2 \frac{\mathrm{d}^2 E^{(1)}}{\mathrm{d} d^2} \left( d_{\mathrm{min}}^{(1)} \right) \right] {\cal N}\left(E_{\mathrm{min}}^{(1)},d_{\mathrm{min}}^{(1)},s(\varphi_0) \right).
	\label{eq:tpsi(s)(-s)2}
\end{equation}
Now we can calculate $G_{ER}^{(0)}(\vec{r},\vec{r})$ using Eq. \eqref{eq:GER(r,r)_psi}. We take into account that $s(\varphi_0) = -\sqrt{r^2 - d_{\mathrm{min}}^{(1)2}}$ and $d(\varphi) - d_{\mathrm{min}}^{(1)} \approx \sqrt{r^2 - d_{\mathrm{min}}^{(1)2}} (\varphi - \varphi_0)$. Then, we obtain the following integral over $\varphi$:
\begin{equation}
	\int\limits_{-\infty}^{+\infty} \frac{d\varphi}{E - E_{\mathrm{min}}^{(1)} - \frac{1}{2} \frac{\mathrm{d}^2 E^{(1)}}{\mathrm{d} d^2} \left( d_{\mathrm{min}}^{(1)} \right) \left(r^2 - d_{\mathrm{min}}^{(1)2} \right)(\varphi - \varphi_0)^2}  = -\pi \sqrt{\frac{2}{\left(E_{\mathrm{min}}^{(1)} - E \right) \frac{\mathrm{d}^2 E^{(1)}}{\mathrm{d} d^2} \left( d_{\mathrm{min}}^{(1)} \right) \left(r^2 - d_{\mathrm{min}}^{(1)2} \right)}},
	\label{eq:phi_int}
\end{equation}
and the Green function takes the form
\begin{equation}
   G_{ER}^{(0)}(\vec{r},\vec{r}) \approx \frac{\pi \nu_0 {\cal N}^{-1} \left(E_{\mathrm{min}}^{(1)},d_{\mathrm{min}}^{(1)},\sqrt{r^2 - d_{\mathrm{min}}^{(1)2}} \right)}{\sqrt{2 \left(E_{\mathrm{min}}^{(1)} - E \right) \frac{\mathrm{d}^2 E^{(1)}}{\mathrm{d} d^2} \left(d_{\mathrm{min}}^{(1)} \right) \left( r^2 - d_{\mathrm{min}}^{(1)2} \right)}}.
	\label{eq:GE(1)_min_R}
\end{equation}
Thus, $G_{ER}^{(0)}(\vec{r},\vec{r}) \to +\infty$ when $E \to E_{\mathrm{min}}^{(1)} - 0$. Similarly, from Eq. \eqref{eq:FE(r,r)_psi} we may obtain
\begin{equation}
	F_E^{\dagger (0)}(\vec{r},\vec{r}) \approx - \cos \left( \frac{\psi_{d_{\mathrm{min}}^{(1)}}\left(\sqrt{r^2 - d_{\mathrm{min}}^{(1)2}} \right) - \psi_{d_{\mathrm{min}}^{(1)}}\left(-\sqrt{r^2 - d_{\mathrm{min}}^{(1)2}} \right)}{2} \right) G_{ER}^{(0)}(\vec{r},\vec{r}).
	\label{eq:FE(0.9777)}
\end{equation}
\end{widetext}
The absolute value of the cosine in the right-hand side here is unity with zero probability, which means that almost certainly
\begin{equation}
	\lim_{E \to E_{\mathrm{min}}^{(1)} - 0} \left[ G_{ER}^{(0)}(\vec{r},\vec{r}) - \abs{F_E^{\dagger (0)}(\vec{r},\vec{r})} \right] = +\infty.
	\label{eq:E_to_Emin(1)}
\end{equation}
As a result, for $d_{\mathrm{min}}^{(1)}<r_i<r_c$ both ${\cal D}_{\uparrow+}(E)$ and ${\cal D}_{\uparrow-}(E)$ tend to $+\infty$ when $E$ tends to $E_{\mathrm{\max}}(r_i)$. Hence, the cases (i) and (ii) from Sec. \ref{sub:Imp_general} are not possible for such positions of the impurity, and there is at least one impurity-induced state. 

\section{Qualitative analysis of the subgap spectral branches}
\label{app:qualitative}

In this Appendix the qualitative structure of the upper spectral branches is derived. 

We start with the case of a coreless vortex. For $\abs{\Delta} = 1$ and $E = 1$ the analytical solution of Eq. \eqref{eq:a} is known. Consider a classical trajectory, such that $\theta (-\infty) = 0$. In our coordinate frame this trajectory is directed towards the $x$ axis. The order parameter on this trajectory equals 
\begin{equation}
	\Delta(s) = - \frac{s + id}{\sqrt{s^2 + d^2}}.
	\label{eq:Delta(s)}
\end{equation}
For such order parameter profile two particular solutions to Eq. \eqref{eq:a} for $E=1$ and $d \neq 1/4$ are \cite{Schopohl98}
\begin{equation}
	a_d^{\pm} = \frac{\pm \sqrt{1-4d} - 2i\sqrt{s^2 + d^2}}{2s - i (2d -1)}.
	\label{eq:ad+-}
\end{equation}
Based on these two solutions we can construct the general solution of the Riccati equaiton:
\[ \ln \left( \frac{a_d - a_d^+}{a_d - a_d^-} \right) = \int e^{-i\theta(s)} [a_d^-(s) - a_d^+(s)] ds, \]
or
\begin{equation}
	\frac{a_d - a_d^+}{a_d - a_d^-} = C G(s),
	\label{eq:ad(E=1)}
\end{equation}
where $C$ is an arbitrary constant, and
\begin{eqnarray}
  & G(s) = \frac{(1-2d) \sqrt{s^2 + d^2} -s\sqrt{1-4d}}{2\sqrt{s^2 + d^2} - i \sqrt{1-4d}} \abs{d}^{-1} & \nonumber \\
	& \times \left( \frac{s + \sqrt{s^2 + d^2}}{\abs{d}} \right)^{\sqrt{1-4d}}. &
	\label{eq:G}
\end{eqnarray}
In the following we will use the functions $\bar{a}_d$ and $\breve{a}_d$, which satisfy Eq. \eqref{eq:a} with $E = 1$ and the intial conditions
\begin{equation}
	\bar{a}_d(0) = 1,
	\label{eq:bar_a_init}
\end{equation}
\begin{equation}
	\breve{a}_d(0) = -1.
	\label{eq:breve_a_init}
\end{equation}
These functions are given by
\begin{equation}
	\bar{a}_d(s) = \frac{a_d^+(s) - G(s) a_d^-(s)}{1 - G(s)},
	\label{eq:bar_a}
\end{equation}
\begin{equation}
	\breve{a}_d(s) = \frac{a_d^+(s) + G(s) a_d^-(s)}{1 + G(s)}.
	\label{eq:breve_a}
\end{equation}
Let us define the functions $\bar{\psi}_d(E,s)$ and $\breve{\psi}_d (E,s)$ as the solutions of Eq. \eqref{eq:psi} with the initial conditions
\begin{equation}
	\bar{\psi}_d(E,0) = 0, \qquad \breve{\psi}_d (E,0) = \pi.
	\label{eq:bar_breve_init}
\end{equation}
For $d>0$ and $E=1$ they can be expressed in terms of $\bar{a}_d$ and $\breve{a}_d$ using Eq. \eqref{eq:a_psi}:
\begin{equation}
	\bar{\psi}_d^c(1,s) = -i \ln \bar{a}_d(s) + \arctan \frac{s}{d},
	\label{eq:bar_psi_ad}
\end{equation}
\begin{equation}
	\breve{\psi}_d^c(1,s) = -i \ln \breve{a}_d(s) + \arctan \frac{s}{d}.
	\label{eq:breve_psi_ad}
\end{equation}
The upper index ``c" here means that these functions correspond to the coreless vortex. Consider $d$ in the range $0<d<1/4$. One can see that
\begin{equation}
	\lim_{s\to -\infty} \breve{a}_d(s) = i,
	\label{eq:lim_a_breve}
\end{equation}
and hence
\[ \lim_{s\to -\infty} \breve{\psi}_d^c(1,s) = 2\pi k, \]
where $k$ is an integer that we will determine now. First, note that $k \leq 0$, since for $E = 1$ the right-hand side of Eq. \eqref{eq:psi} is non-negative, so that $\breve{\psi}_d^c(1,0)>\breve{\psi}_d^c(1,-\infty)$. Let us introduce one more function,
\begin{eqnarray}
	& \psi_d^-(s) = -i \ln a^-_d(s) + \arctan \frac{s}{d} & \nonumber\\
	& = \arctan \frac{2\sqrt{s^2 + d^2}}{\sqrt{1-4d}} + \arctan\frac{1-2d}{-2s} + \arctan \frac{s}{d}. &
	\label{eq:psi^-}
\end{eqnarray}
It satisfies Eq. \eqref{eq:psi} with the initial condition
\begin{equation}
	\psi_d^-(0) = \frac{\pi}{2} + \arctan \frac{2d}{\sqrt{1-4d}}.
	\label{eq:psi^-(0)}
\end{equation}
It follows from Eq. \eqref{eq:psi^-} that $\psi_d^{-}(-\infty) = 0$. Since $\breve{\psi}_d^c(1,0) > \psi_d^-(0)$, for all $s<0$ we have  $\breve{\psi}_d^c(1,s) \geq \psi_d^-(s)$, and thus $\breve{\psi}_d^c(1,-\infty) = 0$. Due to the monotonicity of the right-hand side of Eq. \eqref{eq:psi} in energy, for $E<1$ we find that $\breve{\psi}_d^c(E,s) > \breve{\psi}_d^c(1,s)$ for $s<0$ and hence $\breve{\psi}_d^c(E,-\infty) \geq 0$ for $E<1$. This means that a function $\psi_d(s)$ that satisfies Eqs. \eqref{eq:psi}, \eqref{eq:bound_psi} and Eq. \eqref{eq:psi(0)} with $n=1$ does not exist for $E<1$, and hence there is no spectral branch with $n=1$. Certainly, branches with $n>1$ are absent as well.

Now consider $d>1/4$. One can check then that $a_d^- = 1/a_d^{+*}$ and $\abs{a_d^+} = \abs{G}$. Then
\begin{equation}
	\bar{a}_d(s) = a_d^+ \frac{1-G/\abs{G}^2}{1 - G}.
	\label{eq:d>1/4_abar}
\end{equation}
For $-s \gg d$
\begin{equation}
	G(s) \approx \frac{1 - 2d + i \sqrt{4d -1}}{2d} e^{-i\sqrt{4d-1} \ln \left(\frac{-2s}{d} \right)}.
	\label{eq:G_asympt}
\end{equation}
When $\arg(G) = 0$ we have $\bar{a}_d \approx -a_d^+ \approx -i$. On the other hand, when $\arg(G) = \pi$ one obtains $\bar{a}_d \approx a_d^+ \approx i$. Since with decreasing $s$ the function $G(s)$ goes around the origin in the complex plane an infinite number of times [see Eq. \eqref{eq:G_asympt}], the function $\bar{a}_d(s)$ has no limit when $s \to -\infty$. Then the function $\bar{\psi}_d^c(1,s)$ has no finite limit when $s \to -\infty$. Moreover, it is monotonous in $s$, so that
\begin{equation}
	\lim_{s \to - \infty} \bar{\psi}_d^c(1,s) = -\infty \qquad (d>1/4).
	\label{eq:bar_psi_infty}
\end{equation}
Now we note that the function $\bar{\psi}_d^c(E,s)$ is uniformly continuous in $E$ on any finite interval of the variable $s$. It follows from this that $\bar{\psi}_d^c(E,s)$ reaches arbitrarily large negative values at $s<0$, if the energy is sufficiently close to 1. Then we may obtain $\bar{\psi}_d^c(E,-\infty) = -2\pi k \pm \arccos E$ with arbitrary large $k$, for energies close to 1. Since for every solution $\psi_d(s)$ of Eq. \eqref{eq:psi} $\psi_d(s) + 2\pi k$ is also a solution, for $d>1/4$ we can find a solution of Eqs. \eqref{eq:psi}, \eqref{eq:bound_psi} and \eqref{eq:psi(0)} with arbitrary large $n$. This means that the energy interval $E \in (1-\delta E,1)$ ($\delta E>0$) for $d > 1/4$ contains an infinite amount of spectral branches $E^{(n)}(d)$.

Finally, let us consider $d<0$. We can see that $d\psi_d/ds<0$ for $\psi_d \in (-\arccos(E),0)$. As a consequence, we inevitably have $\psi_d(0)<0$ for $E<1$, if $\psi_d(s)$ satisfies the boundary condition \eqref{eq:bound_psi}. Hence, the spectral branch with $n=1$ is absent, as well as all other higher branches.

Now we will generalize the above consideration for a vortex with core. Let us assume that the order parameter at $r \to \infty$ has the asymptotic behavior given by Eq. \eqref{eq:Delta=1-b/r^2}. We will analyze the behavior of the function $\bar{\psi}_d(1,s)$ when $s \to -\infty$. We introduce the variable $\tilde{s}$ via
\begin{equation}
	\tilde{s} = \int_0^s \abs{\Delta \left( \sqrt{s^{\prime 2} + d^2} \right)} ds'.
	\label{eq:tilde_s'}
\end{equation}
By deviding Eq. \eqref{eq:psi} by $\abs{\Delta \left( \sqrt{s^2 + d^2} \right)}$, we obtain
\begin{equation}
	\frac{\partial \bar{\psi}_d(1,s)}{\partial \tilde{s}} = 2 - 2 \cos \bar{\psi}_d + f(\tilde{s}),
	\label{eq:bar_psi&f}
\end{equation}
where
\begin{equation}
	f(\tilde{s}) = \frac{2 + \frac{d}{d^2+s^2}}{\abs{\Delta \left( \sqrt{s^2 + d^2} \right)}} - 2.
	\label{eq:f}
\end{equation}
In the limit $s \to -\infty$
\begin{equation}
	f(\tilde{s}) = \frac{d + 2h}{\tilde{s}^2} + o(\tilde{s}^{-2}).
	\label{eq:f_asympt}
\end{equation}

Let us take $d + 2h > 1/4$. We choose a number $d'$, such that $1/4< d' < d+2h$. Then, a number $s_0$ exists, such that for $s<s_0$
\begin{equation}
	\frac{d'}{d^{\prime 2} + \tilde{s}(s)^2}< f(\tilde{s}(s)).
	\label{eq:s&f}
\end{equation}
We can compare the function $\bar{\psi}_d(1,s)$ with the function $\bar{\psi}_{d'}^c(1,\tilde{s})$ [defined above by Eq. \eqref{eq:bar_psi_ad}], which satisfies the equation
\begin{equation}
	\frac{\partial \bar{\psi}_{d'}^c(1,\tilde{s})}{\partial \tilde{s}} = 2 - 2 \cos \bar{\psi}_{d'}^c + \frac{d'}{d^{\prime 2} + \tilde{s}^2}.
	\label{eq:bar_psi&d'}
\end{equation}
Note that the functions $\bar{\psi}_{d'}^c(1,\tilde{s}) + 2\pi n$ with any integer $n$ satisfy this equation. For some value of $n$ the inequality $\bar{\psi}_d(1,s_0)< \bar{\psi}_{d'}^c(1,\tilde{s}(s_0)) + 2\pi n$ holds, where $\tilde{s}(s_0)$ is defined by Eq. \eqref{eq:tilde_s'}. By comparing the right-hand sides of Eqs. \eqref{eq:bar_psi&f} and \eqref{eq:bar_psi&d'}, taking into account Eq. \eqref{eq:s&f} we find that $\bar{\psi}_d(1,s)< \bar{\psi}_{d'}^c(1,\tilde{s}(s)) + 2\pi n$ for $s<s_0$. Due to Eq. \eqref{eq:bar_psi_infty} we have
\begin{equation}
	\lim_{s \to - \infty} \bar{\psi}_d(1,s) = -\infty \qquad (d + 2h >1/4).
	\label{eq:bar_psi_infty'}
\end{equation}
This means that for $d + 2h>1/4$ there is an infinite amount of spectral branches with positive energies.

In the case $d + 2h<1/4$ we take the number $d'$ in the interval $d+2h<d'<1/4$. Then, by comparing the functions $\bar{\psi}_{d'}^c(1,\tilde{s}) + 2\pi n$ with a sufficiently large negative $n$ and $\bar{\psi}_d(1,s)$ we find that
\begin{equation}
	\lim_{s \to -\infty} \bar{\psi}_d(1,s) > 	\lim_{s \to -\infty} \bar{\psi}_{d'}^c(1,\tilde{s}(s)) + 2\pi n > -\infty.
	\label{eq:limit_comparison}
\end{equation}
This means that the number of spectral braches with positive energies is finite.

The coefficient $h$ in Eq. \eqref{eq:Delta=1-b/r^2} at temperatures $T$ close to the critical temperature $T_c$ can be determined from the Ginzburg-Landau equation:
\begin{equation}
	-q \nabla^2 \Delta - \Delta + \abs{\Delta}^2 \Delta = 0.
	\label{eq:GL}
\end{equation}
We wrote it in the dimensionless form, where the energy is measured in units \cite{Kopnin-book}
\begin{equation}
	\Delta_{\infty} = \sqrt{\frac{8\pi^2}{7 \zeta(3)} T_c (T_c - T)}.
	\label{eq:GL_Delta}
\end{equation}
The coefficient $q$ in Eq. \eqref{eq:GL} in 3D equals $q_{3D} = 1/6$ \cite{Kopnin-book}. For superconductors with a cylindrical Fermi surface (2D case) one can show that the coefficient $q$ equals $q_{2D} = 3q_{3D}/2 = 1/4$ \cite{GL-derivation}. A simple explanation of this is that $q$ is proportional to $\mean{n_x^2}_{\mathrm{FS}}$, where $n_x$ is the $x$-projection of a unit normal vector to the Fermi surface, and $\mean{...}_{\mathrm{FS}}$ means averaging over the Fermi surface. For a spherical Fermi surface $\mean{n_x^2}_{\mathrm{FS} \, 3D} = 1/3$, while for a cylindrical Fermi surface $\mean{n_x^2}_{\mathrm{FS}  \, 2D} = 1/2$, so that $q_{3D}/q_{2D} = 2/3$.

\begin{figure}[t]
	\centering
		\includegraphics[width = \linewidth]{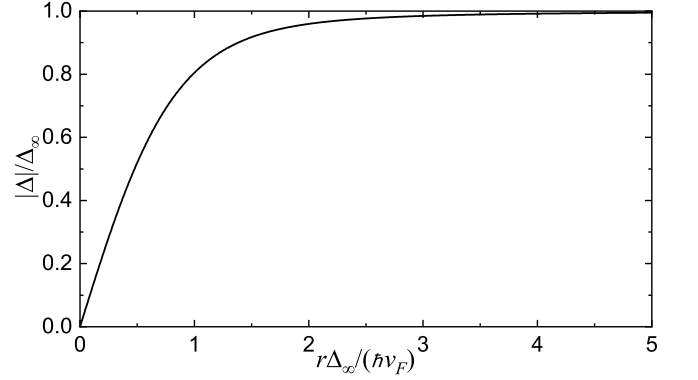}
	\caption{The order parameter profile obtained from the Ginzburg-Landau equation [Eq. \eqref{eq:GL}].}
	\label{fig:Delta}
\end{figure}

After substituting the order parameter given by Eq. \eqref{eq:Delta} into Eq. \eqref{eq:GL} one obtains an equation for $\abs{\Delta(r)}$. The order parameter profile obtained from this equation is shown in Fig. \ref{fig:Delta}. The asymptotic expansion of $\abs{\Delta(r)}$ at $r \to \infty$ is
\begin{equation}
	\abs{\Delta(r)} = 1 - \frac{q}{2r^2} + o(r^{-2}).
	\label{eq:GL_asympt}
\end{equation}
If one compares this with Eq. \eqref{eq:Delta=1-b/r^2}, one can see that in the 2D case $h = 1/8$, and hence an infinite number of spectral branches appear at $d>0$.

\section{Green functions in the presence of a point impurity in 2D}
\label{app:Imp_2D}

In this Appendix we will derive and analyze the Green function of a 2D superconducting system with a point impurity.

Let us first consider a 2D vacuum with a point impurity located at the origin. For a plane wave with a wavefunction $e^{ikx}$ incident on the impurity the whole wavefunction with the scattered wave is
\begin{equation}
	\psi = e^{ikx} + C \mathrm{H}_0^{(1)}(kr),
	\label{eq:psi_2D_scattered}
\end{equation}
where $C$ is some scattering amplitude. Let us take a superposition of such wavefunctions with plane waves propagating in all directions with equal amplitudes:
\begin{eqnarray}
	& \psi_S = \int_{\abs{\vec{n}} = 1} \left[ e^{ik\vec{n} \vec{r}} + C \mathrm{H}_0^{(1)}(kr) \right] \frac{d\vec{n}}{2\pi} & \nonumber \\
	& = \left( C + \frac{1}{2} \right) \mathrm{H}_0^{(1)}(kr) + \frac{1}{2} \mathrm{H}_0^{(1)*}(kr). &
	\label{eq:psi_S}
\end{eqnarray}
Since the probability current through a circle surrounding the origin should vanish, the amplitudes of spherical waves propagating to the origin and from the origin should have equal absolute values, so that
\begin{equation}
	\abs{ C + \frac{1}{2} } = \frac{1}{2}.
	\label{eq:C+1/2}
\end{equation}
Hence
\begin{equation}
	C = i\sin\alpha e^{i\alpha},
	\label{eq:C(alpha)}
\end{equation}
where $\alpha \in [-\pi/2,\pi/2]$ is the scattering phase. Equation \eqref{eq:C(alpha)} is a corollary of the so-called optical theorem.

In the case when the incident wave $\psi_{\mathrm{ext}}(\vec{r})$ is an arbitrary superposition of plane waves, the scattering amplitude depends only on its value at the point where the impurity is located:
\begin{equation}
	\psi(\vec{r}) = \psi_{\mathrm{ext}}(\vec{r}) + \psi_{\mathrm{ext}}(0) ie^{i\alpha} \sin \alpha \mathrm{H}_0^{(1)}(kr).
	\label{eq:psi_ext}
\end{equation}
The last relation can be written in the form
\begin{equation}
	\psi(\vec{r}) = \psi_R(\vec{r}) + \frac{2}{\pi} \ln \left( \frac{2}{kr e^{\gamma}} \right) \psi_R(0) \tan \alpha,
	\label{eq:psi_ext'}
\end{equation}
where
\begin{eqnarray}
	& \hspace{-6cm} \psi_R(\vec{r}) = \psi_{\mathrm{ext}}(\vec{r}) & \nonumber \\
	& + \psi_{\mathrm{ext}}(0) ie^{i\alpha} \sin \alpha \left[ \mathrm{H}_0^{(1)}(kr) + \frac{2i}{\pi} \ln \left( \frac{2}{kr e^{\gamma}} \right) \right]. &
	\label{eq:psi_R}
\end{eqnarray}
Note that $\psi_R(\vec{r})$ is regular at $\vec{r} = 0$.

Relations similar to Eqs. \eqref{eq:psi_ext} and \eqref{eq:psi_ext'} are valid also for Green functions that satisfy Eq. \eqref{eq:Gorkov}, because in the vicinity of the impurity (for $\abs{\vec{r} - \vec{r}_i} \ll \xi$) one may neglect the order parameter, so that the Gor'kov equation reduces to two Schr\"odinger equations. Then we can look for solutions of Eq. \eqref{eq:Gorkov} with spin indices $\uparrow \uparrow$ in the form \cite{Bespalov2018}
\begin{equation}
	G_{E\uparrow\uparrow}(\vec{r},\vec{r}') \! = \! G_E^{(0)}(\vec{r},\vec{r}') + G_E^{(0)}(\vec{r},\vec{r}_i) A_{1\uparrow} + F_{-E}^{\dagger(0)*}(\vec{r},\vec{r}_i) A_{2\uparrow},
	\label{eq:G_upup}
\end{equation}
\begin{equation}
	F_{E\uparrow\uparrow}^{\dagger}(\vec{r},\vec{r}') \! = \! F_E^{\dagger (0)}(\vec{r},\vec{r}') + F_E^{\dagger (0)}(\vec{r},\vec{r}_i) A_{1\uparrow} - G_{-E}^{(0)*}(\vec{r},\vec{r}_i) A_{2\uparrow}.
	\label{eq:F_upup}
\end{equation}
These functions satisfy the Gor'kov equation for all $\vec{r}$ and $\vec{r}'$ that are outside the range of the impurity potentials $V(\vec{r} - \vec{r}_i)$ and $\vec{J}(\vec{r} - \vec{r}_i)$. Using Eq. \eqref{eq:GER_2D} we obtain equations for $A_{1\uparrow}$ and $A_{2\uparrow}$ based on the fact that near the impurity the regular parts of the Green's functions and their logarithmic singularities should be related to each other in accordance with Eq. \eqref{eq:psi_ext'}:
\begin{widetext}
\begin{equation}
	\frac{m A_{1\uparrow}}{\pi \hbar^2} = \left[ G_E^{(0)}(\vec{r}_i,\vec{r}') + G_{ER}^{(0)}(\vec{r}_i,\vec{r}_i) A_{1\uparrow}  + F_{-E}^{\dagger(0)*}(\vec{r}_i,\vec{r}_i) A_{2\uparrow} \right] \frac{2}{\pi} \tan \alpha_{\uparrow},
	\label{eq:A1}
\end{equation}
\begin{equation}
	-\frac{m A_{2\uparrow}}{\pi \hbar^2} = \left[ F_E^{\dagger (0)}(\vec{r}_i,\vec{r}') + F_E^{\dagger (0)}(\vec{r}_i,\vec{r}_i) A_{1\uparrow} - G_{-ER}^{(0)*}(\vec{r}_i,\vec{r}_i) A_{2\uparrow} \right] \frac{2}{\pi} \tan \alpha_{\downarrow}.
	\label{eq:A2}
\end{equation}
We have taken into account here that in the case of a magnetic impurity electrons and holes feel different scattering potentials, and as a result there are two scattering phases -- $\alpha_{\uparrow}$ and $\alpha_{\downarrow}$. The solution of the linear equations \eqref{eq:A1} and \eqref{eq:A2} is straightforward, and after substituting these solutions into Eq. \eqref{eq:G_upup} we obtain 
\begin{equation}
	G_{E\uparrow\uparrow}(\vec{r},\vec{r}') = G_E^{(0)}(\vec{r},\vec{r}') + G_{E\uparrow\uparrow}^{(1)}(\vec{r},\vec{r}'),
	\label{eq:GE_+impurity}
\end{equation}
where
\begin{eqnarray}
	& G_{E\uparrow\uparrow}^{(1)}(\vec{r},\vec{r}') = {\cal D}_{\uparrow}(E + i\eps)^{-1} \left( G_E^{(0)}(\vec{r},\vec{r}_i) \left\{ G_E^{(0)}(\vec{r}_i,\vec{r}') \left[ \frac{m}{2 \hbar^2} \cot \alpha_{\downarrow} - G_{-ER}^{(0)*}(\vec{r}_i,\vec{r}_i) \right] - F_{-E}^{\dagger(0)*}(\vec{r}_i,\vec{r}_i) F_E^{\dagger(0)}(\vec{r}_i,\vec{r}') \right\} \right. & \nonumber \\
	& \left. - F_{-E}^{\dagger(0)*}(\vec{r},\vec{r}_i) \left\{ F_E^{\dagger(0)}(\vec{r}_i,\vec{r}') \left[ \frac{m}{2 \hbar^2} \cot \alpha_{\uparrow} - G_{ER}^{(0)}(\vec{r}_i,\vec{r}_i) \right] + F_E^{\dagger(0)}(\vec{r}_i,\vec{r}_i) G_E^{(0)}(\vec{r}_i,\vec{r}') \right\} \right), &
	\label{eq:GE(1)}
\end{eqnarray}
\begin{equation}
	{\cal D}_{\uparrow}(E) = \left[ \frac{m \cot \alpha_{\uparrow}}{2 \hbar^2} - G_{ER}^{(0)}(\vec{r}_i,\vec{r}_i) \right] \left[ \frac{m \cot \alpha_{\downarrow}}{2\hbar^2}
	- G_{-ER}^{(0)*}(\vec{r}_i,\vec{r}_i) \right] + F_E^{\dagger (0)} (\vec{r}_i,\vec{r}_i) F_{-E}^{\dagger (0)*} (\vec{r}_i,\vec{r}_i).
	\label{eq:D_up_2D}
\end{equation}
\end{widetext}
The Green functions $G_{ER}^{(0)}(\vec{r},\vec{r})$, $G_E^{(0)}(\vec{r},\vec{r}')$ and $F_E^{\dagger (0)} (\vec{r},\vec{r}')$ can be calculated using Eqs. \eqref{eq:GER(r,r)}, \eqref{eq:FE(r,r)}, \eqref{eq:GE_far'} and \eqref{eq:FE_far'}. To obtain $G_{E\downarrow\downarrow}(\vec{r},\vec{r}')$, one should simply swap $\uparrow$ and $\downarrow$ in Eqs. \eqref{eq:GE_+impurity} - \eqref{eq:D_up_2D}.

Of particular interest are impurity-induced poles of the Green function, which correspond to discrete impurity states. It can be seen from Eq. \eqref{eq:GE(1)} that the energies of such states with spin up satisfy the equation ${\cal D}_{\uparrow}(E) = 0$. The function ${\cal D}_{\uparrow}(E)$ is generally complex, unless the density of states  without impurity $\nu(E,\vec{r}_i)$ vanishes. Then, ${\cal D}_{\uparrow}(E)$ becomes real. Indeed, if $\nu(E,\vec{r}') = 0$ for $\vec{r}'$ lying in some area, then the imaginary term $i\eps$ in Eqs. \eqref{eq:G_BdG} and \eqref{eq:F_BdG} can be discarded, and we obtain
\begin{equation}
	G_E^{(0)}(\vec{r}',\vec{r}) = G_E^{(0)*}(\vec{r},\vec{r}'),
	\label{eq:switch_rr'}
\end{equation}
\begin{equation}
	F_{E}^{\dagger{(0)}}(\vec{r},\vec{r}') = F_{-E}^{\dagger{(0)}}(\vec{r}',\vec{r}).
	\label{eq:E_vs_-E:F}
\end{equation}
Using Eq. \eqref{eq:E_vs_-E:G}, which is valid within the quasiclassical approximation, and Eqs. \eqref{eq:D_up_2D} and \eqref{eq:E_vs_-E:F}, taking into account that $G_{ER}^{(0)}(\vec{r}_i,\vec{r}_i)$ is real when $\nu(E,\vec{r}_i) = 0$, we may obtain Eq. \eqref{eq:D_up_2D'}, from which it is obvious that ${\cal D}_{\uparrow}(E)$ is real. Thus, discrete impurity levels should be sought inside the local spectral gap at position $\vec{r}_i$.

Considering energies, for which simultaneously $\nu(E,\vec{r}_i) = 0$ and ${\cal D}_{\uparrow}(E) \neq 0$, we may find that the impurity-induced correction to the Green function with coinciding coordinates, $G_{E\uparrow\uparrow}^{(1)}(\vec{r},\vec{r})$, is real, which can be proven using Eqs. \eqref{eq:switch_rr'} and \eqref{eq:E_vs_-E:F}. For our system with a vortex this means that in the energy range $E \in (E^{(0)}(r_i),E_{\mathrm{max}}(r_i))$ the impurity only induces several discrete states (corresponding to ${\cal D}_{\uparrow}(E) = 0$), and does not affect the continuous spectrum. On the other hand, for $E \notin (E^{(0)}(r_i),E_{\mathrm{max}}(r_i))$ the function ${\cal D}_{\uparrow}(E)$ is complex, and so is $G_{E\uparrow\uparrow}^{(1)}(\vec{r},\vec{r})$, which means a modification of the continuous spectrum in this energy range. In particular, this results in the local spectral gap for all positions $\vec{r}$ lying inside the interval $(E^{(0)}(r_i),E_{\mathrm{max}}(r_i))$.

\section{Wavefunctions of impurity states}
\label{app:wavefunctions}

In this Appendix we will determine the wavefunctions of the impurity states and derive a formula suitable for numerical calculations of these wavefunctions.

To complete our task, we will calculate the Green functions $G_{E\uparrow \uparrow}(\vec{r},\vec{r}')$ and $F_{E\uparrow \uparrow}^{\dagger}(\vec{r},\vec{r}')$ near their poles, corresponding to impurity states.
Let $E = E_{\uparrow i}$ be the energy of a pole of the Green functions, so that ${\cal D}_{\uparrow}(E_{\uparrow i}) = 0$. Then, using Eqs. \eqref{eq:switch_rr'} and \eqref{eq:E_vs_-E:F}, for the function $G_{E\uparrow \uparrow}(\vec{r},\vec{r}')$ [Eqs. \eqref{eq:GE_+impurity} - \eqref{eq:D_up_2D}] at $E \approx E_{\uparrow i}$ we obtain
\begin{equation}
	G_{E\uparrow \uparrow}(\vec{r},\vec{r}') \approx \frac{u_{\uparrow i}(\vec{r}) u^*_{\uparrow i}(\vec{r}')}{E_{\uparrow i} - E - i\eps},
	\label{eq:G_E=Ei}
\end{equation}
where
\begin{equation}
	u_{\uparrow i} (\vec{r}) = A_{\uparrow i} G_{E_{\uparrow i}}^{(0)} (\vec{r},\vec{r}_i) - B_{\uparrow i} F_{-E_{\uparrow i}}^{\dagger (0)*} (\vec{r},\vec{r}_i),
	\label{eq:u_up}
\end{equation}
\begin{equation}
	A_{\uparrow i} = \sqrt{-\left[ G_{E_{\uparrow i} R}^{(0)} (\vec{r}_i,\vec{r}_i) + \frac{m}{2\hbar^2} \cot \alpha_{\downarrow} \right] \left[ \frac{d {\cal D}_{\uparrow}}{dE} (E_{i\uparrow}) \right]^{-1} },
	\label{eq:A_up}
\end{equation}
\begin{eqnarray}
	&B_{\uparrow i} = -F_{E_{\uparrow i}}^{\dagger (0)} (\vec{r}_i,\vec{r}_i) \sgn \left(\frac{d {\cal D}_{\uparrow}}{dE} (E_{i\uparrow}) \right)  & \nonumber \\
	& \times \left\{ -\left[ G_{E_{\uparrow i} R}^{(0)} (\vec{r}_i,\vec{r}_i) + \frac{m}{2\hbar^2} \cot \alpha_{\downarrow} \right] \frac{d {\cal D}_{\uparrow}}{dE} (E_{i\uparrow}) \right\}^{-1/2} \!\!\!\!\!\!\!\!\!\!\!\!. &
	\label{eq:B_up}
\end{eqnarray}
Since an expansion of the form \eqref{eq:G_BdG} is also valid for $G_{E\uparrow \uparrow}(\vec{r},\vec{r}')$, we conclude that $u_{\uparrow i} (\vec{r})$ is the electron component of the wavefunction of an impurity state. The hole component can be obtained from an expansion of the form \eqref{eq:G_E=Ei} for the function $F_{E\uparrow\uparrow}^{\dagger}(\vec{r},\vec{r}')$:
\begin{equation}
	v_{\uparrow i} (\vec{r}) = A_{\uparrow i} F_{E_{\uparrow i}}^{\dagger (0)} (\vec{r},\vec{r}_i) + B_{\uparrow i} G_{-E_{\uparrow i}}^{(0)*} (\vec{r},\vec{r}_i).
	\label{eq:v_up}
\end{equation}
The functions $G_{E}^{(0)} (\vec{r},\vec{r}_i)$ and $F_{E}^{\dagger (0)} (\vec{r},\vec{r}_i)$ can be calculated using Eqs. \eqref{eq:GE_far'} and \eqref{eq:FE_far'}. 
Taking into account the symmetry relations
\begin{equation}
	\tilde{g}_{-E}(\vec{r}',-\vec{n},-s) = - \tilde{g}_E (\vec{r}',\vec{n},s),
	\label{eq:tg_symmetry}
\end{equation}
\begin{equation}
	\tilde{f}_{-E}^{\dagger} (\vec{r}',-\vec{n},-s) = \tilde{f}^{\dagger}_E (\vec{r}',\vec{n},s),
	\label{eq:tf_symmetry}
\end{equation}
which follow from Eqs. \eqref{eq:Andreev_tg} and \eqref{eq:Andreev_tf}, we obtain
\begin{widetext}
\begin{eqnarray}
	& u_{\uparrow i}(\vec{r}) = \frac{mi}{\hbar^2} \sqrt{\frac{1}{2\pi k_F \abs{\vec{r} - \vec{r}_i}}} \left\{ \left[ A_{\uparrow i} \tilde{g}_{E_{\uparrow i}}(\vec{r}_i,\vec{n},\abs{\vec{r}-\vec{r}_i}) + B_{\uparrow i} \tilde{f}_{E_{\uparrow i}}^{\dagger *} (\vec{r}_i,\vec{n},\abs{\vec{r}-\vec{r}_i}) \right] e^{ik_F\abs{\vec{r} - \vec{r}_i} - i\frac{\pi}{4}} \right. & \nonumber \\
	& \left. + \left[ A_{\uparrow i} \tilde{g}_{E_{\uparrow i}}(\vec{r}_i,-\vec{n},-\abs{\vec{r}-\vec{r}_i}) + B_{\uparrow i} \tilde{f}_{E_{\uparrow i}}^{\dagger *} (\vec{r}_i,-\vec{n},-\abs{\vec{r}-\vec{r}_i}) \right] e^{-ik_F\abs{\vec{r} - \vec{r}_i} + i\frac{\pi}{4}} \right\}, & \label{eq:u_up_qq} \\
	& v_{\uparrow i}(\vec{r}) = \frac{mi}{\hbar^2} \sqrt{\frac{1}{2\pi k_F \abs{\vec{r} - \vec{r}_i}}} \left\{ \left[ A_{\uparrow i} \tilde{f}_{E_{\uparrow i}}^{\dagger}(\vec{r}_i ,\vec{n},\abs{\vec{r}-\vec{r}_i}) + B_{\uparrow i} \tilde{g}_{E_{\uparrow i}}^* (\vec{r}_i,\vec{n},\abs{\vec{r}-\vec{r}_i}) \right] e^{ik_F\abs{\vec{r} - \vec{r}_i} - i\frac{\pi}{4}} \right. & \nonumber \\
	& \left. + \left[ A_{\uparrow i} \tilde{f}_{E_{\uparrow i}}^{\dagger}(\vec{r}_i,-\vec{n},-\abs{\vec{r}-\vec{r}_i}) + B_{\uparrow i} \tilde{g}_{E_{\uparrow i}}^* (\vec{r}_i,-\vec{n},-\abs{\vec{r}-\vec{r}_i}) \right] e^{-ik_F\abs{\vec{r} - \vec{r}_i} + i\frac{\pi}{4}} \right\}, & \label{eq:v_up_qq} 
\end{eqnarray}
where $\vec{n} = (\vec{r}-\vec{r}_i)/|(\vec{r}-\vec{r}_i)|$. One can see that the functions $u_{\uparrow i}(\vec{r})$ and $v_{\uparrow i}(\vec{r})$ oscillate in space with a period of the order of $k_F^{-1}$. After averaging their squared absolute values over an oscillation period we have
\begin{eqnarray}
	& \mean{\abs{u_{\uparrow i}(\vec{r})}^2} = \frac{m^2}{2\pi \hbar^4 k_F \abs{\vec{r} - \vec{r}_i}} \left[ \abs{ A_{\uparrow i} \tilde{g}_{E_{\uparrow i}}(\vec{r}_i,\vec{n},\abs{\vec{r}-\vec{r}_i}) + B_{\uparrow i} \tilde{f}_{E_{\uparrow i}}^{\dagger *} (\vec{r}_i,\vec{n},\abs{\vec{r}-\vec{r}_i})}^2 \right. & \nonumber \\
	& \left. + \abs{ A_{\uparrow i} \tilde{g}_{E_{\uparrow i}}(\vec{r}_i,-\vec{n},-\abs{\vec{r}-\vec{r}_i}) + B_{\uparrow i} \tilde{f}_{E_{\uparrow i}}^{\dagger *} (\vec{r}_i,-\vec{n},-\abs{\vec{r}-\vec{r}_i})}^2 \right], & \label{eq:mean(u2)} \\
	& \mean{\abs{v_{\uparrow i}(\vec{r})}^2} = \frac{m^2}{2\pi \hbar^4 k_F \abs{\vec{r} - \vec{r}_i}}  \left[ \abs{ A_{\uparrow i} \tilde{f}_{E_{\uparrow i}}^{\dagger}(\vec{r}_i,\vec{n},\abs{\vec{r}-\vec{r}_i}) + B_{\uparrow i} \tilde{g}_{E_{\uparrow i}}^* (\vec{r}_i,\vec{n},\abs{\vec{r}-\vec{r}_i}) }^2 \right. & \nonumber \\
	& \left. + \abs{ A_{\uparrow i} \tilde{f}_{E_{\uparrow i}}^{\dagger}(\vec{r}_i,-\vec{n},-\abs{\vec{r}-\vec{r}_i}) + B_{\uparrow i} \tilde{g}_{E_{\uparrow i}}^* (\vec{r}_i,-\vec{n},-\abs{\vec{r}-\vec{r}_i}) }^2 \right]. & \label{eq:mean(v2)} 
\end{eqnarray}
\end{widetext}
It follows from Eqs. \eqref{eq:Andreev_tg} and \eqref{eq:Andreev_tf} that
\begin{equation}
	\frac{\partial}{\partial s} \left[ \abs{\tilde{g}_E(\vec{r}',\vec{n},s)}^2 - \abs{\tilde{f}_E^{\dagger}(\vec{r}',\vec{n},s)}^2 \right] = 0
	\label{eq:|g|2-|f|2}
\end{equation}
for $s \neq 0$. Due to the vanishing of $\tilde{g}$ and $\tilde{f}^{\dagger}$ at $s \to \pm \infty$, we have
\begin{equation}
	\abs{\tilde{g}_E(\vec{r}',\vec{n},s)} = \abs{\tilde{f}_E^{\dagger}(\vec{r}',\vec{n},s)}.
	\label{eq:|g|=|f|}
\end{equation}
Equations \eqref{eq:mean(u2)} and \eqref{eq:mean(v2)} together with Eq. \eqref{eq:|g|=|f|} yield Eq. \eqref{eq:mean=mean}.

The remainder of this Appendix is purely technical and is devoted to numerical calculations of $\mean{\abs{u_{\uparrow i}(\vec{r})}^2}$.

For a start, let us write the main relations in dimensionless form. Like in Sec. \ref{sec:clean}, we use $\Delta_{\infty}$ as energy units and $\hbar v_F/\Delta_{\infty}$ as units of length. The functions $G_E$ and $F_E^{\dagger}$ will be written in units of $\pi \nu_0$, and $\mean{\abs{u_{\uparrow i}(\vec{r})}^2}$ -- in units of $\Delta_{\infty}^2/(\pi \hbar^2 v_F^2)$. In the equations for $\tilde{g}_E(\vec{r}_i,\vec{n},s)$ and $\tilde{f}_E^{\dagger}(\vec{r}_i,\vec{n},s)$ let us shift the origin, so that $s=0$ corresponds to the point on the trajectory that is closes to the vortex center (like in Sec. \ref{sec:clean}). Then we have the following set of equations:
\begin{widetext}
\begin{eqnarray}
	& \mean{\abs{u_{\uparrow i}(\vec{r})}^2} = \frac{1}{\abs{\vec{r} - \vec{r}_i}} \left[ \abs{ A_{\uparrow i} \tilde{g}_{E_{\uparrow i}}(\vec{r}_i,\vec{n},s_i(\vec{n})+\abs{\vec{r}-\vec{r}_i}) + B_{\uparrow i} \tilde{f}_{E_{\uparrow i}}^{\dagger *} (\vec{r}_i,\vec{n},s_i(\vec{n}) + \abs{\vec{r}-\vec{r}_i})}^2 \right. & \nonumber \\
	& \left. + \abs{ A_{\uparrow i} \tilde{g}_{E_{\uparrow i}}(\vec{r}_i,-\vec{n},s_i(-\vec{n}) - \abs{\vec{r}-\vec{r}_i}) + B_{\uparrow i} \tilde{f}_{E_{\uparrow i}}^{\dagger *} (\vec{r}_i,-\vec{n},s_i(-\vec{n}) -\abs{\vec{r}-\vec{r}_i})}^2 \right], & \label{eq:mean(u2)'}
\end{eqnarray}
\end{widetext}
\begin{equation}
	- i \frac{\partial \tilde{g}_E}{\partial s} - E \tilde{g}_E + \Delta(\vec{r}_i + (s-s_i) \vec{n}) \tilde{f}^{\dagger}_E = -i \delta(s-s_i),
	\label{eq:Andreev_tg'}
\end{equation}
\begin{equation}
	i\frac{\partial \tilde{f}_E^\dagger}{\partial s} - E \tilde{f}_E^\dagger + \Delta^*(\vec{r}' + (s-s_i)\vec{n}) \tilde{g}_E = 0,
	\label{eq:Andreev_tf'}
\end{equation}
where $s_i(\vec{n})$ is the coordinate of the impurity on the trajectory, and the coefficients $A_{\uparrow i}$ and $B_{\uparrow i}$ are in the dimensionless form.

It follows from Eqs. \eqref{eq:Andreev_tg'} and \eqref{eq:Andreev_tf'} that for $s\neq s_i$ the ratio $i \tilde{g}_E/\tilde{f}_E^{\dagger}$ satisfies Eq. \eqref{eq:a}, and the ratio $i \tilde{f}_E^{\dagger}/\tilde{g}_E$ satisfies Eq. \eqref{eq:b}. Moreover, from Eqs. \eqref{eq:tg+_bound} and \eqref{eq:tf+_bound} we obtain
\begin{equation}
	i \frac{\tilde{f}_E^{\dagger}(\vec{r}_i,\vec{n},s_i)}{\tilde{g}_E(\vec{r}_i,\vec{n},s_i+0)} = i \frac{f_E^{\dagger}(\vec{r}_i,\vec{n})}{g_E(\vec{r}_i,\vec{n})+1} = b(s_i).
	\label{eq:b(si)}
\end{equation}
Similarly, one finds that
\begin{equation}
	i \frac{\tilde{g}_E(\vec{r}_i,\vec{n},s_i-0)}{\tilde{f}_E^{\dagger}(\vec{r}_i,\vec{n},s_i)} = a(s_i).
	\label{eq:a(si)}
\end{equation}
Then, by virtue of the uniqueness theorem for the solution of the Cauchy problem for ordinary differential equations,
\begin{equation}
	i \frac{\tilde{g}_E(\vec{r}_i,\vec{n},s)}{\tilde{f}_E^{\dagger}(\vec{r}_i,\vec{n},s)} = a(s) \qquad \mbox{for } s<s_i,
	\label{eq:tilde_vs_a}
\end{equation}
\begin{equation}
	i \frac{\tilde{f}_E^{\dagger}(\vec{r}_i,\vec{n},s)}{\tilde{g}_E(\vec{r}_i,\vec{n},s)} = b(s) \qquad \mbox{for } s>s_i.
	\label{eq:tilde_vs_b}
\end{equation}
It turns out that to calculate the density of states, it is enough to calculate only the function $\tilde{g}_E$. Indeed, from Eqs. \eqref{eq:Andreev_tg'} and \eqref{eq:Andreev_tf'} with Eq. \eqref{eq:|g|=|f|} we obtain
\begin{equation}
	\frac{\partial}{\partial s} \left( \frac{\tilde{g}_E}{\tilde{f}_E^{\dagger *}} \right) = \frac{\delta(s)}{\tilde{f}_E^{\dagger *}}.
	\label{eq:g/f*}
\end{equation}
Hence,
\begin{equation}
	\frac{\tilde{f}_E^{\dagger *}(\vec{r}_i,\vec{n},s)}{\tilde{g}_E (\vec{r}_i,\vec{n},s)} = \frac{\tilde{f}_E^{\dagger *}(\vec{r}_i,\vec{n},s_i)}{\tilde{g}_E (\vec{r}_i,\vec{n},s_i+0)} = \frac{f_E^{\dagger *}(\vec{r}_i,\vec{n})}{1 + g_E(\vec{r}_i,\vec{n})} 
	\label{eq:f*/g(+0)}
\end{equation}
for $s>s_i$, and
\begin{equation}
	\frac{\tilde{f}_E^{\dagger *}(\vec{r}_i,\vec{n},s)}{\tilde{g}_E (\vec{r}_i,\vec{n},s)} = \frac{\tilde{f}_E^{\dagger *}(\vec{r}_i,\vec{n},s_i)}{\tilde{g}_E (\vec{r}_i,\vec{n},s_i-0)} = \frac{f_E^{\dagger *}(\vec{r}_i,\vec{n})}{-1 + g_E(\vec{r}_i,\vec{n})}
	\label{eq:f*/g(-0)}
\end{equation}
for $s<s_i$. By substituting $\tilde{f}_E^{\dagger *}(\vec{r}_i,\vec{n},s)$ from Eqs. \eqref{eq:f*/g(+0)} and \eqref{eq:f*/g(-0)} into Eq. \eqref{eq:mean(u2)'} we have
\begin{widetext}
\begin{eqnarray}
	& \mean{\abs{u_{\uparrow i}(\vec{r})}^2} = \frac{1}{\abs{\vec{r} - \vec{r}_i}} \left[ \abs{ A_{\uparrow i}  + B_{\uparrow i} \frac{f_{E_{\uparrow i}}^{\dagger *}(\vec{r}_i,\vec{n})}{1 + g_{E_{\uparrow i}}(\vec{r}_i,\vec{n})}}^2 \abs{\tilde{g}_{E_{\uparrow i}}(\vec{r}_i,\vec{n},s_i(\vec{n}) +\abs{\vec{r}-\vec{r}_i})}^2 \right. & \nonumber \\
	& \left. + \abs{ A_{\uparrow i}  + B_{\uparrow i} \frac{f_{E_{\uparrow i}}^{\dagger *}(\vec{r}_i,-\vec{n})}{-1 + g_{E_{\uparrow i}}(\vec{r}_i,-\vec{n})}}^2 \abs{\tilde{g}_{E_{\uparrow i}}(\vec{r}_i,-\vec{n},s_i(-\vec{n}) - \abs{\vec{r}-\vec{r}_i})}^2 \right]. & \label{eq:mean(u2)''}
\end{eqnarray}
\end{widetext}

Let us focus on calculating the function $\tilde{g}_E$. We put the impurity in a position with coordinates $\vec{r}_i = (r_i,0)$. Then, the coordinate of the impurity on a trajetory with direction vector $\vec{n}$ and the impact parameter of this trajectory are [see Fig. \ref{fig:2DFrame}]
\begin{equation}
	s_i(\varphi) \equiv s_i(\vec{n}) = -r_i \cos \varphi, \qquad  d(\varphi) \equiv d(\vec{n}) = r_i \sin \varphi.
	\label{eq:sidi}
\end{equation}
For brevity, we will perform further calculations for a coreless vortex: $\abs{\Delta(r)} = \mathrm{const}$. All the following considerations are easily generalized to the case of a vortex with core.

Using Eqs. \eqref{eq:a_psi} and \eqref{eq:tilde_vs_a}, for $s<s_i$ we may rewrite Eq. \eqref{eq:Andreev_tg'} in the form
\begin{equation}
	-i \frac{\partial \tilde{g}_E}{\partial s} - E \tilde{g}_E + e^{-i\psi_d(s)} \tilde{g}_E = 0.
	\label{eq:tg_psi}
\end{equation}
For $\abs{\tilde{g}_E}^2$ we obtain
\begin{equation}
	\frac{\partial \abs{\tilde{g}_E}^2}{\partial s} + 2 \sin(\psi_d(s)) \abs{\tilde{g}_E}^2 = 0.
	\label{eq:abs(tg)_psi}
\end{equation}
The solution of this equation has the form
\begin{eqnarray}
	& \abs{\tilde{g}_E (s)}^2 = \abs{\tilde{g}_E (s_i-0)}^2 \exp \left( - 2\int_{s_i}^s \sin \psi_d(s') ds' \right) & \nonumber \\
	& = \frac{\exp \left( - 2\int_{s_i}^s \sin \psi_d(s') ds' \right)}{4 \sin^2 \left( \frac{\psi_d(s_i) + \psi_d(-s_i)}{2} \right)} \qquad \mbox{for } s<s_i. &
	\label{eq:abs(tg)_s<s_i}
\end{eqnarray}
For $s>s_i$ using Eqs. \eqref{eq:b_psi} and \eqref{eq:tilde_vs_b} we rewrite Eq. \eqref{eq:Andreev_tg'} in the form
\begin{equation}
	-i \frac{\partial \tilde{g}_E}{\partial s} - E \tilde{g}_E + e^{i\psi_d(-s)} \tilde{g}_E = 0.
	\label{eq:tg_psi(-s)}
\end{equation}
From this we find
\begin{equation}
	\abs{\tilde{g}_E (s)}^2 = \frac{\exp \left( 2\int_{s_i}^s \sin \psi_d(-s') ds' \right)}{4 \sin^2 \left( \frac{\psi_d(s_i) + \psi_d(-s_i)}{2} \right)} \qquad \mbox{for } s>s_i.
	\label{eq:abs(tg)_>s_i}
\end{equation}
Now, in Eq. \eqref{eq:mean(u2)''} we can express all Green functions in terms of $\psi_d(s)$:
\begin{widetext}
\begin{eqnarray}
	& \mean{\abs{u_{\uparrow i}(\vec{r})}^2} = \frac{1}{\abs{\vec{r} - \vec{r}_i}} \left[ \frac{\abs{ A_{\uparrow i}  - B_{\uparrow i} e^{i \psi_{d}(s_i)}}^2}{4 \sin^2 \left( \frac{\psi_{d}(s_i) + \psi_{d}(-s_i)}{2} \right)}  \exp \left( 2\int\limits_{s_i}^{s_i + \abs{\vec{r} - \vec{r}_i}} \sin \psi_{d}(-s') ds' \right) \right. & \nonumber \\
	& \left. + \frac{\abs{ A_{\uparrow i}  - B_{\uparrow i} e^{-i \psi_{-d}(s_i)}}^2}{4 \sin^2 \left( \frac{\psi_{-d}(s_i) + \psi_{-d}(-s_i)}{2} \right)} \exp \left( -2\int\limits_{-s_i}^{-s_i - \abs{\vec{r} - \vec{r}_i}} \sin \psi_{-d}(s') ds' \right) \right]. & \label{eq:mean(u2)_psi}
\end{eqnarray}
We imply here $d = d(\vec{n})$, $s_i = s_i(\vec{n})$, and we took into account that $d(-\vec{n}) = -d(\vec{n})$, $s_i(-\vec{n}) = -s_i(\vec{n})$. Finally, substituting here explicit expressions for $A_{\uparrow i}$ and $B_{\uparrow i}$ [Eqs. \eqref{eq:A_up} and \eqref{eq:B_up}], taking into account that $F_E^{\dagger(0)}(\vec{r}_i,\vec{r}_i)$ is real in our case, we obtain
\begin{eqnarray}
	& \mean{\abs{u_{\uparrow i}(\vec{r})}^2} = \frac{1}{4 \abs{\vec{r} - \vec{r}_i} \frac{d {\cal D}_{\uparrow}}{dE}(E_{\uparrow i})} \left[ \frac{\cot \alpha_{\uparrow} - \cot \alpha_{\downarrow} - 2 G_{E_{\uparrow i} R}^{(0)}(\vec{r}_i,\vec{r}_i) + 2F_{E_{\uparrow i}}^{\dagger(0)}(\vec{r}_i,\vec{r}_i) \cos \psi_d(s_i)}{\sin^2 \left( \frac{\psi_{d}(s_i) + \psi_{d}(-s_i)}{2} \right)}  \exp \left( 2\int\limits_{-s_i - \abs{\vec{r} - \vec{r}_i}}^{-s_i} \sin \psi_{d}(s') ds' \right) \right. & \nonumber \\
	& \left. + \frac{\cot \alpha_{\uparrow} - \cot \alpha_{\downarrow} - 2 G_{E_{\uparrow i} R}^{(0)}(\vec{r}_i,\vec{r}_i) + 2F_{E_{\uparrow i}}^{\dagger(0)}(\vec{r}_i,\vec{r}_i) \cos \psi_{-d}(s_i)}{\sin^2 \left( \frac{\psi_{-d}(s_i) + \psi_{-d}(-s_i)}{2} \right)} \exp \left( 2\int\limits_{-s_i - \abs{\vec{r} - \vec{r}_i}}^{-s_i} \sin \psi_{-d}(s') ds' \right) \right]. & \label{eq:mean(u2)_final}
\end{eqnarray}
This equation has been used for numerical calculations of the wavefunction of the impurity state.
\end{widetext}

%\bibliographystyle{aipnum4-1}
%\bibliography{Vortex}
%

\end{document}